\documentclass{jfm}
\usepackage{graphicx}
\usepackage{epstopdf, epsfig}

\usepackage{amsmath,amsfonts,mathtools}
\usepackage{float}
\usepackage{subfigure}
\usepackage{dcolumn}
\usepackage{bm}
\usepackage[utf8]{inputenc}
\usepackage[T1]{fontenc}
\usepackage{breqn}
\usepackage{bigints}
\usepackage[normalem]{ulem}
\usepackage{color}
\usepackage[colorlinks=true,linkcolor=blue,allcolors=blue]{hyperref}
\usepackage{array}
\newcolumntype{P}[1]{>{\centering\arraybackslash}p{#1}}
\usepackage{makecell}
\usepackage[table]{xcolor}

\shorttitle{Dilute dispersion of compound particles: deformation dynamics and rheology}
\shortauthor{Pavan Kumar Singeetham, Chaithanya K. V. S. and Sumesh P. Thampi}

\title{Dilute dispersion of compound particles: deformation dynamics and rheology}

\author{Pavan Kumar Singeetham,
	Chaithanya K. V. S., 
	\and Sumesh P. Thampi \corresp{\email{sumesh@iitm.ac.in}} }

\affiliation{Department of Chemical Engineering, Indian Institute of Technology Madras, Chennai-36, India.}

\begin{document}
	
	\maketitle
	\begin{abstract}
		Compound particles are a class of composite systems in which solid particles encapsulated in a fluid droplet are suspended in another fluid. They are encountered in various natural and biological processes, for \textit{e.g.}, nucleated cells, hydrogels, microcapsules etc. Generation and transportation of such multiphase structures in microfluidic devices is associated with several challenges because of the poor understanding of their structural stability in a background flow and the rheological characteristics of their dispersions. Hence, in this work, we analyze the flow in and around a concentric compound particle and investigate the deformation dynamics of the confining drop and its stability against breakup in imposed linear flows. In the inertia-less limit (Reynolds number, $\Rey<<1$) and assuming that the surface tension force dominates the viscous forces (low capillary number, $Ca$, limit), we obtain analytical expressions for the velocity and pressure fields upto $\textit{O}(Ca)$ for a compound particle subjected to a linear flow using a domain perturbation technique. Simultaneously, we determine the deformed shape of the confining drop correct upto $\textit{O}(Ca^2)$ facilitating the following. (i) Since $\textit{O}(Ca^2)$ calculations account for the rotation of the anisotropically deformed interface, the reorientation dynamics of the deformed compound particles is determined. (ii) Calculations involving $\textit{O}(Ca^2)$ shape of the confining interface are found to be important for compound particles as $\textit{O}(Ca)$ calculations make qualitatively different predictions in generalised extensional flows. (iii) An $\textit{O}(Ca)$ constitutive equation for the volume-averaged stress for a dilute dispersion of compound particles was developed to study both shear and extensional rheology in a unified framework. Our analysis shows that the presence of an encapsulated particle always enhances all the measured rheological quantities such as the effective shear viscosity, extensional viscosity, and normal stress differences. (iv) Moreover, linear viscoelastic behavior of a dilute dispersion of compound particles is characterized in terms of complex modulus by subjecting the dilute dispersion to a small amplitude oscillatory shear (SAOS) flow. (v) Various expressions pertaining to a suspension of particles, drops, and particles coated with a fluid film are also derived as limiting cases of compound particles. 
		\end{abstract}
	
	
	\section{Introduction}
	\label{intro}
	Compound particles are particles confined in a fluid drop. Suspensions of these complex and multiphase structures are encountered in petroleum, food, and pharmaceutical processing industries \citep{bird1987,barnes1989,tadros2011,jia2020}, in  biological and soft matter systems \citep{choe2018, gasperini2014, wen2015, zhang2017, reigh2017, wisdom2013,somerville2020}. While several classical and seminal works have focused on the rheological characterisation of a suspension of particles \citep{einstein1906,einstein1911} or fluid droplets \citep{taylor1932,stone1990a,stone1994}, relatively less attention has been given to study the stability and rheology of composite systems, namely multiphase structures consisting of solid particles and fluid droplets together such as compound particles. Therefore, in this work, we theoretically investigate the fluid flow in and around a compound particle, the deformation dynamics of the confining drop and its stability against breakup in an imposed flow. Further, we characterize a dilute dispersion of compound particles by determining its rheology in terms of the effective viscosity, normal stress differences and complex modulus. 
	
	Several numerical \citep{kawano1997,smith2004,chen2013,chen2015a,chen2015b,kim2017,hua2014,Patlazhan2015} and theoretical \citep{mori1978,johnson1981,harper1982,johnson1985,sadhal1997,sagis2013} studies have addressed the hydrodynamics and the deformation dynamics of composite systems in an imposed flow.  \cite{johnson1981} investigated the creeping flow past a rigid sphere coated with a thin film. Later, \cite{sadhal1983,johnson1983} extended this study to a fluid droplet coated with a thin film and calculated the modification to the drag force due to a thin coating. \cite{rushton1983} theoretically investigated the translational dynamics of a compound droplet, a multiphase system where a droplet is confined in another droplet, either concentrically or eccentrically \citep{sadhal1985,qu2012}. The dynamics of a compound drop is also influenced by externally imposed flows. For example, \cite{davis1981} investigated the steady state deformation dynamics of a concentric compound particle in a shear flow. \cite{stone1990a} generalised this study to understand compound droplets. They also performed boundary integral method based numerical simulations and analyzed the breakup of compound droplets in a linear flow. \cite{kim2017} numerically investigated the time evolution of eccentric compound droplets subjected to a simple shear flow. The effects of inertia \citep{bazhlekov1995}, viscoelasticity \citep{zhou2006}, surfactant laden interfaces \citep{xu2013,zhang2015,hamedi2010,srinivasan2014,mandal2016}, confinement \citep{song2010} and electric field \citep{soni2018,santra2020,santra2020a} etc., on compound drops have also been addressed and thus, the literature on compound drops is aplenty. However the presence of three fluids and multiple deforming interfaces make the analysis of compound drops cumbersome. On the other hand, analytical calculations with compound particles though a subset of compound drops, (i) are  tractable relatively easily and (ii) demonstrate the strong hydrodynamic response such as interface deformation that results from the interaction between the encapsulated solid particle and the confining fluid interface. Therefore, the analysis of compound particles provide useful physical insights into the deformation dynamics of the interface and rheological response of dispersions containing multiphase structures as we illustrate in this work. Table \ref{table1} summarizes and distinguishes the present work from the previous theoretical studies which analyzed the deformation dynamics of composite systems in an imposed flow. The bottom up approach of microstructure based rheology determination dates back to the classical calculation of effective viscosity of dilute suspensions of single phase constituents such as particles and droplets \citep{einstein1906,einstein1911,taylor1932}. Further modifications, for example, to analyze the effect of insoluble surfactants on the rheology of dilute emulsions have been addressed numerically \citep{li1997}. Several works to incorporate such effects \citep{pal1996a,zhao2002,vlahovska2009,mandal2017} and to relax the assumption of diluteness \citep{loewenberg1996,loewenberg1998,jansen2001,golemanov2008} to predict the rheology of concentrated emulsions are available in literature but attempts to understand the rheological behavior of dispersions of compound particles or compound droplets are scarce \citep{johnson1985,stone1990a,pal1996b,pal2007,pal2011,mandal2016,santra2020a,das2020}. 

	\begin{table}\scriptsize
    \centering
    \begin{tabular}{P{1.75cm}  P{1.75cm}   P{1.75cm}   P{1.75cm}  P{1.75cm}  P{2.5cm} }
	Author (year)& System & Background flow & \makecell{Accuracy \\of  flow field,\\ $\bm{u}$,  $p$ }&  \makecell[l]{Rheology }& Remarks / Comments \\ \hline
		\cite{davis1981} &   Compound particle & Shear flow& $O(1)$  & Newtonian & \makecell[l]{$O(Ca)$ steady state\\ deformation dynamics}  \\ \hline
		\cite{stone1990a} &   Compound  droplet & Linear flow &  $O(1)$ & Newtonian &   \makecell[l]{$O(Ca)$ steady state\\ deformation dynamics}\\
	   	\hline
		\cite{mandal2016} & Compound  droplet (Concentric) & Arbitrary background flow + Effects of surfactants  &  $O(1)$  &  Newtonian &  \makecell[l]{$O(Ca)$ steady state\\ deformation dynamics}		\\ 
		\\
		&  Eccentric &  Poiseuille  & $O(1)$  & $\rm{X}$   & \makecell{Migration dynamics}
		\\ \hline
		\cite{chaithu2019} &   Compound particle &  Linear flow&   $O(1)$  & \rm{X} & \makecell[l]{The rotational and\\ translational dynamics, \\and transient $O(Ca)$\\ deformation dynamics}  \\ \hline
		
		\cite{santra2020a}  &   Compound droplet & Uniaxial flow + Effects of electric field& $O(Ca)$ & non-Newtonian & \makecell[l]{$O(Ca^2)$ steady state \\deformation dynamics, \\ and extensional rheology}	\\ \hline
	\cite{das2020}  &   Compound droplet & Linear flow + Effects of surfactants & $O(1)$ & Newtonian & \makecell[l]{$O(Ca)$ steady state\\ deformation dynamics }  \\ \hline
		\textbf{Present study}&   Compound particle & Linear flow &  $O(Ca)$    & non-Newtonian  & \makecell[l]{The transient $O(Ca^2)$\\deformation dynamics,\\and linear viscoelastic\\behavior}  \\ \hline
		\end{tabular}
\caption{A summary of theoretical works in the past that analyzed the deformation dynamics of composite systems in a background flow, and the rheology of their dilute dispersion.}\label{table1}
\end{table} 
	
	In the present work, we analyze the deformation dynamics of a concentric compound particle in an imposed linear flow using an asymptotic expansion in capillary number, $Ca$. This is interesting because deformation dynamics is usually analyzed only up to $\textit{O}(Ca)$ and it may be interesting to understand the deviations predicted by $\textit{O}(Ca^{2})$ calculations. The $O(1)$ velocity field and $O(Ca)$ deformation dynamics of a concentric compound particle in a linear flow is already reported in \cite{chaithu2019}. In contrast, the present work investigates the $O(Ca)$ velocity field, and $O(Ca^2)$ deformation dynamics of a concentric compound particle. It has been found that the higher order calculations are particularly relevant for compound particles, as they predict qualitatively different behavior compared to the leading order calculations as discussed in \S\ref{example}. The present work is also able to analyze the effect of rotation of the deformed compound particle on the deformation dynamics, which is neglected in \cite{chaithu2019}. Moreover, using these calculations we develop a constitutive equation for volume averaged stress and characterize the rheological response of a dilute dispersion of compound particles. Both shear and extensional rheology are analyzed in a single framework by quantifying the effective shear viscosity, extensional viscosity, and normal stress differences.  We also analyze, for the first time, the linear viscoelastic behavior (of the dilute dispersion) in terms of complex modulus by subjecting the compound particle to a small amplitude oscillatory shear (SAOS) flow. The approach followed in this paper is similar to the work by \cite{leal2007,arun2012} in the context of a fluid droplet in a linear flow. 
	
	This paper is organized as follows. In \S\ref{mathform}, we present the mathematical formulation and present the analytical solutions using a domain perturbation technique with capillary number, $Ca$ as the small parameter. We thus calculate the velocity and pressure fields upto $\textit{O}(Ca)$ using the standard technique of superposition of vector harmonics and then determine the consequence of this flow field, namely the shape of the confining drop corrected upto $\textit{O}(Ca^2)$. In \S\ref{example}, we analyze the deformation of the confining drop, and contrast the dynamics obtained by the $\textit{O}(Ca)$ and $\textit{O}(Ca^2)$ calculations. Then in \S\ref{rheology}, we determine the rheology of a dilute dispersion of compound particles in terms of shear and extensional viscosities, normal stress differences, and the complex modulus. 

	\section{Mathematical formulation}\label{mathform}
\begin{figure}
		\centering
		\includegraphics[scale=0.45]{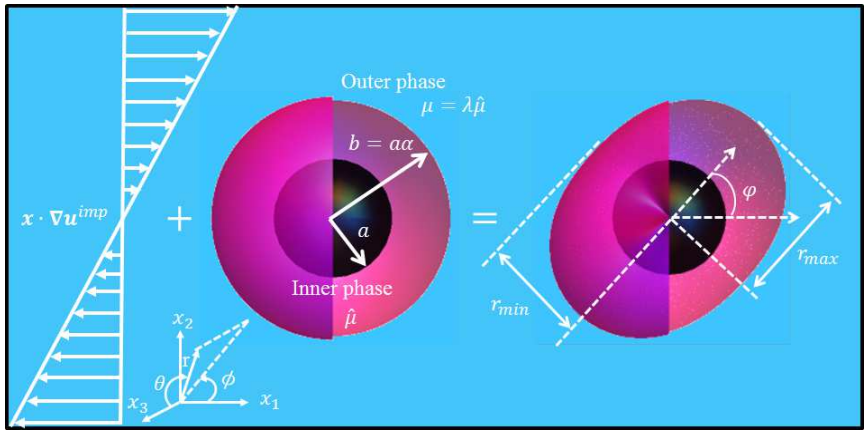}
		\caption{A schematic of the compound particle consisting of a solid sphere of radius $a$ encapsulated in a drop of radius $b = a \alpha$. The confining interface deforms when subjected to an imposed flow $\bm{u}^{imp}$. The longest and shortest dimensions of the deformed interface are respectively indicated by $r_{max}$ and $r_{min}$. If the imposed flow is a simple shear flow, the alignment angle $\varphi$ is defined as the angle between the elongated direction of the confining drop and the direction of the imposed flow.}\label{fig1}
	\end{figure}

	We consider a drop of radius $b$ that concentrically encapsulates a solid particle of radius $ a = b/\alpha$ as shown in figure~\ref{fig1}, and $\alpha$ is referred to as the size ratio. The inner and outer fluids are assumed to be Newtonian having viscosities $\hat{\mu}$ and $\mu = \lambda \hat{\mu}$ respectively, and $\lambda$ is referred to as the viscosity ratio. The typical applications in which the compound particles are encountered are associated with small length and velocity scales, thus the viscous effects are dominant compared to the inertial effects. Hence, we assume that the Reynolds number is small and solve for the Stokes' equations in the inner and outer fluids \citep{russel1989},
	\begin{eqnarray}\label{eq1}
		(1/\lambda)\nabla^{2}\bm{\hat{u}}-\nabla \hat{p}=0; \quad \nabla \cdot \bm{\hat{u}}=0,\\ \label{eq2}
		\nabla^{2}\bm{u}-\nabla p=0; \quad \nabla \cdot \bm{u}=0,
	\end{eqnarray}
	where, $\bm{\hat{u}}$ and $\bm{u}$ are the velocity fields in the inner and outer fluids respectively. The corresponding pressure fields are $\hat{p}$ and $p$. The variables are non-dimensionalised by using the confining drop size $b$ as the characteristic length. Let $\mathcal{G}$ be the characteristic strain rate associated with the compound particle and therefore, $b\mathcal{G}$ and  $\mu \mathcal{G}$ are chosen as the characteristic velocity and pressure respectively for obtaining non-dimensional quantities.
	
	Equations (\ref{eq1})-(\ref{eq2}) are solved using the standard technique of superposition of vector harmonics subjected to the following boundary conditions:
	
	\begin{enumerate}		
		\item In the far field, the velocity of the outer fluid approaches the imposed velocity field, \textit{i.e.},
		\begin{equation}\label{eq3}
			\bm{u} \rightarrow {\bm{u}}^{\infty} \quad \text{as} \quad \bm{x} \rightarrow  \infty.
		\end{equation}
		We consider the ambient flow to be linear, of the form,
		\begin{equation}\label{eq4}
			{\bm{u}}^{\infty}=(\bm{E}+\bm{\Omega}) \cdot \bm{x},
		\end{equation} 
		where, $\bm{x}$ is the position vector, $\bm{E}$ and $\bm{\Omega}$ are respectively the symmetric and anti-symmetric parts of the velocity gradient tensor, $\nabla {\bm{u}}^{\infty}$. 
		\item No slip boundary condition is imposed on the surface of the encapsulated particle,
		\begin{equation}\label{eq5}
			\bm{\hat{u}}=\bm{\Omega} \cdot \bm{x} \quad \text{at} \quad |\bm{x}| = 1/\alpha.
		\end{equation}
		\item Continuity in velocity and stresses are maintained on the confining interface,
		\begin{equation}\label{eq6}
			\bm{u}-\bm{\hat{u}}=\bm{0},
		\end{equation}
		\begin{equation}\label{eq7}
			(\bm{\sigma}-\frac{1}{\lambda}\bm{\hat{\sigma}})\cdot \bm{n} = \frac{1}{Ca}(\nabla \cdot \bm{n})\bm{n}, 
		\end{equation} 
		where $Ca=\mu \mathcal{G} b/\gamma$ is the capillary number, $\bm{n}$ is the unit normal vector to the confining interface,   and $\bm{\sigma}=-p \bm{I}+ (\nabla \bm{u}+\nabla \bm{u}^{T})$ and $(1/\lambda)\bm{\hat{\sigma}}=-\hat{p} \bm{I}+(1/\lambda)(\nabla \bm{\hat{u}}+\nabla \bm{\hat{u}}^{T})$ are the stress tensors in the outer and inner fluids respectively. 
		\item Under the action of the imposed flow, the confining interface of the compound particle deforms and does not remain spherical. With the intention of using domain perturbation technique as a method to solve the problem which can be done by considering $Ca$ as a small parameter, it is convenient to describe the shape of the confining interface using a scalar function, $F(\bm{x}_{s},t)=r-\left(1+Ca f \left(\bm{x}_{s},t \right)\right)=0$, where $r = |\bm{x}|$ and $\bm{x}_{s}$ is the position of the interface. In terms of this scalar function, the kinematic boundary condition on the interface is given by \citep{leal2007,arun2012}
		\begin{equation}\label{eq14}
			\frac{1}{\left| \nabla F \right|} \frac{\partial F}{\partial t }+ Ca (\bm{u} \cdot \bm{n})=0,
		\end{equation} 
		where $\bm{u}$ is the fluid velocity evaluated at the interface, either approached from the inner or the outer fluid and $t$ is non-dimensionalised by the interface relaxation time scale $Ca \mathcal{G}^{-1}$. The normal vector $\bm{n}$ on the interface and the curvature $\nabla \cdot \bm{n}$ of the interface may be obtained as
		\begin{equation}\label{eq15}
			\bm{n}=\frac{\nabla F}{\left| \nabla F \right|}=\frac{1}{\left| \nabla F \right|} \left( \frac{\bm{x}}{r} -Ca \nabla f \right) \quad \textnormal{and}
		\end{equation} 
		\begin{equation}\label{eq16}
			\begin{aligned}
				\nabla \cdot \bm{n}=&\frac{1}{\left| \nabla F \right|} \left( \frac{2}{r} - Ca \nabla^2 f \right)-  Ca^2 \frac{1}{\left| \nabla F \right|^3} \left( \frac{\bm{x}}{r} - Ca \nabla f \right) \cdot (\nabla(\nabla f) \cdot \nabla f),
			\end{aligned}
		\end{equation}   
		respectively. Here, $\left| \nabla F \right|=\sqrt{1+Ca^2 |\nabla f|^2}$. Further, the kinematic boundary condition (\ref{eq14}) may be simplified as,
		\begin{equation}\label{eq17}
			\frac{\partial f}{\partial t}=\left( \bm{u}|_{r=(1+Ca f)} \cdot \bm{n}\right) \sqrt{1+  Ca^{2} |\nabla f|^2}.
		\end{equation}
	\end{enumerate}
	
	Even though the governing  equations are linear, the problem is non-linear because of the coupling between the unknown confining drop shape ($f$) and the flow field. However, the problem can be solved analytically, by assuming that the droplet deformations are small, \textit{i.e.}, $Ca<<1$. We solve the governing equations by expanding all the variables via domain perturbation approach in terms of capillary number, $Ca$, \citep{stone1990a} as,
	\begin{equation}\label{eq18}
		\begin{gathered}
			f(\bm{x},t)=f^{(0)}(\bm{x},t)+Ca~ f^{(1)}(\bm{x},t)+Ca^{2}~ f^{(2)}(\bm{x},t)+\cdots,\\
			p(\bm{x},t)=p^{(0)}(\bm{x},t)+Ca~ p^{(1)}(\bm{x},t)+Ca^{2}~ p^{(2)}(\bm{x},t)+\cdots,\\
			\bm{u}(\bm{x},t)=\bm{u}^{(0)}(\bm{x},t)+Ca~ \bm{u}^{(1)}(\bm{x},t)+Ca^{2}~ \bm{u}^{(2)}(\bm{x},t)+\cdots,\\  
			\hat{p}(\bm{x},t)=\hat{p}^{(0)}(\bm{x},t)+Ca~ \hat{p}^{(1)}(\bm{x},t)+Ca^{2}~ \hat{p}^{(2)}(\bm{x},t)+\cdots,\\
			\bm{\hat{u}}(\bm{x},t)=\bm{\hat{u}}^{(0)}(\bm{x},t)+Ca~ \bm{\hat{u}}^{(1)}(\bm{x},t)+Ca^{2}~ \bm{\hat{u}}^{(2)}(\bm{x},t)+\cdots.\\  
		\end{gathered}
	\end{equation}
	This asymptotic expansion assumes that the capillary number calculated based on either of the fluids is small, \textit{i.e.}, $Ca << 1$ and $Ca/\lambda << 1$. In the domain perturbation approach, the interfacial quantities at $r= (1 +Ca f)$ can be obtained by using Taylor's series expansion in terms of quantities calculated at the droplet surface $r=1$. For \textit{e.g.}, the velocity at the confining (deformed) interface can be obtained as 
	\begin{equation}\label{eq19}
		\begin{aligned}
			&\bm{u}|_{r= 1+Ca f}=\bm{u}^{(0)}|_{r=1}
			+  Ca \Big( \bm{u}^{(1)}|_{r=1} + \bm{x} \cdot \nabla \bm{u}^{(0)}|_{r=1} f^{(0)}  \Big)+\cdots.
		\end{aligned}
	\end{equation}
	\noindent Using this approach, we solve for the velocity and pressure fields upto $\textit{O}(Ca)$  and simultaneously determine the $\textit{O}(Ca)$ and $\textit{O}(Ca^2)$ correction to the confining spherical drop shape.
	
	\section{Hydrodynamics of the compound particle in a linear flow}
	In this section, we  determine the velocity and pressure fields by solving (\ref{eq1}) and (\ref{eq2}) along with the boundary conditions (\ref{eq3})-(\ref{eq14}) using domain perturbation technique explained above and the standard solution methodology of superposition of vector harmonics to solve Stokes' equations.
	\subsection{Leading order solution $\textit{O}(1)$}\label{order1}
	By substituting (\ref{eq18}) in (\ref{eq1}) and (\ref{eq2}), the $\textit{O}(1)$ governing equations and the corresponding boundary conditions can be obtained. The leading order solution is reported and analysed in \cite{chaithu2019}, so below we just provide the final expressions.
	
	Using the standard technique of superposition of vector harmonics \citep{leal2007,chaithu2019}, we may write the velocity and pressure fields in the outer fluid as a linear combination of $\bm{E}$ and $\bm{\Omega}$ as: 
	\begin{align}\label{eq8a}
		p^{(0)}&=c_{1} \bm{D}_{(2)}\bm{: E},\\ \label{eq9a}
		\bm{u}^{(0)}&=\bm{(E+\Omega) \cdot x}+\frac{1}{2}p^{(0)}\bm{x}+c_{2}\bm{D}_{(1)} \bm{\cdot E }+c_{3} \bm{D}_{(3)} \bm{: E}, 
	\end{align}
	where $\bm{D}_{(n)}$ represents the decaying spherical harmonics of order $n$, defined as $n^{th}$ order gradients of fundamental solution of Laplace's equation, $1/r$. Similarly, the pressure and velocity fields in the inner fluid are:
	\begin{align}
		\hat{p}^{(0)}&=d_{1} \bm{G}_{(2)} \bm{: E}+e_{1} \bm{D}_{(2)}\bm{: E}, \label{eq10a}  \\ 
		\hat{\bm{u}}^{(0)}&=\frac{\lambda}{2}\bm{x}\hat{p}^{(0)}+\bm{ \Omega \cdot}  \bm{x}+d_{2}\bm{G}_{(1)} \bm{\cdot E}+d_{3}\bm{G}_{(3)} \bm{: E }+e_{2}\bm{D}_{(1)} \bm{\cdot E }+e_{3} \bm{D}_{(3)} \bm{: E },  \label{eq11a} 
	\end{align}
	where, $\bm{G}_{(n)}$ represents the growing spherical harmonics of order $n$ and it is related to the decaying harmonics as, $\bm{G}_{(n)}=r^{2n+1}\bm{D}_{(n)}$.
	Here $c_i$, $d_i$, and $e_i$ are the constants that are linear function of shape parameter $b_1$, which is related to the leading order shape function as,
		\begin{equation}\label{eq6ab}
		f^{(0)}=b_{1}\bm{E:}\frac{\bm{xx}}{r^2}.
	\end{equation}
	The expressions for the constants  $c_i$, $d_i$, and $e_i$ are given in the Appendix \ref{appA}.
	Using (\ref{eq6ab}) and leading order form of (\ref{eq17}), we obtain the temporal evolution of the shape parameter,
		\begin{equation}\label{eq15a}
		\begin{aligned}
			b_{1}(t)=b_{1}^{*} \bigg(1-\exp\Big(-\frac{t}{t_{c}}\Big)\bigg),
		\end{aligned}
	\end{equation}
	indicating that the relaxation process of the confining interface is exponential with $b_1^{*}$ as the steady state shape parameter and $t_{c}$ as the relaxation time scale.

	\subsection{First order solution $\textit{O}(Ca)$}\label{order2}
	In this section, we solve for the $\textit{O}(Ca)$  pressure and velocity fields. The governing equations at $\textit{O}(Ca)$ are given by 
	\begin{equation}\label{eq16a}
		\begin{gathered}
			(1/\lambda)	\nabla^{2}\bm{\hat{u}}^{(1)}-\nabla \hat{p}^{(1)}=0; \quad \nabla \cdot \bm{\hat{u}}^{(1)}=0,\\
			\nabla^{2}\bm{u}^{(1)}-\nabla p^{(1)}=0; \quad \nabla \cdot \bm{u}^{(1)}=0.
		\end{gathered}
	\end{equation}
	The corresponding boundary conditions are,
	\begin{equation}\label{eq17a}
		\bm{u}^{(1)} \rightarrow 0 \quad  \text{as} \quad \bm{x} \rightarrow  \infty,
	\end{equation}
	the no slip boundary condition on the interface of the encapsulated particle,
	\begin{equation}\label{eq18a}
		\bm{\hat{u}}^{(1)}=0 \quad \text{at} \quad r=1/\alpha,
	\end{equation}  
	the continuity of normal and tangential velocities on the confining interface,
	\begin{equation}\label{eq19a}
		\begin{gathered}
			\left( \left(\bm{u}-\bm{\hat{u}}\right)  \cdot  \bm{n} \right)^{(1)}_{r= ( 1+Ca f^{(0)})} =0, \\
			\left(\left( \bm{I}-\bm{n} \bm{n}\right) \cdot \left(\bm{u} -\bm{\hat{u}} \right)\right)^{(1)}_{r= ( 1+Ca f^{(0)})} =0,
		\end{gathered}
	\end{equation}
	and the stress balance at the interface,
	\begin{subequations}\label{eq20a}
		\begin{equation}\label{eq20aa}
			\bigg(\left(\bm{\sigma} -\frac{1}{\lambda}\bm{\hat{\sigma}} \right) \colon \bm{n n} \bigg)^{(1)}_{r=  ( 1+Ca f^{(0)})} = \left(\nabla \cdot \bm{n}\right)^{(2)}, 
		\end{equation}
		\begin{equation}\label{eq20ab}
			\bigg( \left(  \left( \bm{I}-\bm{n} \bm{n} \right) \cdot  \left( \bm{\sigma}-\frac{1}{\lambda} \bm{\hat{\sigma}} \right) \right)  \cdot \bm{n} \bigg)^{(1)}_{r= ( 1+Ca f^{(0)})}=0.
		\end{equation}
	\end{subequations}
	Similar to the previous section, (\ref{eq16a}) are solved using the technique of superposition of vector harmonics. The pressure and velocity fields in the outer fluid are at most quadratically dependent on $\bm{E}$ and $\bm{\Omega}$. Thus,
	\begin{align}\label{eq21a}
		p^{(1)}&=c_{4}\bm{D}_{(2)}\bm{: E}+c_{5}\bm{D}_{(2)}\bm{:} (\bm{E \cdot E})+c_{6}(\bm{D}_{(4)}\bm{:E})\bm{:E}
		+c_{7}(\bm{E:E})\nonumber \\ 
		&+c_{8}\bm{D}_{(2)}\bm{:} (\bm{\Omega \cdot E})+c_{9}\bm{D}_{(2)}\bm{:}(\bm{E\cdot \Omega})+c_{10}(\bm{\Omega: \Omega}), \\ \label{eq22a}
		\bm{u}^{(1)}&=\frac{1}{2}\bm{x}p^{(1)}+c_{11}\bm{D}_{(1)} \bm{\cdot E}+ c_{12}\bm{D}_{(3)} \bm{: E}+c_{13}( \bm{E \cdot E}) \bm{\cdot} \bm{D}_{(1)}+c_{14}\bm{E \cdot}(\bm{D}_{(3)}\bm{: E}) \nonumber \\
		&+c_{15}\bm{D}_{(1)}(\bm{E : E})+c_{16}\bm{D}_{(3)}\bm{:}(\bm{E \cdot E})
		+c_{17}\bm{\Omega \cdot}\bm{D}_{(1)}+c_{18}(\bm{\Omega \cdot E}) \bm{\cdot D}_{(1)}\nonumber \\
		&+c_{19} (\bm{E \cdot \Omega}) \bm{\cdot D}_{(1)}+c_{20}\bm{\Omega \cdot}( \bm{D}_{(3)}\bm{: E })+c_{21}(\bm{\Omega \cdot \Omega} )\bm{\cdot D}_{(1)}+c_{22}(\bm{\Omega : \Omega} )\bm{D}_{(1)}\nonumber\\
		&+c_{23}\bm{E :}(\bm{E:D}_{(5)})+c_{24} (\bm{D}_{(3)} \bm{\cdot \Omega} )\bm{: \Omega}+c_{25} (\bm{D}_{(3)} \bm{\cdot \Omega} )\bm{: E}. 
	\end{align}
	Similarly, the pressure and velocity fields in the inner fluid are,
	\begin{align}\label{eq23a}
		\hat{p}^{(1)}&=d_{4}\bm{G}_{(2)}\bm{: E}+d_{5}\bm{G}_{(2)}\bm{:} (\bm{E \cdot E})+d_{6}(\bm{G}_{(4)}\bm{:E})\bm{:E}+d_{7}(\bm{E:E})+d_{8}\bm{G}_{(2)}\bm{:} (\bm{\Omega \cdot E})\nonumber\\
		&+d_{9}\bm{G}_{(2)}\bm{:}(\bm{E\cdot \Omega})+d_{10}(\bm{\Omega: \Omega})+e_{4}\bm{D}_{(2)}\bm{: E}+e_{5}\bm{D}_{(2)}\bm{:} (\bm{E \cdot E})+e_{6}(\bm{D}_{(4)}\bm{:E})\bm{:E}\nonumber\\
		&+e_{7}(\bm{E:E})+e_{8}\bm{D}_{(2)}\bm{:} (\bm{\Omega \cdot E}) +e_{9}\bm{D}_{(2)}\bm{:}(\bm{E\cdot \Omega})+e_{10}(\bm{\Omega: \Omega}), \\\label{eq24a}
		\bm{\hat{u}}^{(1)}&=\frac{\lambda}{2}\bm{x}\hat{p}^{(1)}+d_{11}\bm{G}_{(1)} \bm{\cdot E}+ d_{12}\bm{G}_{(3)} \bm{: E}+d_{13}( \bm{E \cdot E}) \bm{\cdot} \bm{G}_{(1)}+d_{14}\bm{E \cdot}(\bm{G}_{(3)}\bm{: E})\nonumber\\&
		+d_{15}\bm{G}_{(1)}(\bm{E : E})+d_{16}\bm{G}_{(3)}\bm{:}(\bm{E \cdot E})+d_{17}\bm{\Omega \cdot}\bm{G}_{(1)}+d_{18}(\bm{\Omega \cdot E}) \bm{\cdot G}_{(1)}\nonumber\\
		&+d_{19} (\bm{E \cdot \Omega}) \bm{\cdot G}_{(1)}+d_{20}\bm{\Omega \cdot}( \bm{G}_{(3)}\bm{: E })+d_{21}(\bm{\Omega \cdot \Omega} )\bm{\cdot G}_{(1)}+c_{22}(\bm{\Omega : \Omega} )\bm{G}_{(1)}\nonumber\\
		&+d_{23}\bm{E :}(\bm{E:G}_{(5)})+d_{24} (\bm{G}_{(3)} \bm{\cdot \Omega} )\bm{: \Omega}+d_{25} (\bm{G}_{(3)} \bm{\cdot \Omega} )\bm{: E}+e_{11}\bm{D}_{(1)} \bm{\cdot E}\nonumber\\
		&+e_{12}\bm{D}_{(3)} \bm{: E}+e_{13}( \bm{E \cdot E}) \bm{\cdot} \bm{D}_{(1)}+e_{14}\bm{E \cdot}(\bm{D}_{(3)}\bm{: E})+e_{15}\bm{D}_{(1)}(\bm{E : E})\nonumber\\
		&+e_{16}\bm{D}_{(3)}\bm{:}(\bm{E \cdot E})+e_{17}\bm{\Omega \cdot}\bm{D}_{(1)}+e_{18}(\bm{\Omega \cdot E}) \bm{\cdot D}_{(1)}+e_{19} (\bm{E \cdot \Omega}) \bm{\cdot D}_{(1)}\nonumber\\
		&+e_{20}\bm{\Omega \cdot}( \bm{D}_{(3)}\bm{: E })+e_{21}(\bm{\Omega \cdot \Omega} )\bm{\cdot D}_{(1)}+e_{22}(\bm{\Omega : \Omega} )\bm{D}_{(1)}+e_{23}\bm{E :}(\bm{E:D}_{(5)})\nonumber\\
		&+e_{24} (\bm{D}_{(3)} \bm{\cdot \Omega} )\bm{: \Omega}+e_{25} (\bm{D}_{(3)} \bm{\cdot \Omega} )\bm{: E},
	\end{align}
	where $c_{i}$, $d_{i}$, and $e_{i}$ for $i=4 ~\text{to}~ 25$ are the constants determined from the boundary conditions as follows.
	As earlier, the shape function $f^{(1)}$ may be expressed as quadratic combinations of $\bm{E}$ and $\bm{\Omega}$ as, 
	\begin{equation}\label{eq25a}
		\begin{aligned}
			f^{(1)}=b_{2}\frac{\bm{x \cdot E \cdot x}}{r^2}+b_{3}\bm{E : E}+b_{4}\frac{\bm{x \cdot (E \cdot E) \cdot x}}{r^2}+b_{5}\frac{(\bm{x \cdot E \cdot x})^{2}}{r^{4}}\\+b_{6} \frac{\bm{x \cdot (\Omega \cdot E) \cdot x}}{r^2}
			+b_{7} \bm{\Omega : \Omega}+b_{8}\frac{\bm{x \cdot (\Omega \cdot \Omega) \cdot x}}{r^2},
		\end{aligned}
	\end{equation}
	where $b_{j}$, $j=2 ~\text{to}~ 8$ are unknown constants. In other words, we have 7 additional constants along with $b_{1}$ discussed in the previous section to describe the deformed drop shape. As the volume of the inner fluid in the compound particle should remain constant, 
	\begin{equation}\label{eq26a}
		\begin{aligned}
			\frac{1}{3} \int_{S_{D}}   \left(1+Ca f^{(0)}+Ca^2 f^{(1)}\right)^{3} d\Omega-\frac{1}{3} \int_{S_{P}}  (1/\alpha)^{3} d\Omega  =	\frac{4\upi  }{3}-\frac{4\upi (1/\alpha)^{3}}{3},
		\end{aligned}
	\end{equation} 
	where $S_{D}$ and $S_{P}$ are the surfaces of the confining drop and the encapsulated particle, respectively. Simplifying (\ref{eq26a}), we get 
	\begin{equation}\label{eq27a}  
		\int_{S_{D}} \left(1+Ca f^{(0)}+Ca^2 f^{(1)}\right)^{3} d\Omega =4\upi,
	\end{equation}
	and this gives the following relations between the constants describing the interface
	\begin{equation}\label{eq28a}
		\begin{gathered}
			b_{3}+\frac{b_{4}}{3}+\frac{2 b_{5}}{15}+\frac{2 b_{1}^{2}}{15}=0, \quad 
			b_{7}+\frac{b_{8}}{3}=0.
		\end{gathered}
	\end{equation}
	Enforcing the equation of continuity on the velocity field results additional relations between the unknown constants,
	\begin{equation}\label{eq29a}
		\begin{gathered}
			c_{6}=-c_{14},\quad  c_{7}=0,\quad c_{10}=0, \quad c_{11}=0,	\quad c_{13}=0, \quad c_{18}=c_{19},\quad  c_{21}=0,\\
			\lambda d_{4}=-\frac{126 d_{12}}{5},\quad  \lambda d_{6}=-\frac{110 d_{23}}{7},\quad 
			\lambda d_{9}=-\frac{42d_{24}}{5}, \quad -\frac{5\lambda d_{5}}{2}+6d_{14}+21d_{16}=0,\\
			\frac{3 \lambda d_{7}}{2}+d_{13}+3d_{15}=0,
			\quad		\frac{3 \lambda d_{10}}{2}+3d_{22}-d_{21}=0, \quad
			\frac{5 \lambda d_{8}}{2}+14d_{20}+21d_{25}=0,\\
			\lambda e_{6}=-e_{14},\quad  e_{7}=0, \quad
			e_{10}=0, \quad  e_{11}=0, \quad e_{13}=0, \quad e_{18}=e_{19},\quad  e_{21}=0.
		\end{gathered}
	\end{equation}
	In order to impose the continuity in velocity and stress boundary conditions on the deformed interface, we evaluate those quantities at $r = (1+Ca f^{(0)})$ as,
	\begin{equation}\label{eq30a}
		\begin{aligned}
			&	\big(\bm{u}\big|_{r= (1+Ca f^{(0)})}\big)^{(1)}=\bm{u}^{(1)}\big|_{r=1}+ f^{(0)} \nabla \bm{u}^{(0)} \cdot \bm{x}\big|_{r=1} ,\\
			&	\big(\bm{\sigma} \cdot \bm{n} \big)^{(1)}_{r= (1+Ca f^{(0)})} =\bm{\sigma}^{(1)}\big|_{r=1} \cdot \bm{n}^{(0)}+\bm{\sigma}^{(0)}\big|_{r=1} \cdot \bm{n}^{(1)} +  f^{(0)}\left(  \nabla \bm{\sigma}^{(0)} \cdot \bm{x}\big|_{r=1} \right) \cdot \bm{n}^{(0)}.  
		\end{aligned}
	\end{equation}
	From (\ref{eq15}), (\ref{eq16}) and (\ref{eq25a}), we obtain the normal vector $\bm{n}^{(1)}$ and the curvature of the interface $(\nabla \bm{\cdot n})^{(2)}$ at $\textit{O}(Ca)$ as
	\begin{align}\label{eq31a}
		\bm{n}^{(1)}=&-\nabla f^{(0)}=2b_{1}\bigg(\frac{\bm{x}\left(\bm{x}\cdot \bm{E} \cdot \bm{x} \right)}{r^{4}}-\frac{\bm{E} \cdot \bm{x}}{r^{2}}  \bigg)_{r=1} \quad \textnormal{and}\\ \nonumber
		(\nabla \bm{\cdot n})^{(2)}=&2\left(\left(f^{(0)}\right)^{2}-f^{(1)}\right)-\left( \nabla^{2} f\right)^{(1)}-\left(\nabla f^{(0)}  \right)^{2} -\bm{x \cdot}\left( \nabla f^{(0)} \bm{\cdot} \nabla \nabla f^{(0)}  \right)\\ \nonumber
		=&(-10b_{1}^{2}+18b_{5})\left( \bm{x \cdot E \cdot x}  \right)^{2}+4b_{2}  \bm{x \cdot E \cdot x} -(2b_3+2b_4) \bm{E : E} \\ \nonumber	
		&+ (4b_{4}-8b_{5}) \bm{x \cdot } (\bm{E \cdot E})\bm{ \cdot x}+4b_{6} \bm{x \cdot } (\bm{\Omega \cdot E})\bm{ \cdot x}-(2b_{7}+2b_{8})\bm{\Omega : \Omega}  \\ \label{eq32a}
		&+4b_{8}  \bm{x \cdot } (\bm{\Omega \cdot \Omega})\bm{ \cdot x}, 
	\end{align}
	respectively.
	The system of equations obtained by imposing the boundary conditions are given in the appendix \ref{appB}. In the limit $\alpha\to\infty$, the expressions for the constants (refer (\ref{eqB15})-(\ref{eqB29}) in appendix \ref{appB}) are consistent with the $\mathcal{O}(Ca)$ calculations for a drop \citep{arun2012}. As done earlier, the kinematic boundary condition may be used to evaluate the shape function, $f^{(1)}$ as,
	\begin{equation}\label{eq33a}
		\frac{\partial f^{(1)}}{\partial t}=(\bm{u} \bm{\cdot n})^{(1)}_{r=(1+Ca f^{(0)})},
	\end{equation}
	which further reduce to the description of temporal evolution of the deformed interface as,
	\begin{subequations}\label{eq34a}
		\begin{equation}\label{eq34aa}
			\frac{\partial b_{2}}{\partial t}=\frac{c_{4}}{2  }+  9c_{12} , 
		\end{equation}
		\begin{equation}\label{eq34ab}
			\begin{aligned}
				\frac{\partial b_{4}}{\partial t}=-\frac{3c_{5}}{2 }- 24c_{6} + 9c_{16} 
				- 300c_{23} -b_{1}\left(2- 12c_{3} \right),
			\end{aligned}
		\end{equation}
		\begin{equation}\label{eq34ac}
			\frac{\partial b_{5}}{\partial t}=\frac{75c_{6}}{2 }+ 525c_{23} +b_{1}\left(3-c_{1} - 48c_{3} \right),
		\end{equation}
		\begin{equation}\label{eq34ad}
			\frac{\partial b_{6}}{\partial t}=-\frac{3c_{8}}{2 }- 6c_{20} -9c_{25} +2b_{1},
		\end{equation}
		\begin{equation}\label{eq34ae}
			\frac{\partial b_{8}}{\partial t}=-\frac{3c_{9}}{2 }-9c_{24}.
		\end{equation} 
	\end{subequations}
	Here the variables $c_i$ are linear functions of $b_j$ where $j$ varies from $2$ to $8$. We solve the above set of equations (\ref{eq34aa})-(\ref{eq34ae}) along with (\ref{eq28a}) to obtain the shape parameters $b_{j}$. The temporal evolution of the shape is illustrated using plots in the next section and the steady state shape values of $b_{j}$ are given in the appendix \ref{appC}. The expressions for the shape functions in the limit $\alpha \to \infty$ are also given in the appendix \ref{appC} (refer (\ref{eqC2})-(\ref{eqC7})), they are consistent with the calculations for a drop \citep{arun2012}.
	
	Hence, using a domain perturbation approach and standard technique of superposition of vector harmonics, we calculated the pressure and velocity fields upto $\textit{O}(Ca)$ in both inner and outer fluids for a compound particle when subjected to an imposed linear flow. Along with these, we determined the time evolution of the deforming interface of the confining drop upto $\textit{O}(Ca^2)$. These solutions are illustrated in the next section for various linear flows.
	
	\section{Deformation dynamics of a compound particle in a general linear flow}\label{example}
	In the previous section, we described the $\textit{O}(1)$ flow field \citep{chaithu2019}, and then derived the $\textit{O}(Ca)$ flow field generated by a compound particle and the corresponding consequences, namely the $\textit{O}(Ca)$ and $\textit{O}(Ca^2)$ corrections to the confining drop shape. We now analyze these results for different linear flows (i) a simple shear flow, and (ii) extensional (both uniaxial and biaxial) flows. We then expand the discussion on interface deformation dynamics for generalised shear and generalized extensional flows in this section.
	
	\subsection{Simple shear flow}\label{simple}
	Consider a simple shear flow of the form $\bm{u}=  x_{2} \mathbf{i}_{1}$, where $x_{2}$ is the component of position vector $\bm{x}$ and $i_{1}$ is the unit normal vector associated with $x_{1}$ in the chosen coordinate system. The symmetric and anti-symmetric parts of the  velocity gradient tensor of the imposed flow are,
	\begin{equation}\label{eq1b}
		\bm{E}^{S}=\frac{1}{2}\begin{bmatrix}
			0 & 1 & 0 \\
			1 & 0 & 0\\
			0 & 0 & 0
		\end{bmatrix}, \quad \bm{\Omega}^{S}=\frac{1}{2}\begin{bmatrix}
			0 & 1& 0 \\
			-1 & 0 & 0\\
			0 & 0 & 0
		\end{bmatrix}.
	\end{equation}
	Using (\ref{eq1b}), the shape of the confining drop may be obtained as,
	\begin{equation}\label{eq2b}
		\begin{aligned}
			r(\theta, \phi)=
			\Bigg(1+ b_{1} Ca\left(\frac{ x_{1}x_{2}}{r^{2}}\right)+Ca^{2}\bigg(b_{2}\frac{ x_{1}x_{2}}{r^{2}}+\frac{b_{3}-b_{7}}{2} +\frac{b_{4}-b_{8}}{4}\left(\frac{x_{1}^{2}+x_{2}^{2}}{r^{2}}\right)\\
			+b_{5}\frac{x_{1}^{2}x_{2}^{2}}{r^{4}}+\frac{b_{6}}{4}\left(\frac{x_{1}^{2}-x_{2}^{2}}{r^{2}}\right)\bigg)+\textit{O}(Ca^3)\Bigg),
		\end{aligned}
	\end{equation}
	where 
	$	x_{1}=r\sin{\theta}\cos{\phi}$,   $x_{2}=r\sin{\theta}\sin{\phi}$ and $x_{3}=r\cos{\theta}$, 
	$r$ is the magnitude of the position vector $\bm{x}$, $\theta$ 
	is the polar angle measured from the $x_{3}$ axis
	($0\leq \theta \leq \upi $) and $\phi$ is the azimuthal angle measured around the $x_3$ axis ($0 \leq \phi \leq 2\upi$).
	
	\begin{figure}
		\centering
		\subfigure[]{
			\includegraphics[height=5cm]{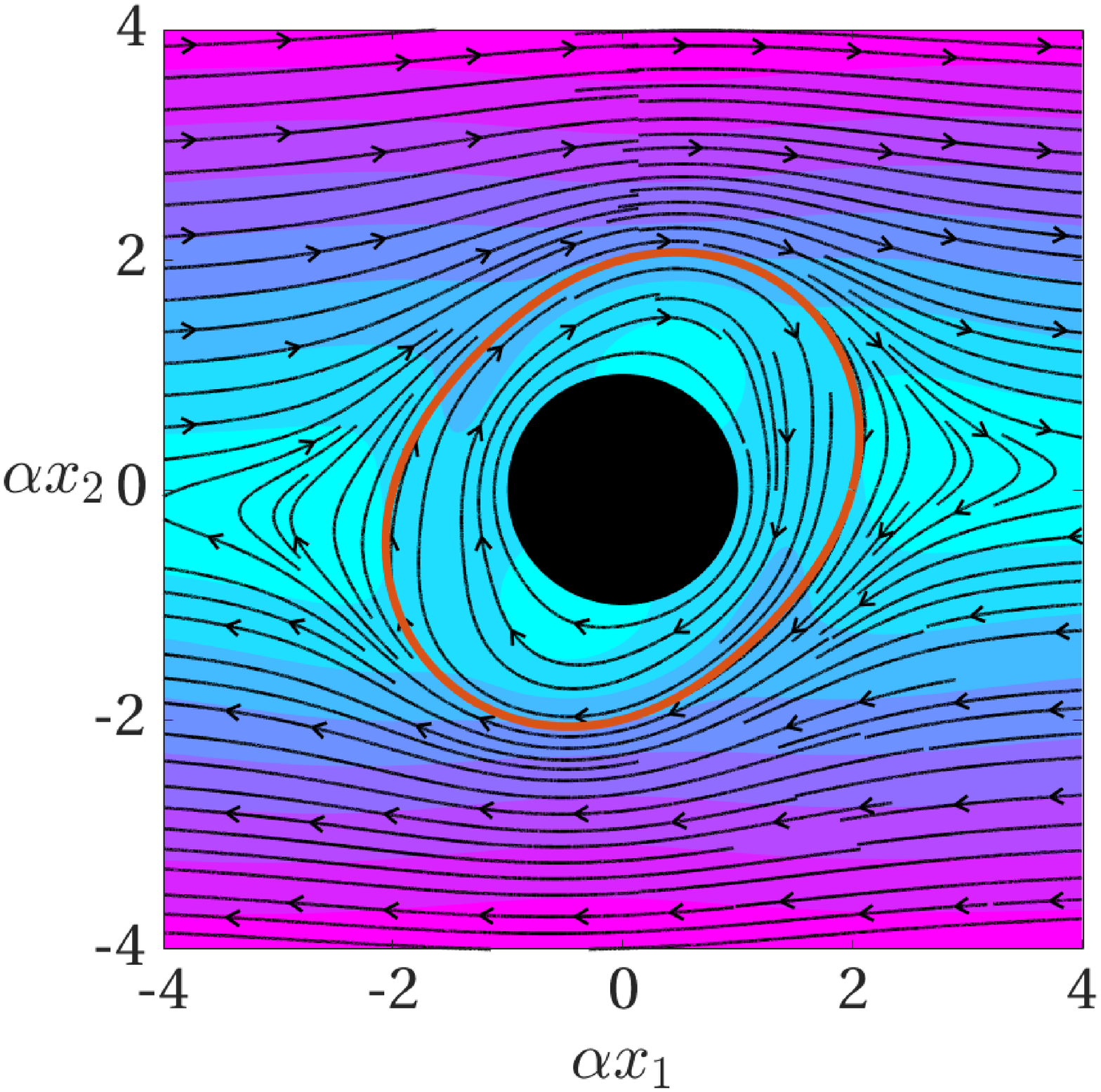}\label{fig2a}}\quad 		
		\subfigure[]{
			\includegraphics[height=5cm]{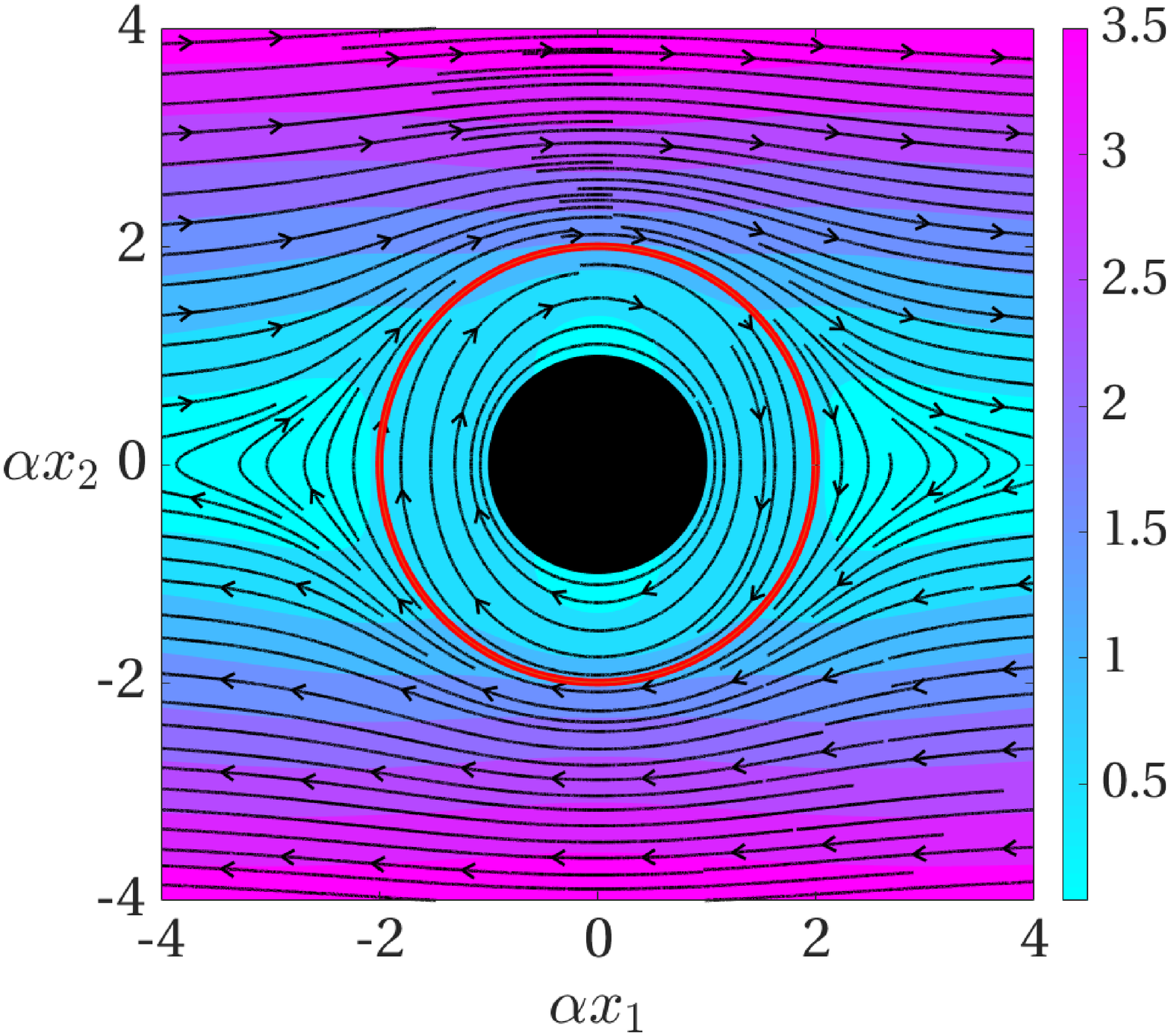}\label{fig2b}}\quad
		\caption{(a) Velocity field, represented as streamlines, in and around a compound particle in an imposed shear flow in the flow - gradient ($x_{1}$ - $x_{2}$) plane, for $\alpha=2$, $Ca=0.1$ and $\lambda=1$. The $\textit{O}(1)$ flow field is shown in (b) for comparison. The color field shows the magnitude of velocity. The filled circle is the encapsulated solid particle and the solid red line shows the confining drop interface.} 
		\label{fig2}
	\end{figure}
	
	The steady state flow fields in and around a deformed compound particle are illustrated in figure~\ref{fig2a}. Equations (\ref{eqA2}), (\ref{eqA4}), (\ref{eqB2}), (\ref{eqB4}), and (\ref{eqC1}) have been used to calculate the velocity field and the corresponding confining drop shape. For comparison, $\textit{O}(1)$ velocity field is shown in figure~\ref{fig2b} \citep{chaithu2019}. In the imposed simple shear flow, the encapsulated rigid particle rotates with an angular velocity $\mathcal{G}/2$, and the curved streamlines inside the confining drop illustrate the recirculating fluid flow around this rotating rigid particle. Similarly, the outer fluid close to the interface also has curved streamlines. However, compared to that of a spherical compound particle (figure~\ref{fig2b}) or a particle without a confining drop \citep{leal2007,sadhal1997}, the streamlines around the compound particle show distortions corresponding to the deformed interface. 
	
	\begin{figure}
		\centering
		\subfigure[]{
			\includegraphics[height=5cm]{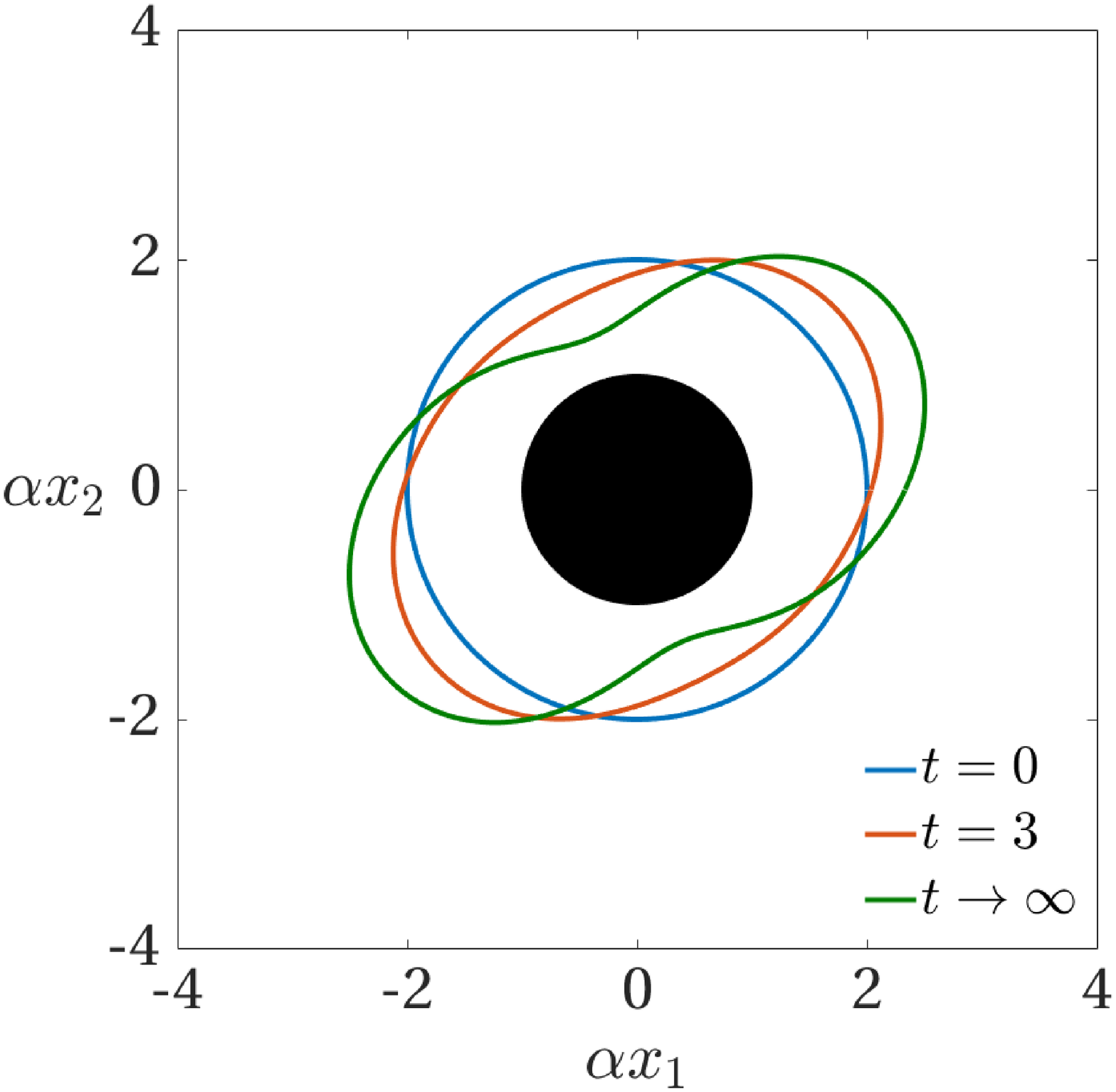}\label{fig3a}}\quad
		\subfigure[]{
			\includegraphics[height=5cm]{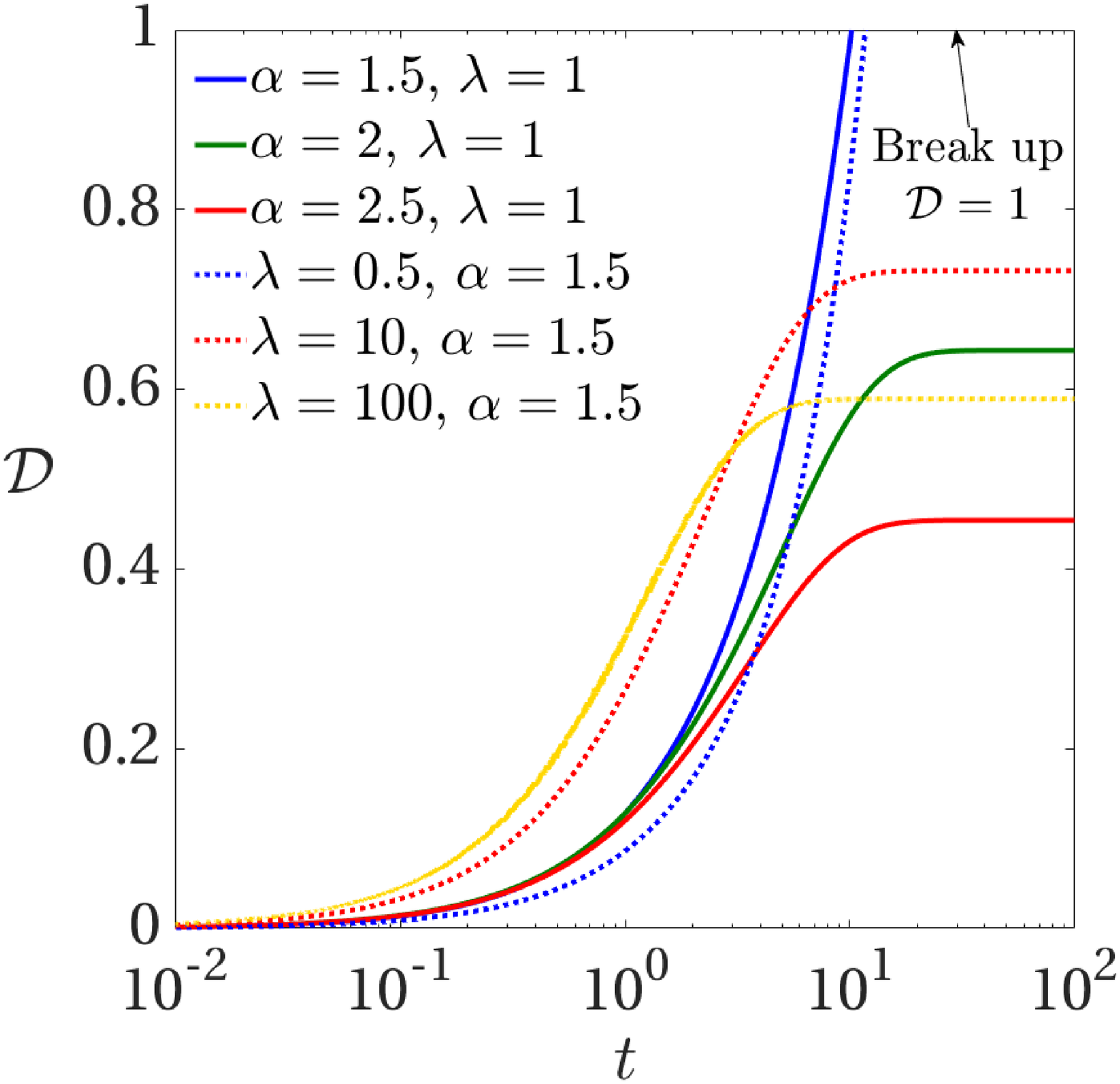}\label{fig3b}}\quad 
		\subfigure[]{
			\includegraphics[height=5cm]{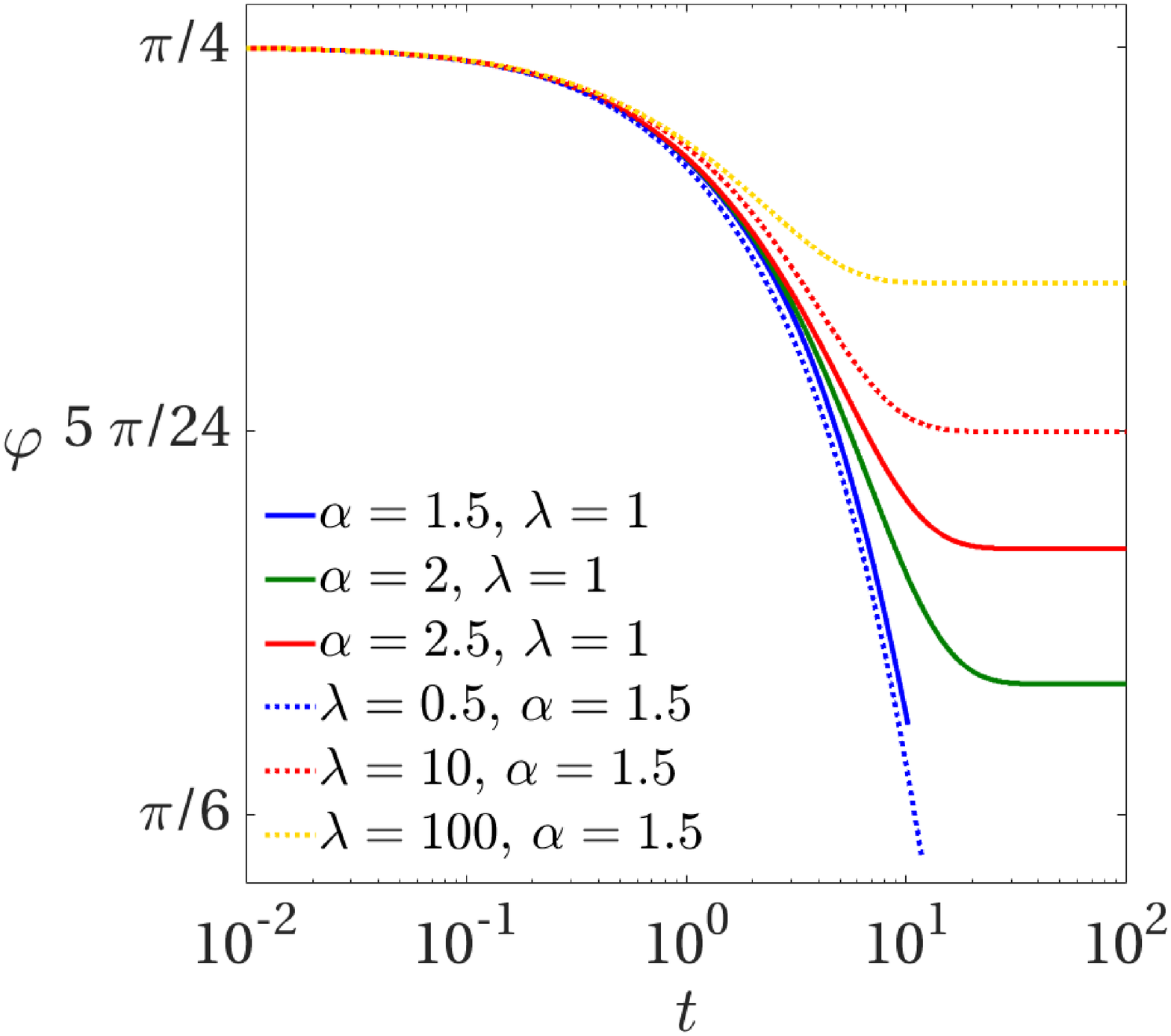}\label{fig3c}}\quad 
		\caption{The time evolution of (a) the confining drop shape in a simple shear flow for $\lambda=1$, $\alpha=2$ and $Ca=0.2$   (equations (\ref{eq15a}), (\ref{eq34a}) along with (\ref{eq28a})). The black patch at the center indicates the encapsulated solid particle. (b) The deformation parameter $\mathcal{D}$ determined by (\ref{eq3b}), and (c) the alignment angle $\varphi$, for various size ratio $\alpha$ and viscosity ratio $\lambda$. The solid lines correspond to $\mathcal{D}$ (and $\varphi$) obtained for various $\alpha$ and the dotted lines correspond to those obtained for various $\lambda$.} 
		\label{timedrop}
	\end{figure}
	
	Now we proceed to analyze the confining drop shape and its evolution. The time evolution of the confining drop shape (upto $\mathcal{O}(Ca^{2})$) in a simple shear flow for $\alpha=2$, $\lambda=1$ and $Ca=0.2$ is depicted in figure \ref{fig3a}. At $t=0$, the confining drop is spherical, but the imposed flow deforms the interface with time. Finally, the confining drop attains a steady shape that is elongated in a direction close to the extensional axis of the imposed flow and compressed in the orthogonal direction. 
	
	Similar to the deformation parameter defined by \cite{taylor1932} to analyze the shape of drops, we define a deformation parameter $\mathcal{D}$ as
	\begin{equation}\label{eq3b}
		\mathcal{D}=\frac{(r_{max}- a)-(r_{min}-a)}{(r_{max}-a)+(r_{min}-a)},
	\end{equation}
	that quantifies the extent of deformation of the compound particle as done in \cite{chaithu2019}. Here, $r_{max}$ and $r_{min}$ are the longest and shortest dimensions of the deformed interface as shown in figure~\ref{fig1}. The limits, $\mathcal{D}=0$ and $\mathcal{D}=1$ respectively correspond to the case of an undeformed spherical interface and the case where the confining interface touches the encapsulated solid particle. The latter case can be regarded as the onset of break-up of the confining drop since the interface comes into contact with the encapsulated solid particle. Therefore, unlike that of a simple drop,  large values of deformation parameter doesn't necessarily mean a large deformation of the confining drop. This is especially the case in the limit $\alpha \to 1$. In other words, in the limit $\alpha \to 1$, breakup ($\mathcal{D} = 1$) may occur even with a weak deformation of the confining drop, and therefore the perturbation approach used in this work remains valid all the way upto the confining drop breakup.
		
	The time evolution of deformation parameter $\mathcal{D}$  for $Ca=0.2$ but for various values of $\alpha$ and $\lambda$ are plotted in figure~\ref{fig3b}.  In all cases, $\mathcal{D}$ increases with time indicating the progression of deformation. Finally $\mathcal{D}$ reaches a plateau corresponding to the steady state shape of the confining drop. In some cases, $\mathcal{D}$ reaches $1$ representing the break up of the confining interface. Of course the progression of deformation and the final value of $\mathcal{D}$ depend upon the particular values of $\alpha$ and $\lambda$, and this dependence is discussed later in this section.
	
	Another consequence of calculations at $\textit{O}(Ca^2)$ is its ability to predict the orientation of the elongated interface in a simple shear flow. The $\textit{O}(Ca)$ correction to the confining drop shape shows that the elongated direction of the confining drop aligns with the extensional axis of the flow, \textit{i.e.}, $45^\circ$ with the flow direction in the flow - gradient plane as reported in \cite{chaithu2019}. However, $\textit{O}(Ca^2)$ calculation takes into account the effect of vorticity of the imposed flow, which then predicts that the elongated interface does not orient along the extensional axis of the imposed flow. We define an alignment angle $\varphi$, as shown in figure~\ref{fig1}, as the angle between the elongated direction of the confining drop and the flow direction ($x_1$) in a simple shear flow.
	
	The time evolution of alignment angle $\varphi$ at $Ca=0.2$ for various values of $\alpha$ and $\lambda$ is plotted in figure~\ref{fig3c}. The alignment angle $\varphi$ decreases with time before reaching a steady value, indicating that the deformed drop rotates towards the flow direction as dictated by the imposed vorticity, finally attaining an orientation that is in between the flow direction and the extensional axis of the imposed shear flow. 
	
	\begin{figure}
		\centering
		\subfigure[]{
			\includegraphics[height=5cm]{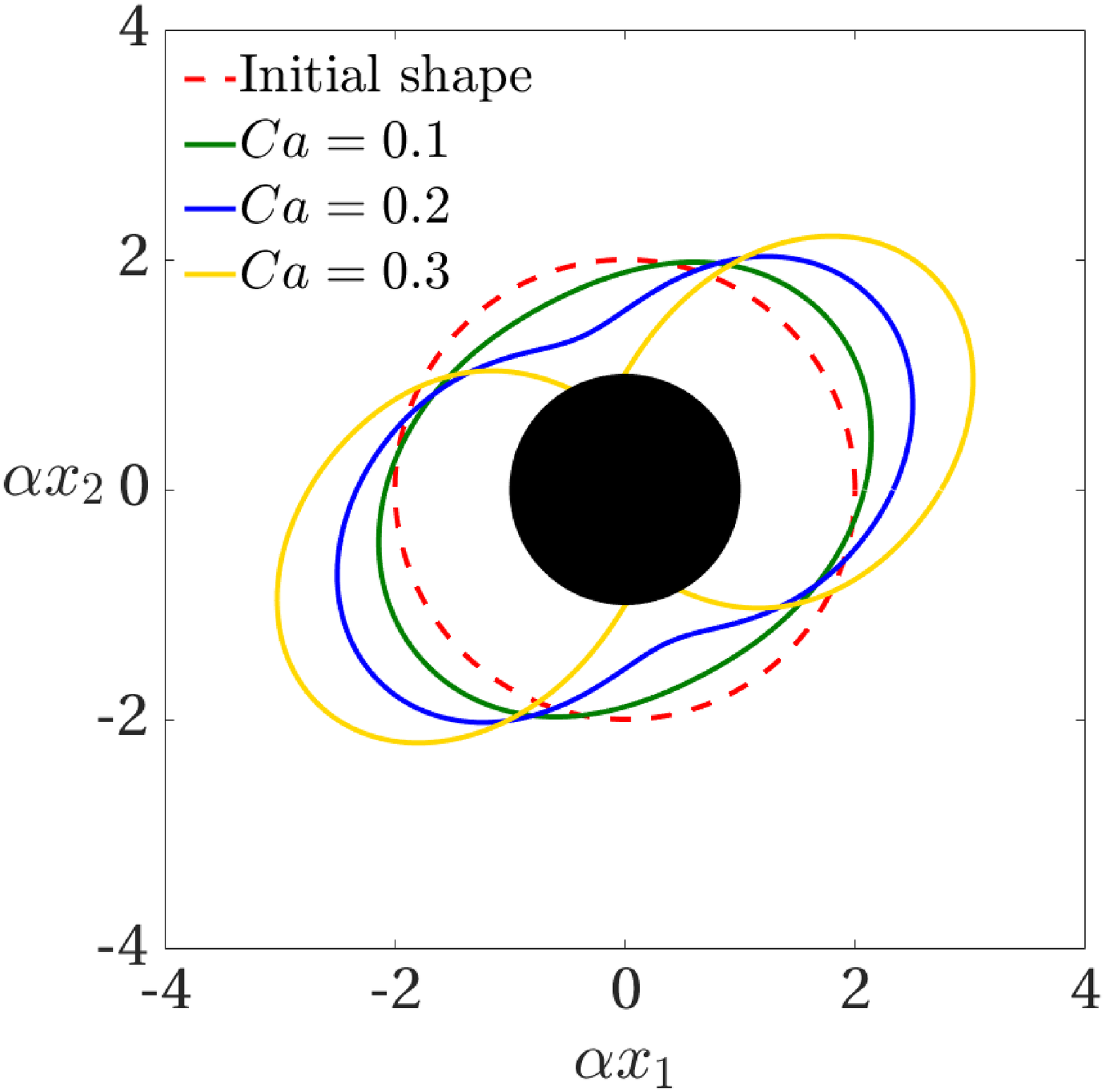}\label{fig4a}}\quad 
		\subfigure[]{
			\includegraphics[height=5.0cm]{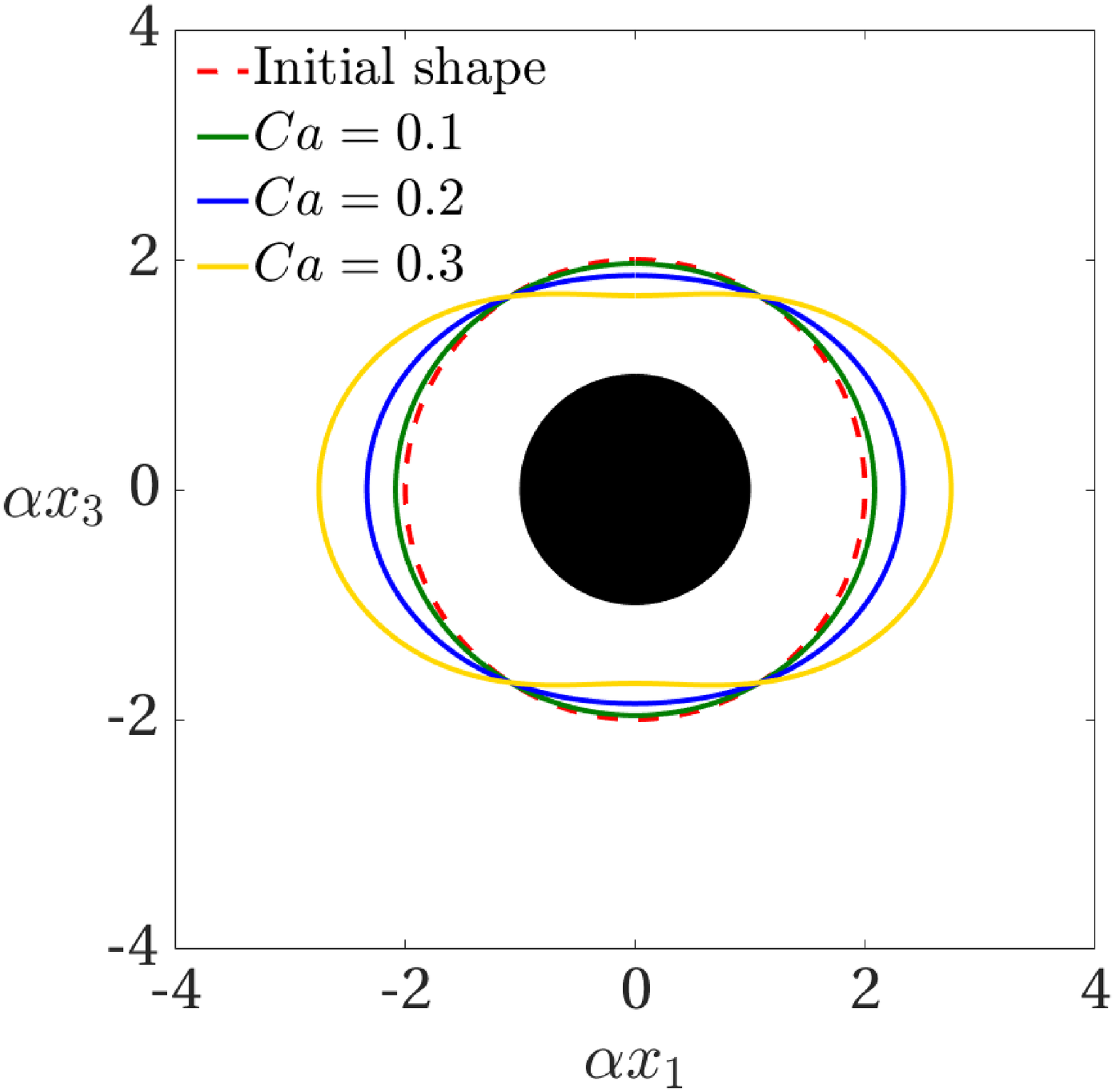}\label{fig4b}}\quad 
		\subfigure[]{
			\includegraphics[height=5.0cm]{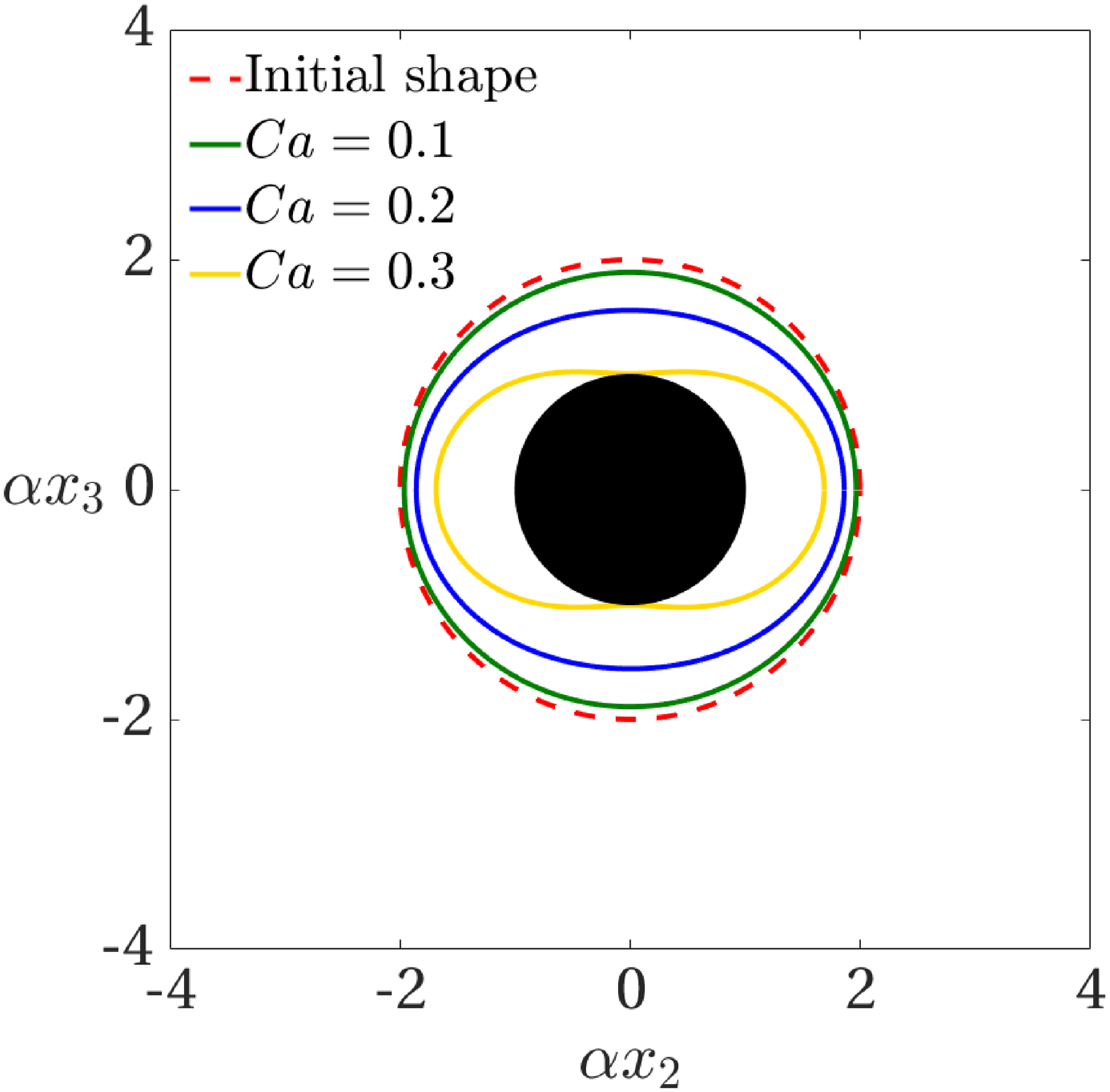}\label{fig4c}}\quad 
		\caption{The deformed shape (see (\ref{eq2b})) of the confining drop viewed from various planes: (a) in flow - gradient ($x_{1}$ - $x_{2}$) plane, (b) in flow - vorticity ($x_{1}$ - $x_{3}$) plane, and (c) in gradient - vorticity ($x_{2}$ - $x_{3}$) plane for different capillary numbers. The dashed line represents the initial, undeformed shape of the confining drop and the filled circle is the encapsulated solid particle.}
		\label{fig4}
	\end{figure}
	
	The extent of deformation of the confining interface of the compound particle that resulted from the imposed shear flow is different in different directions. The steady state shape of the confining drop (see~(\ref{eq2b})), when viewed in three different planes, namely the flow - gradient ($ x_{1}$ - $ x_{2}$), flow - vorticity ($ x_{1}$ - $x_{3}$), and gradient - vorticity ($x_{2}$ - $x_{3}$) planes are shown in figures~\ref{fig4a}-\ref{fig4c}. The interface deformation is largest in the flow - gradient plane with the confining drop elongated in a direction described by $\varphi$. Confining drop is lesser deformed in the flow - vorticity plane and the gradient - vorticity plane, with the interface elongated in the flow direction in the former and compressed along the vorticity direction in the latter.
	
	Figures~\ref{fig4a}-\ref{fig4c} also illustrate the deformed shape of the confining drop for various capillary numbers. Increase in $Ca$ signifies increasingly dominant shear force over the resisting interfacial tension and therefore, the extent of deformation of the confining interface increases with increase in $Ca$. This is evident from the confining drop shapes shown in various planes (figures~\ref{fig4a}-\ref{fig4c}). For the largest capillary number $Ca = 0.3$ considered in this figure, dimples form on the interface, as evident in the flow - gradient plane (figure~\ref{fig4a}). The extent of compression of the interface is not same in all directions, with the maximum compression observed along the vorticity direction. Incidentally, in this particular case, the deformed interface comes in contact with the encapsulated solid particle (figure~\ref{fig4c}) and marks the break up of the confining drop.

	\begin{figure}
		\centering
		\subfigure[]{
			\includegraphics[height=5.0cm]{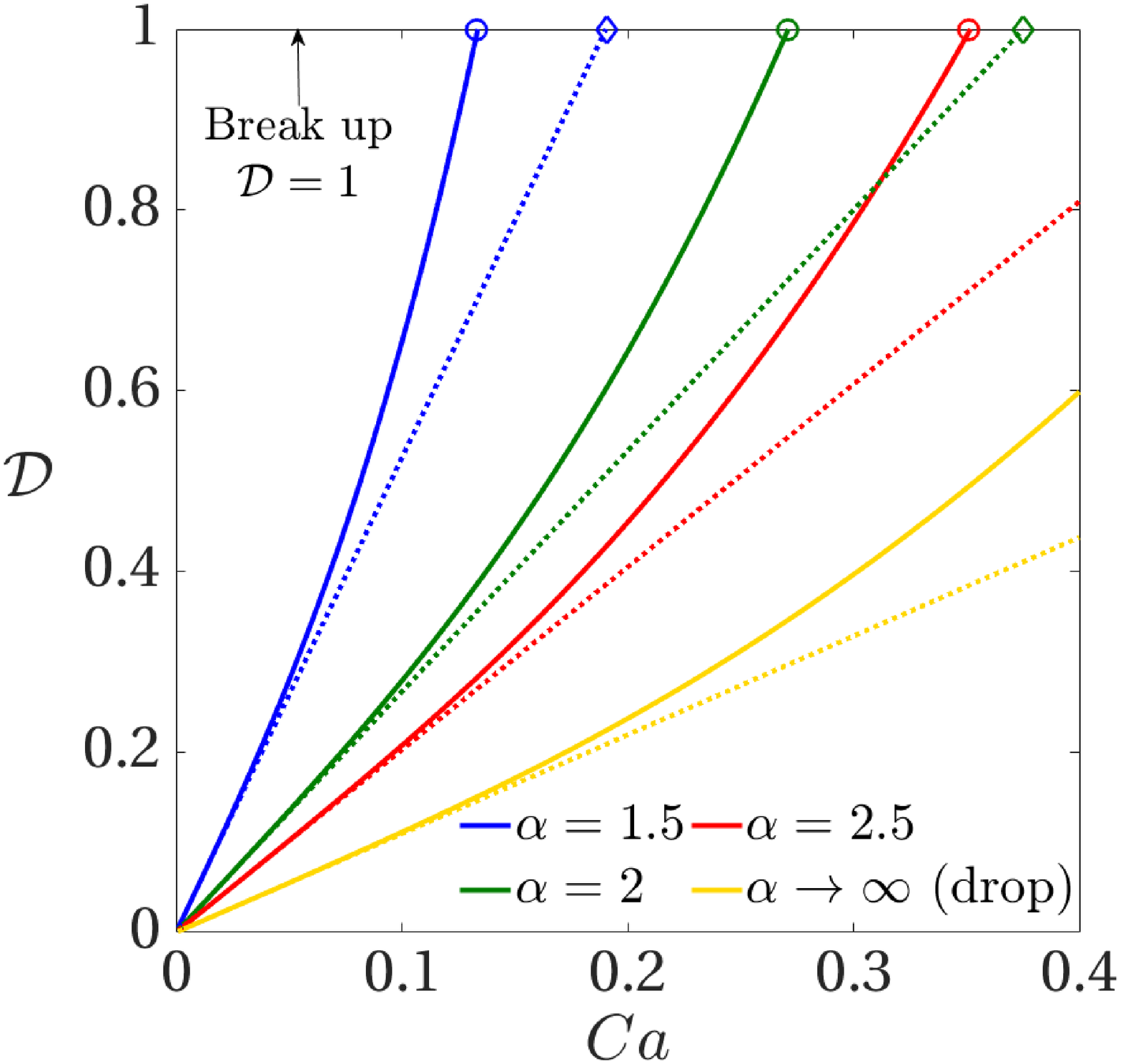}\label{fig5a}}\quad
		\subfigure[]{
			\includegraphics[height=5.0cm]{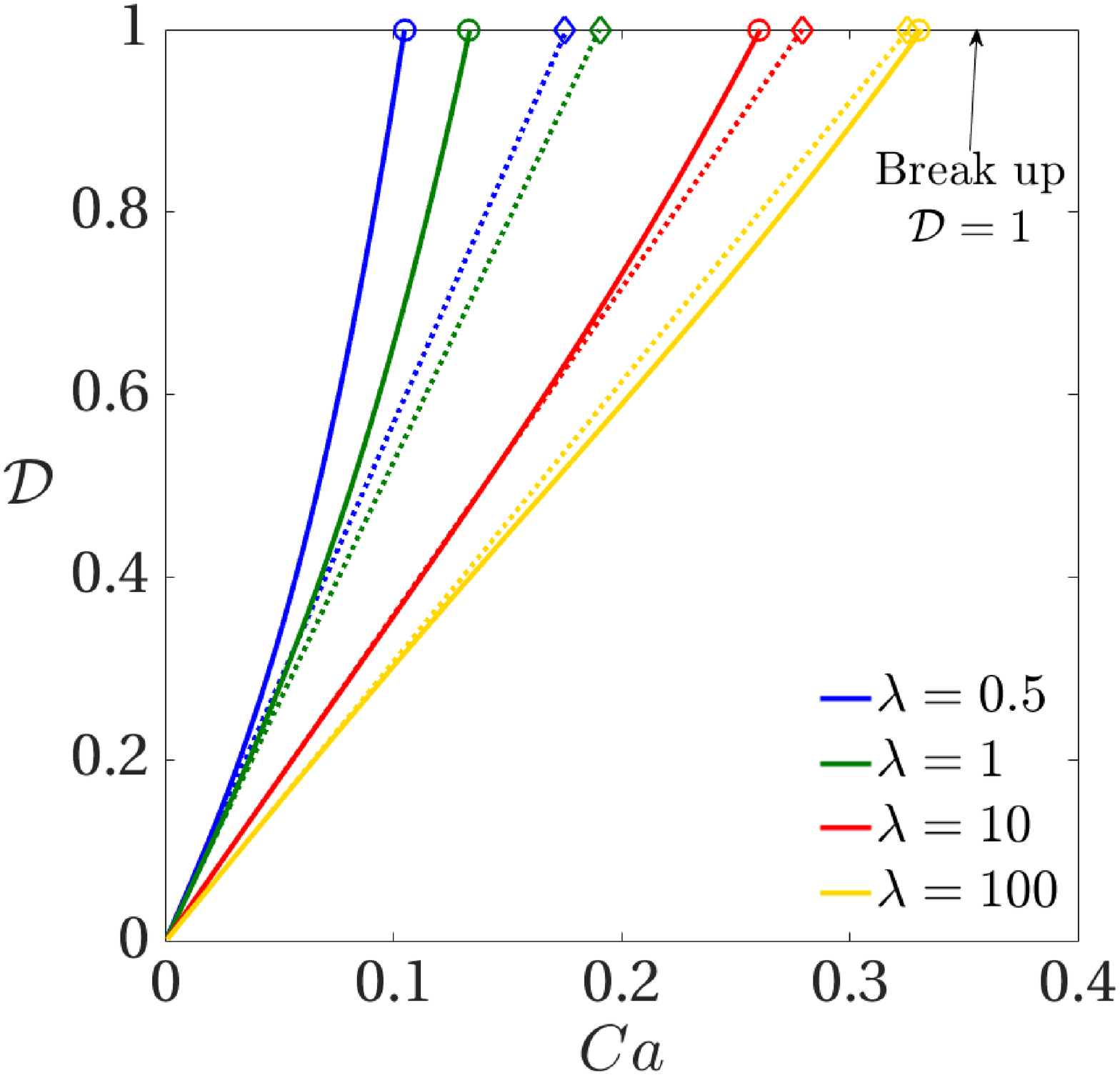}\label{fig5b}}\quad 
		\subfigure[]{
			\includegraphics[height=5.0cm]{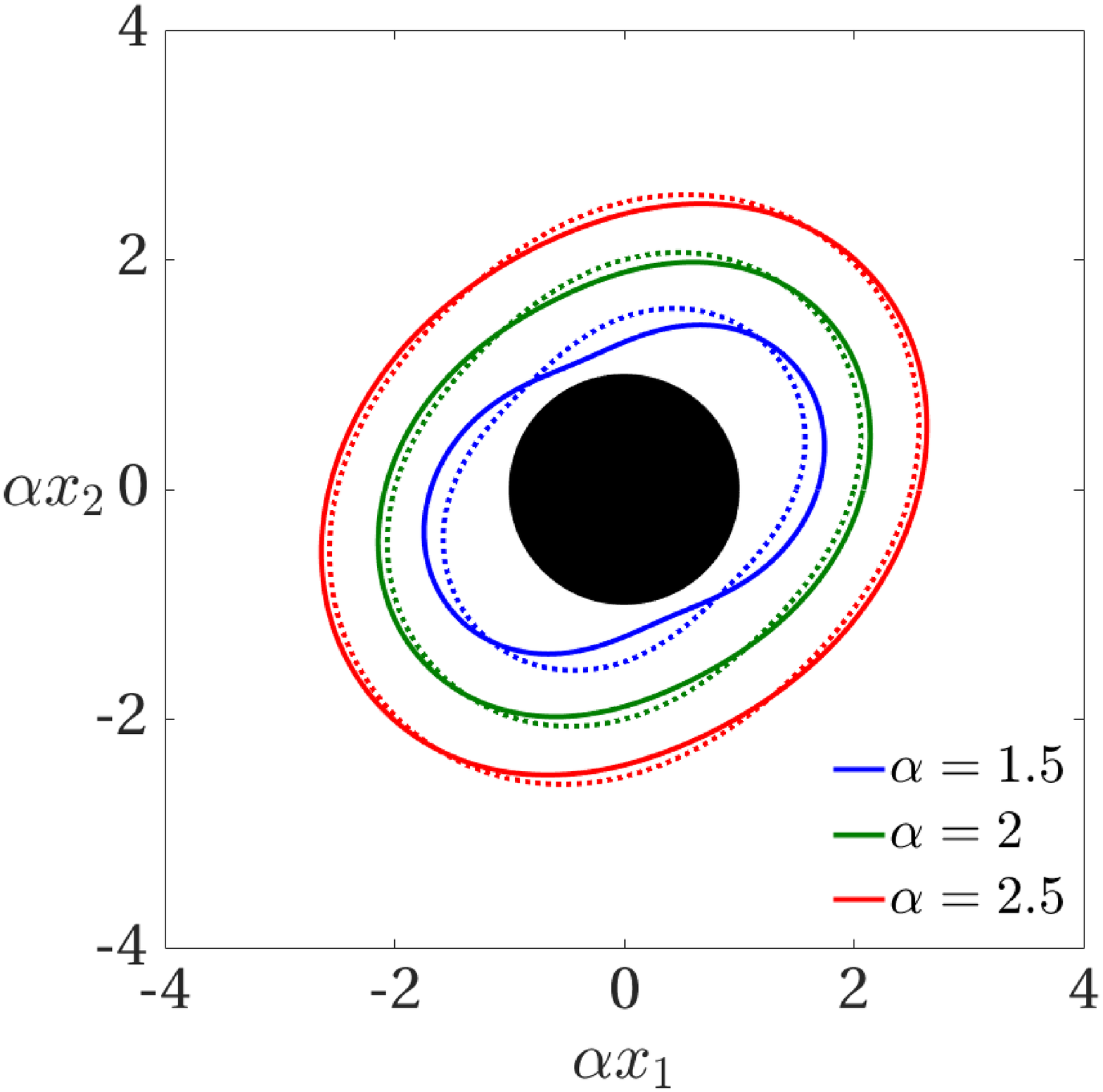}\label{fig5c}}\quad 
		\subfigure[]{
			\includegraphics[height=5.0cm]{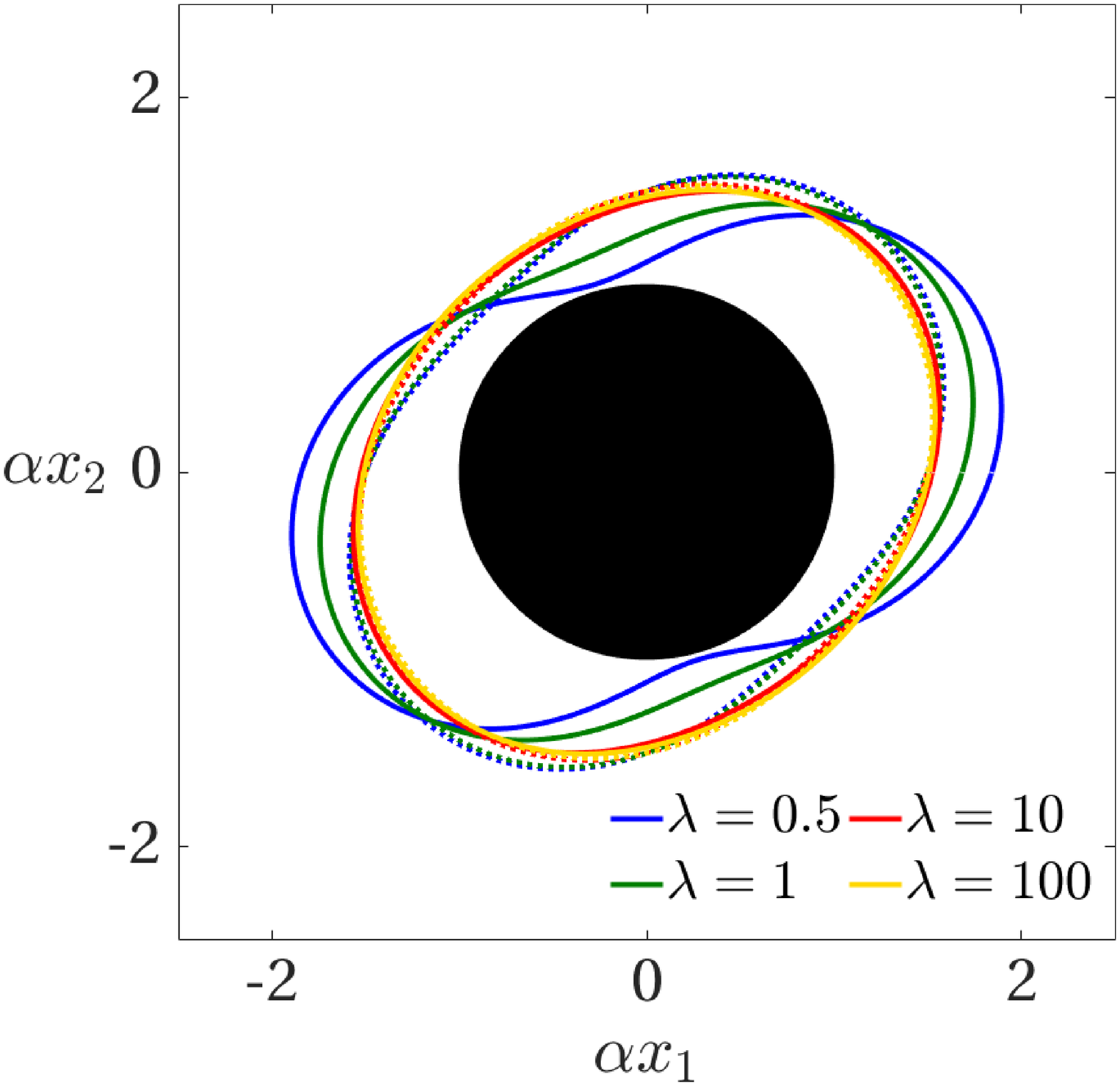}\label{fig5d}}\quad 
		\caption{The effect of $Ca$ on the steady state deformation parameter $\mathcal{D}$ (a) for various size ratio $\alpha$, at $\lambda=1$ and (b) for various viscosity ratio $\lambda$,  at $\alpha=1.5$. The markers in (a) and (b) indicates $Ca_{crit}$, the capillary number at which $D = 1$, which signifies breakup of the confining drop as the interface comes in contact with the encapsulated rigid particle. The corresponding steady state confining drop shapes at $Ca=0.1$ and (c) for various $\alpha$ and (d) for various $\lambda$. The solid and the dotted lines correspond to $\textit{O}(Ca^2)$ and $\textit{O}(Ca)$ calculations respectively. }
		\label{fig5}
	\end{figure}

	The extent of deformation of the confining drop is mainly dependent upon the three non-dimensional numbers, namely the capillary number $Ca$, the size ratio $\alpha$, and the viscosity ratio $\lambda$, as illustrated comprehensively in figure~\ref{fig5}. The steady state deformation parameter $\mathcal{D}$ as a function of $Ca$ for various $\alpha$ and $\lambda$ is shown in figures~\ref{fig5a} and \ref{fig5b} respectively. The corresponding steady state shapes of the deformed interface of the compound particle at a fixed $Ca =0.1$ but for various $\alpha$ and $\lambda$ are shown in figures~\ref{fig5c} and \ref{fig5d} respectively. The deformation parameter (or deformation of the confining drop) increases with increase in capillary number $Ca$, and decreases with increase in size ratio $\alpha$ and viscosity ratio $\lambda$. This dependency of steady state deformation parameter on $Ca$, $\alpha$, and $\lambda$ based on leading order calculations is reported by \cite{chaithu2019}, and they are shown as dotted lines in figure~\ref{fig5}. The $O(Ca^2)$ calculations are shown as solid lines in figure~\ref{fig5}. 

	As expected $\textit{O}(Ca)$ and  $\textit{O}(Ca^2)$ theory respectively show a linear and a quadratic dependence of $\mathcal{D}$ on $Ca$. More importantly, $\textit{O}(Ca)$ calculations underpredict the deformation parameter. (However, at very large viscosity ratios, $\textit{O}(Ca)$ calculations overpredict the value of $\mathcal{D}$.) Interestingly, the difference between $\textit{O}(Ca^2)$ and $\textit{O}(Ca)$ calculations decreases with decrease in $\alpha$, and with increase in $\lambda$ at a fixed capillary number. This is because, small $\alpha$ or large $\lambda$ correspond to cases where the deformed interface of the confining drop deviates least from the spherical shape. The degree of underprediction in $\mathcal{D}$ by linear theory is proportional to  $Ca$, showing that $\textit{O}(Ca^2)$ theory significantly deviates from $\textit{O}(Ca)$ predictions as capillary number increases. This observation also assumes importance because, compared to $\textit{O}(Ca^2)$ calculations the linear theory will also be overpredicting $Ca_{crit}$, the capillary number at which break up of the confining interface of the compound particle occurs. 
	
	
	\begin{figure}
		\centering
		\subfigure[]{
			\includegraphics[height=5cm]{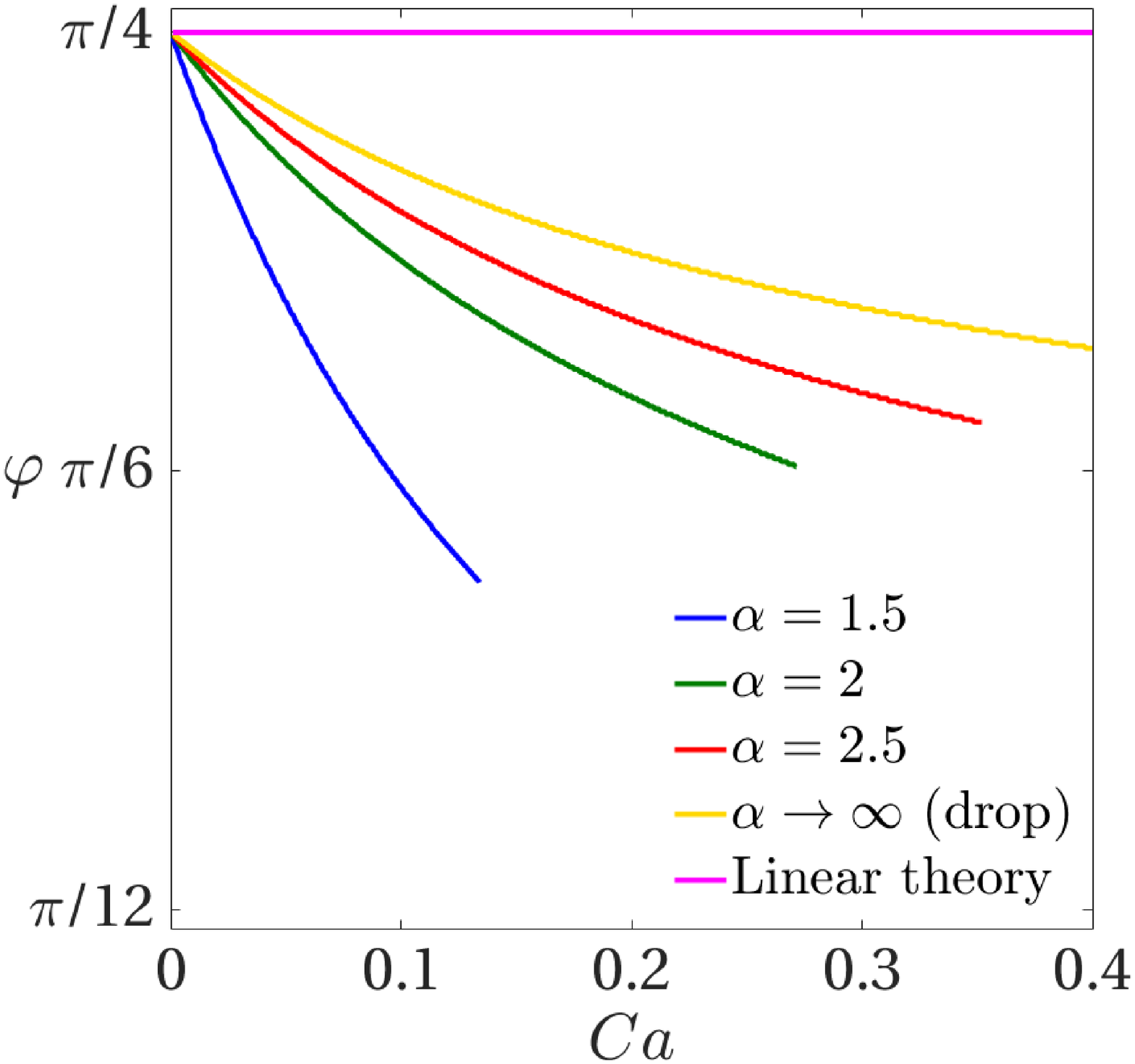}\label{fig6a}}\quad
		\subfigure[]{
			\includegraphics[height=5cm]{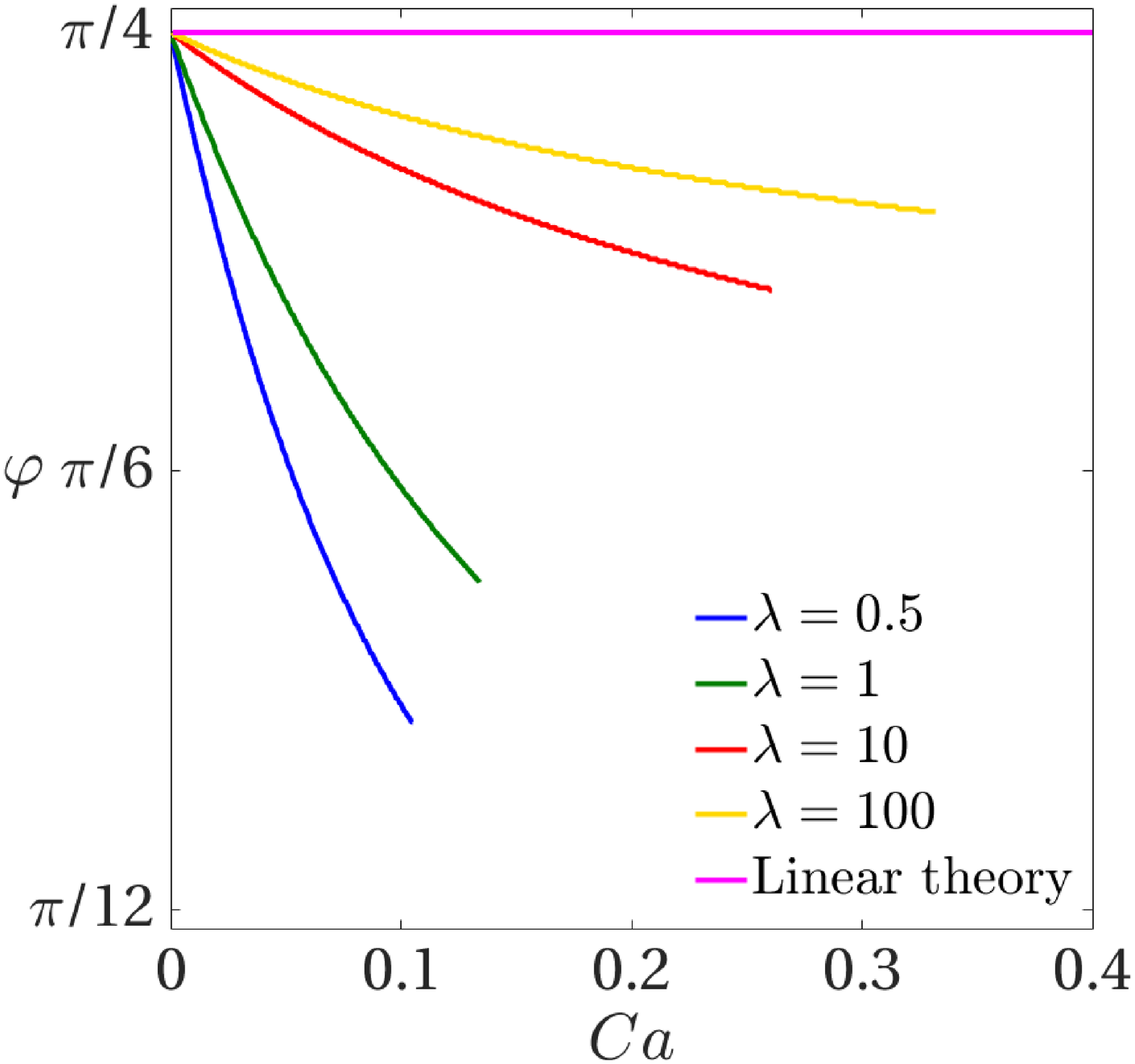}\label{fig6b}}\quad 
		\caption{The effect of capillary number on the steady state alignment angle, $\varphi$ calculated from (\ref{eq2b})  in a simple shear flow (a) for various size ratio at $\lambda = 1$ and (b) for various viscosity ratio at $\alpha = 1.5$.}
		\label{fig6}
	\end{figure}
	
	Similar to $\mathcal{D}$, the alignment angle also depends upon the capillary number, the size ratio and the viscosity ratio. Figures \ref{fig6a} and  \ref{fig6b} show the dependency of steady state $\varphi$ on $Ca$ for various size ratios and viscosity ratios. In all cases, the alignment angle decreases with increase in $Ca$, making the elongated interface to orient closer to the flow direction of the imposed shear flow. The vorticity associated with the imposed flow acts to rotate the deformed drop towards the flow axis, while the surface tension force acts to revert it back to the spherical shape, and this competition determines the alignment angle. If the surface tension forces dominate the shear forces (smaller Ca), the confining drop aligns with the extensional axis, otherwise the confining drop aligns more with the flow axis. Similarly, $\varphi$ decreases with decrease in $\alpha$ or $\lambda$ at a given $Ca$. Thus, the deformed drop aligns closer to the flow axis in strong shear flows, and when the encapsulated particle is bigger and the viscosity of the confining fluid is larger.   This behavior of alignment angle following that of deformation parameter occurs as larger deformation results in more anisotropic shapes that orient closer to the flow direction.

	To summarize this section, we have analyzed fluid flows in and around a compound particle as well as the consequent deformation of the confining interface when it is subjected to an imposed shear flow. The $\textit{O}(Ca^2)$ calculation of the deformed shape of the confining drop shows that the deformation of the interface increases with  increase in $Ca$ and this larger deformation results in orienting the deformed interface to align close to the flow axis in a simple shear flow. Further, the steady state deformation parameter and the critical capillary number increase while the alignment angle decreases with increase in $\alpha$ and $\lambda$ due to increased hydrodynamic interaction between the enclosed solid and the confining interface.
	
	\subsection{Extensional flows}\label{extension}
	In this section, we analyze the deformation of the confining drop when the compound particle is subjected to an extensional flow. This analysis is important since earlier investigations, both experimental  \citep{taylor1934,grace1982,bai1979,mietus2002,mulligan2011} and theoretical or numerical \citep{taylor1934,qu2012,bai1979}, have concluded that the deformation and breakup of a simple drop (without encapsulated particles) differ considerably when subjected to shear and extensional flows.
	
	Consider the two extensional flows namely, the uniaxial and biaxial flows described by the strain rate tensor,
	\begin{equation}\label{eq4b}
		\bm{E}^{E}=\frac{1}{2}\begin{bmatrix}
			\mp1 & 0 & 0 \\
			0 & \mp1 & 0\\
			0 & 0 & \pm2
		\end{bmatrix}, \quad \bm{\Omega}^{E}=\mathbf{0},
	\end{equation}
	where the first and second sign indicate uniaxial and biaxial flows respectively. We follow the same convention throughout this section. Then the shape of the confining drop for uniaxial and biaxial flows can be obtained as
	\begin{equation}\label{eq5b}
		\begin{aligned}
			r(\theta,  \phi)= \bigg(
			1+b_{1} Ca  \Big(\mp\frac{1}{2}\frac{(x_{1}^{2}+x_{2}^{2})}{r^{2}}\pm\frac{x_{3}^{2}}{r^{2}}\Big)+Ca^{2}\bigg(b_{2}\Big(\mp\frac{1}{2}\frac{(x_{1}^{2}+x_{2}^{2})}{r^{2}}\pm\frac{x_{3}^{2}}{r^{2}}\Big)\\
			+\frac{3}{2}b_{3} +b_{4} \Big(\frac{1}{4}\frac{(x_{1}^{2}+x_{2}^{2})}{r^{2}}+\frac{x_{3}^{2}}{r^{2}}\Big)  
			+b_{5} \Big(\frac{1}{2}\frac{(x_{1}^{2}+x_{2}^{2})}{r^{2}}-\frac{x_{3}^{2}}{r^{2}}\Big)^{2}\bigg)+\textit{O}(Ca^3) \bigg).
		\end{aligned}
	\end{equation}
	\begin{figure}
		\centering
		\subfigure[]{
			\includegraphics[height=4.5cm]{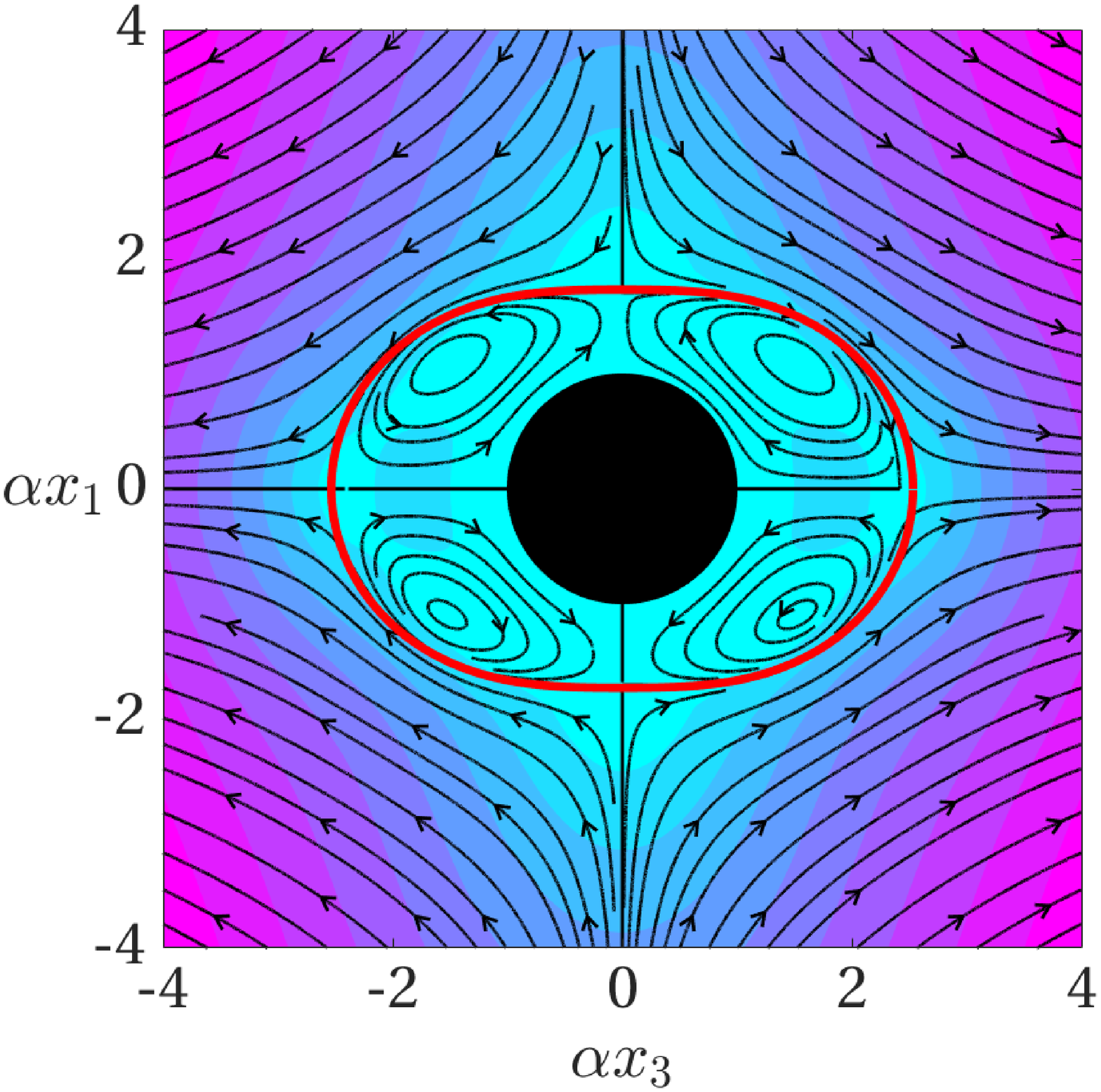}\label{fig7a}}\quad
		\subfigure[]{
			\includegraphics[height=4.5cm]{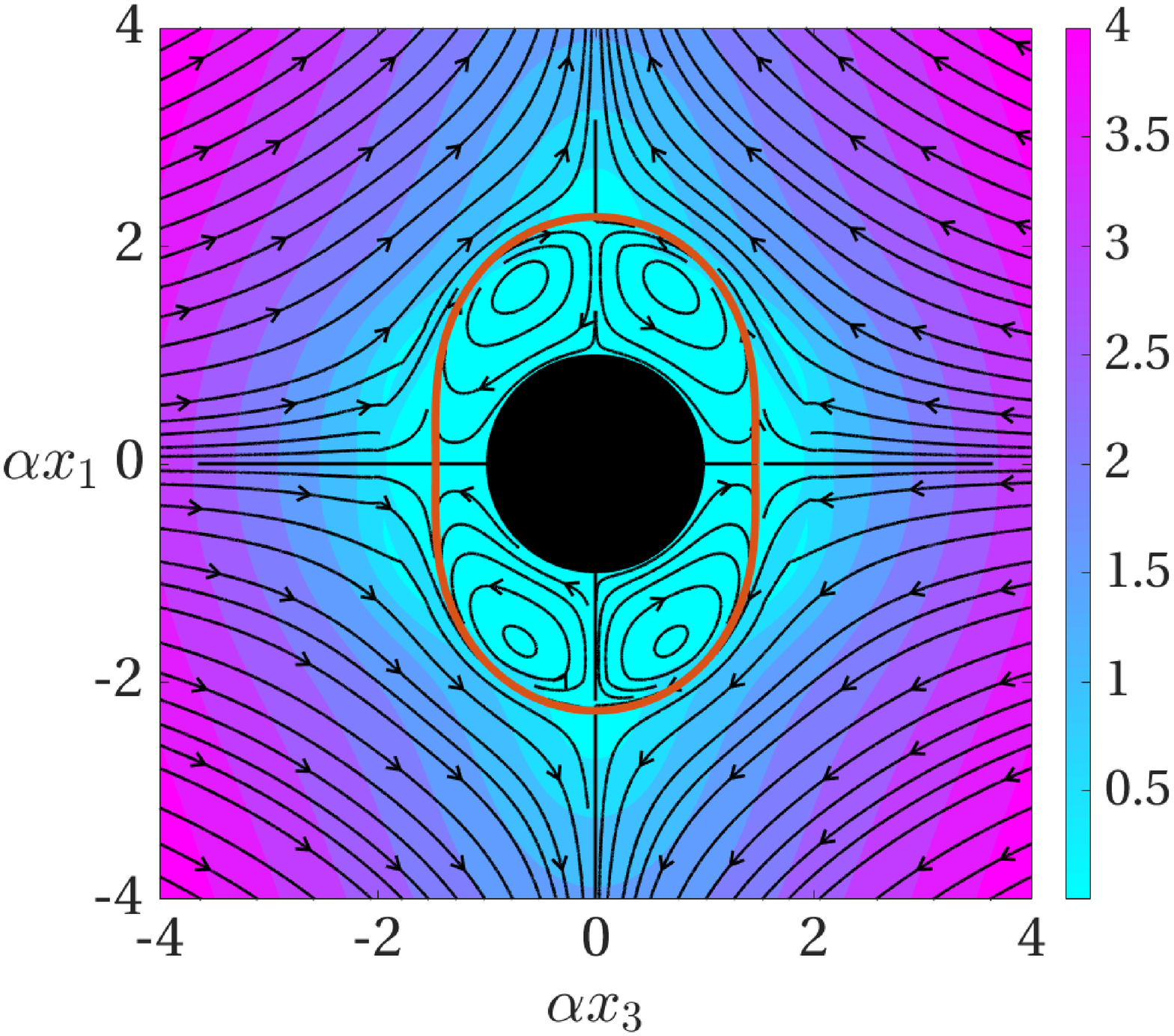}\label{fig7b}}\quad \quad
		\subfigure[]{
			\includegraphics[height=5cm]{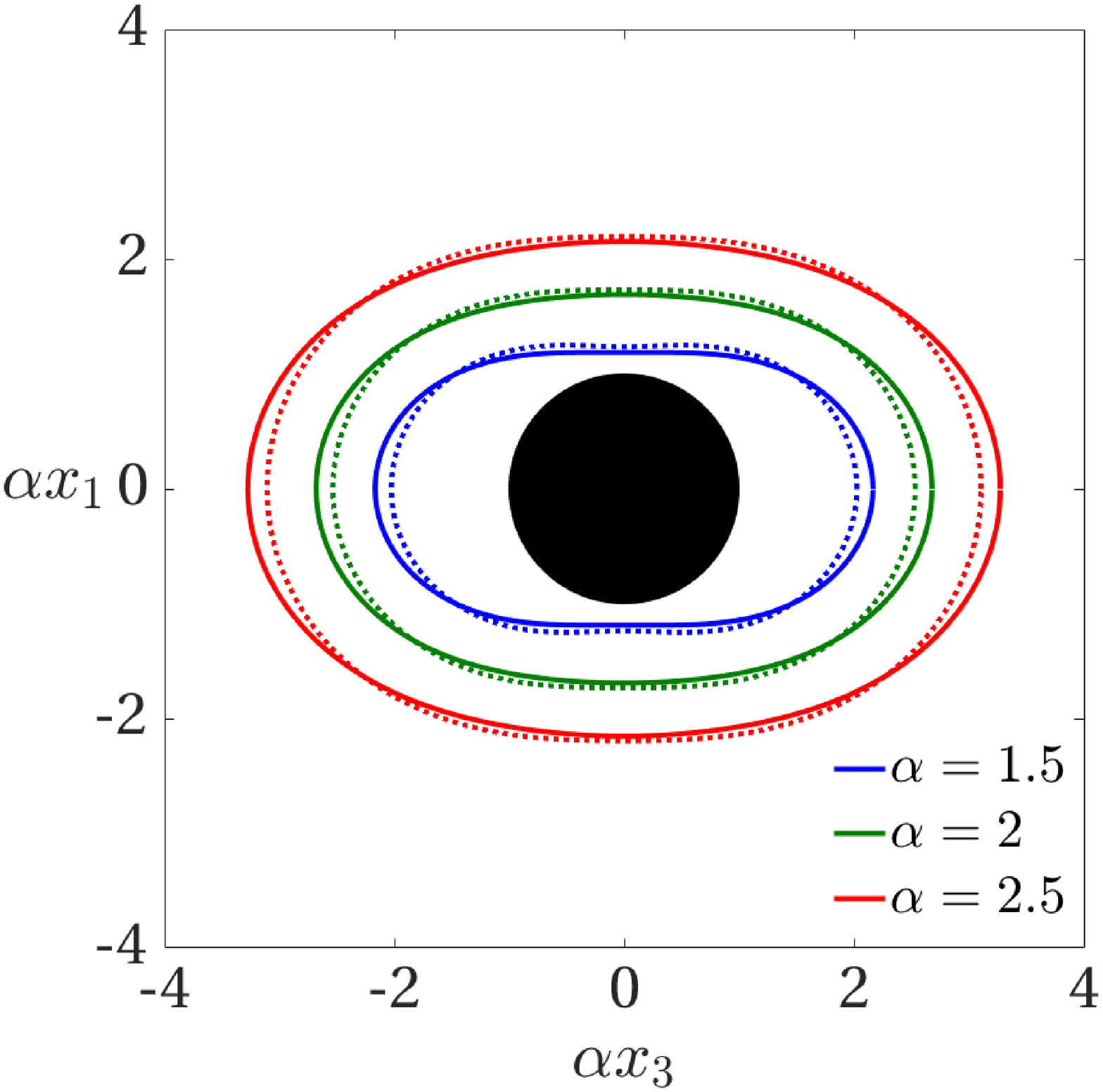}\label{fig7c}}\quad 
		\subfigure[]{
			\includegraphics[height=5cm]{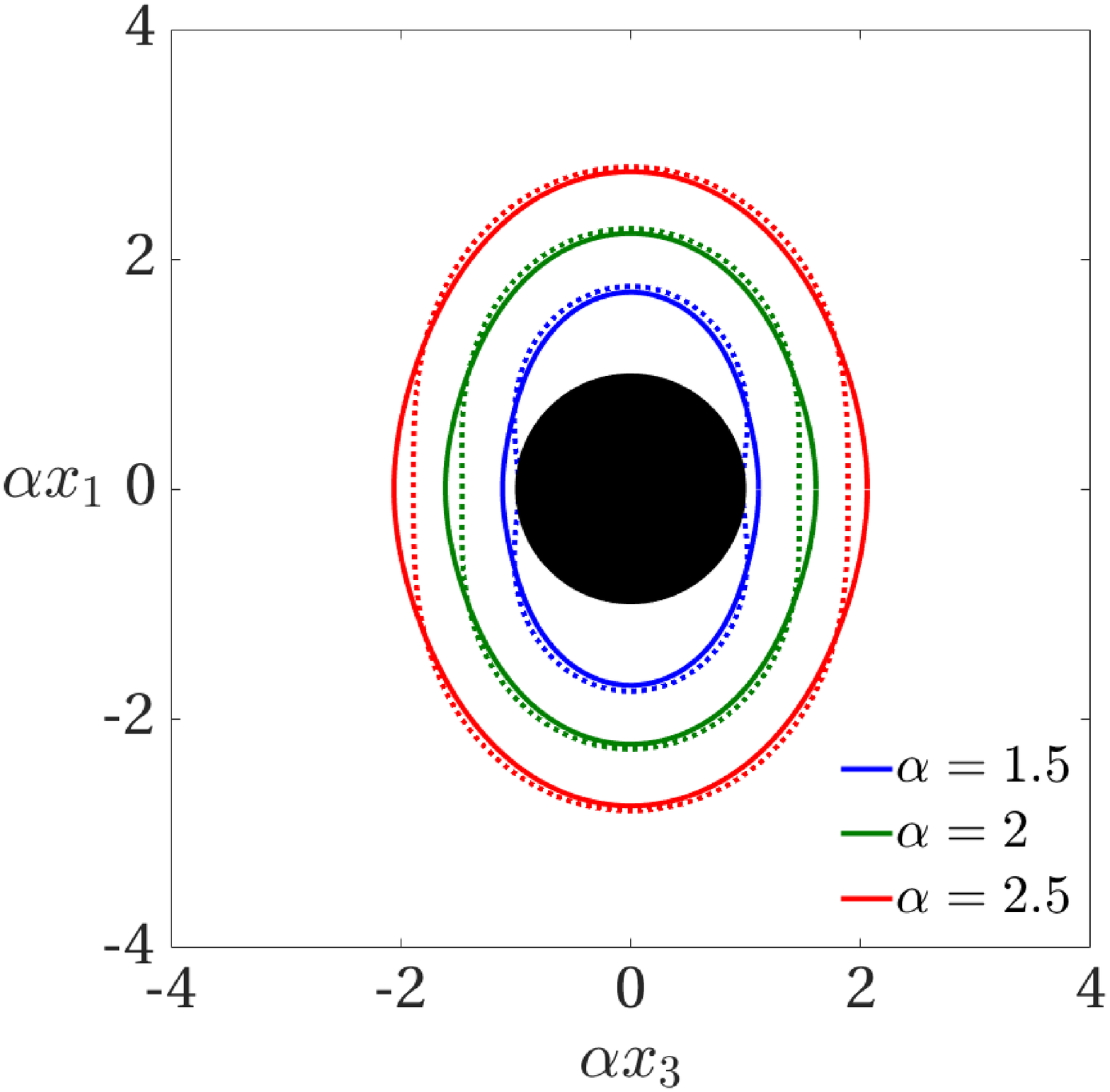}\label{fig7d}}\quad 
		\subfigure[]{
			\includegraphics[height=5cm]{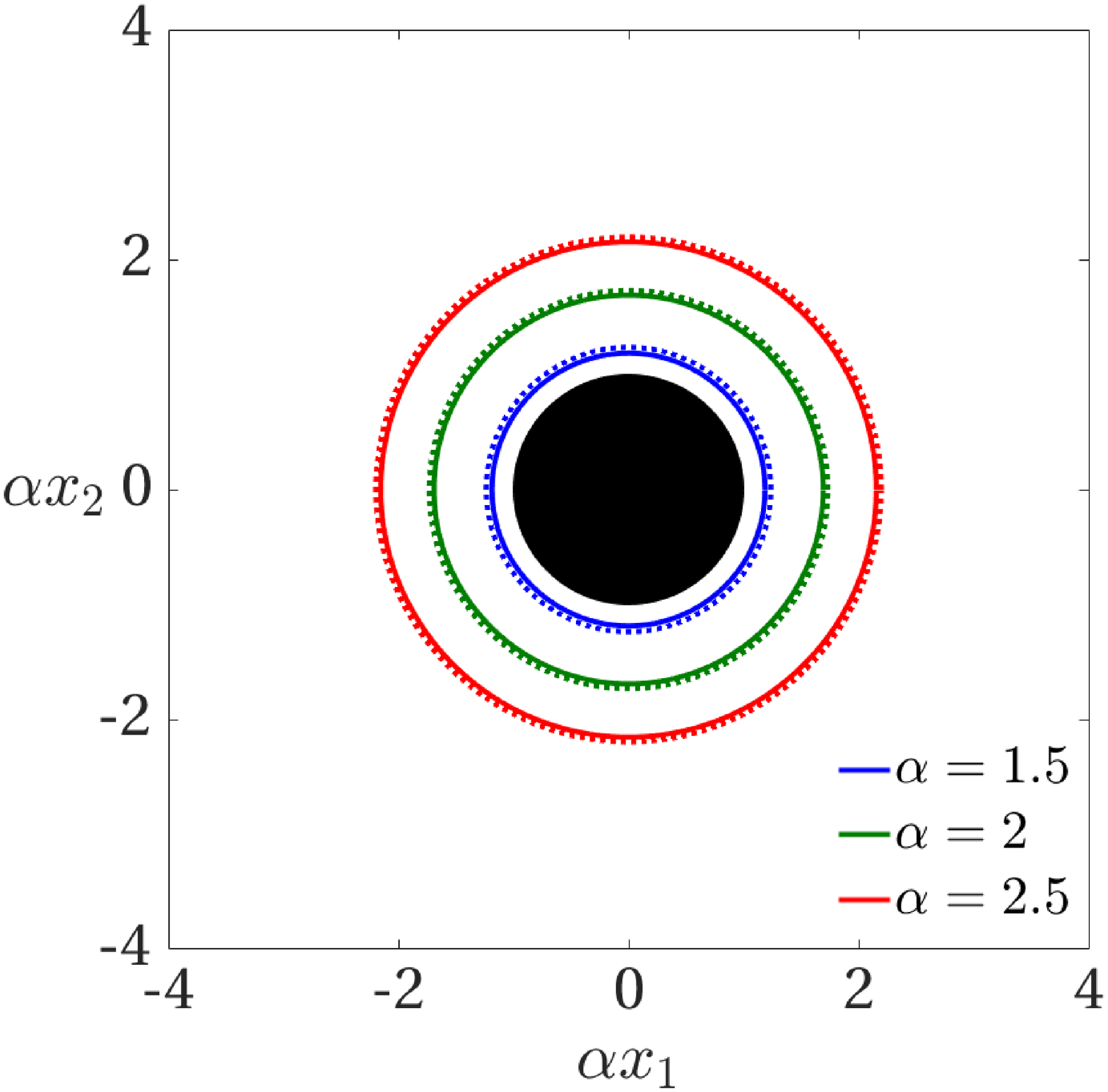}\label{fig7e}}\quad 
		\subfigure[]{
			\includegraphics[height=5cm]{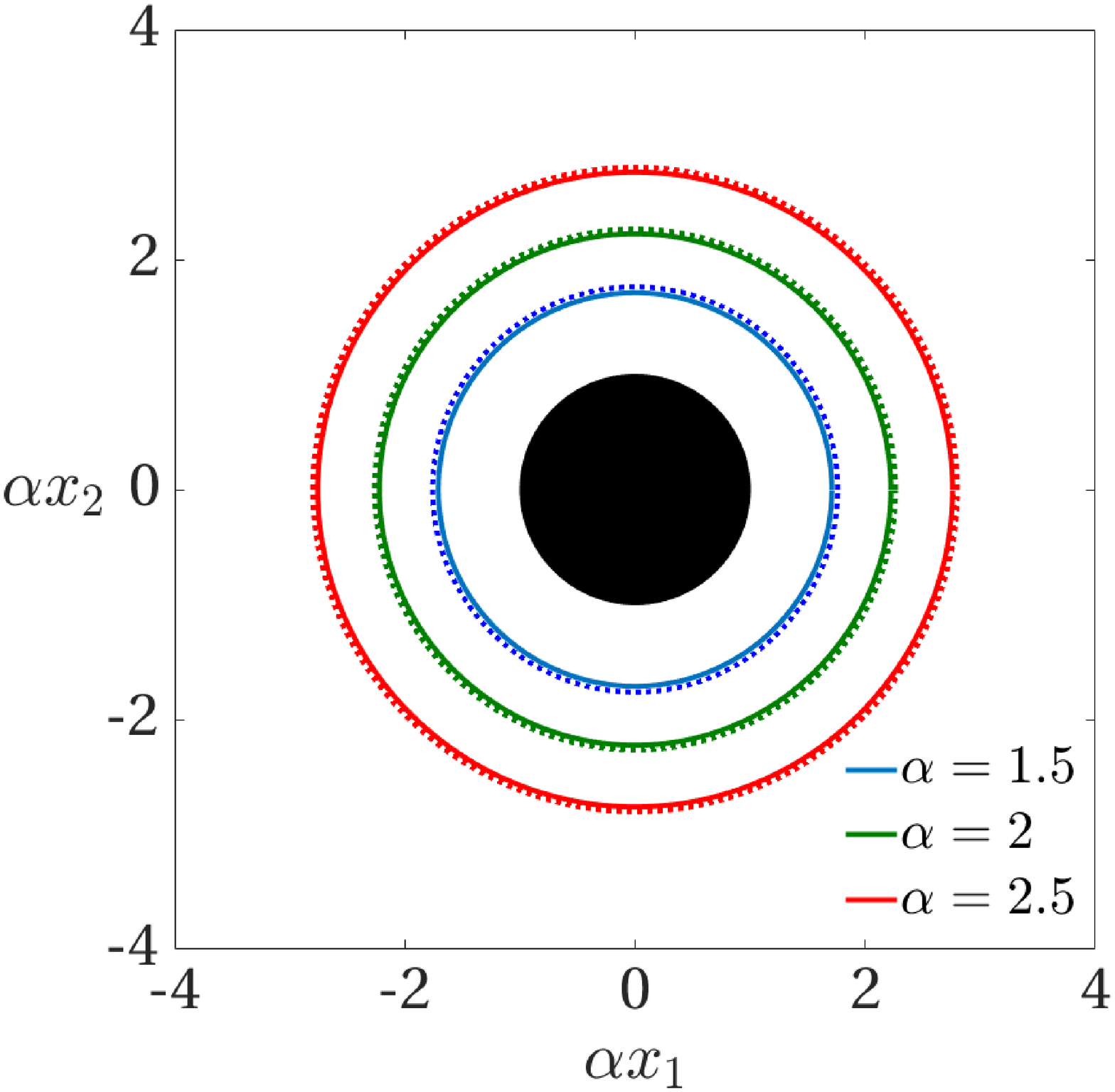}\label{fig7f}}\quad 
		\caption{Velocity field in and around a compound particle in an imposed (a) uniaxial and (b) biaxial flow. Here, $\alpha=2$, $Ca=0.1$ and $\lambda=1$. The color bar represents the magnitude of the velocity. Effect of the size ratio $\alpha$ on the deformation of a compound particle (c) in $x_{1}$-$x_{3}$ plane and (e) in $x_{1}$-$x_{2}$ plane for uniaxial flow, and (d) in $x_{1}$-$x_{3}$ plane and (f) in $x_{1}$-$x_{2}$ plane for biaxial flow. The solid and dotted lines represent the drop shape at $\textit{O}(Ca^2)$ and $\textit{O}(Ca)$ \citep{chaithu2019} respectively. }\label{fig7}
	\end{figure}
	
	Figures~\ref{fig7a} and \ref{fig7b} show the velocity fields in and around a compound particle obtained for uniaxial and biaxial flows respectively. The deformed shape of the confining drop is also plotted (solid red line). Clearly, the velocity fields are similar in both cases with recirculating fluid flows appearing in each quadrant. The orientation of the elongated interface is along the extensional axis of the flow but the extent of the deformation is different for uniaxial and biaxial flows. For uniaxial flows, extensional axis of the imposed flow is along the $x_3$ axis while for biaxial flows extensional flow is in the radial direction in $x_1-x_2$ plane. This difference results in the confining interface adopting a prolate spheroid like shape for uniaxial flows and an oblate spheroid like shape for biaxial flows. These shapes, for various $\alpha$, when viewed in two different planes, $ x_1 -  x_3$ plane and $ x_1 - x_2$ plane are shown in  figures~\ref{fig7c}-\ref{fig7f}. The axisymmetry of the deformed shapes is apparent in the $ x_1 -  x_2 $ plane in all cases. 
	
	As earlier, the deformation parameter $\mathcal{D}$ can also be calculated using (\ref{eq3b}) for uniaxial and biaxial flows as
	\begin{equation}\label{eq6b}
		\begin{aligned}
			\mathcal{D}=\frac{3\alpha}{4(\alpha-1)}\Bigg( b_{1} Ca+Ca^{2}\bigg(b_{2}\pm\frac{b_{4}}{2}\pm\frac{b_{5}}{2}\mp\frac{\alpha b_{1}^{2}}{4(\alpha-1)} \bigg)  \Bigg) +\textit{O}(Ca^3).
		\end{aligned}
	\end{equation}
	In the limit of $\lambda \rightarrow \infty$ and ($\alpha - 1) \rightarrow 0$ such that the hydrodynamic interaction between the encapsulated particle and the interface is a dominant effect, (\ref{eq6b}) reduces to give the deformation parameter as,
	\begin{equation}\label{eq7b}
		\mathcal{D}=
		\frac{630\alpha(\alpha-1)Ca\pm\alpha(671\alpha-986)Ca^{2}}{420 (\alpha-1)^2}.
	\end{equation}
	\noindent Thus, $\mathcal{D}$ is same for uniaxial and biaxial flows upto $\textit{O}(Ca)$ but differ at $\textit{O}(Ca^2)$. This difference is also illustrated in figures~\ref{fig7c}-\ref{fig7f}. The deformation calculated at $\textit{O}(Ca^2)$ (solid lines) is more compared to that at $\textit{O}(Ca)$ (dotted lines) for uniaxial flows. On the other hand, for biaxial flows, drop deformation and thus the predicted value $\mathcal{D}$ are less at $\textit{O}(Ca^2)$ compared to that at $\textit{O}(Ca)$. In other words,  $\textit{O}(Ca)$ calculation underpredicts deformation for uniaxial flows and overpredicts it for biaxial flows.
	
\begin{figure}
 		\centering
 		\includegraphics[height=5cm]{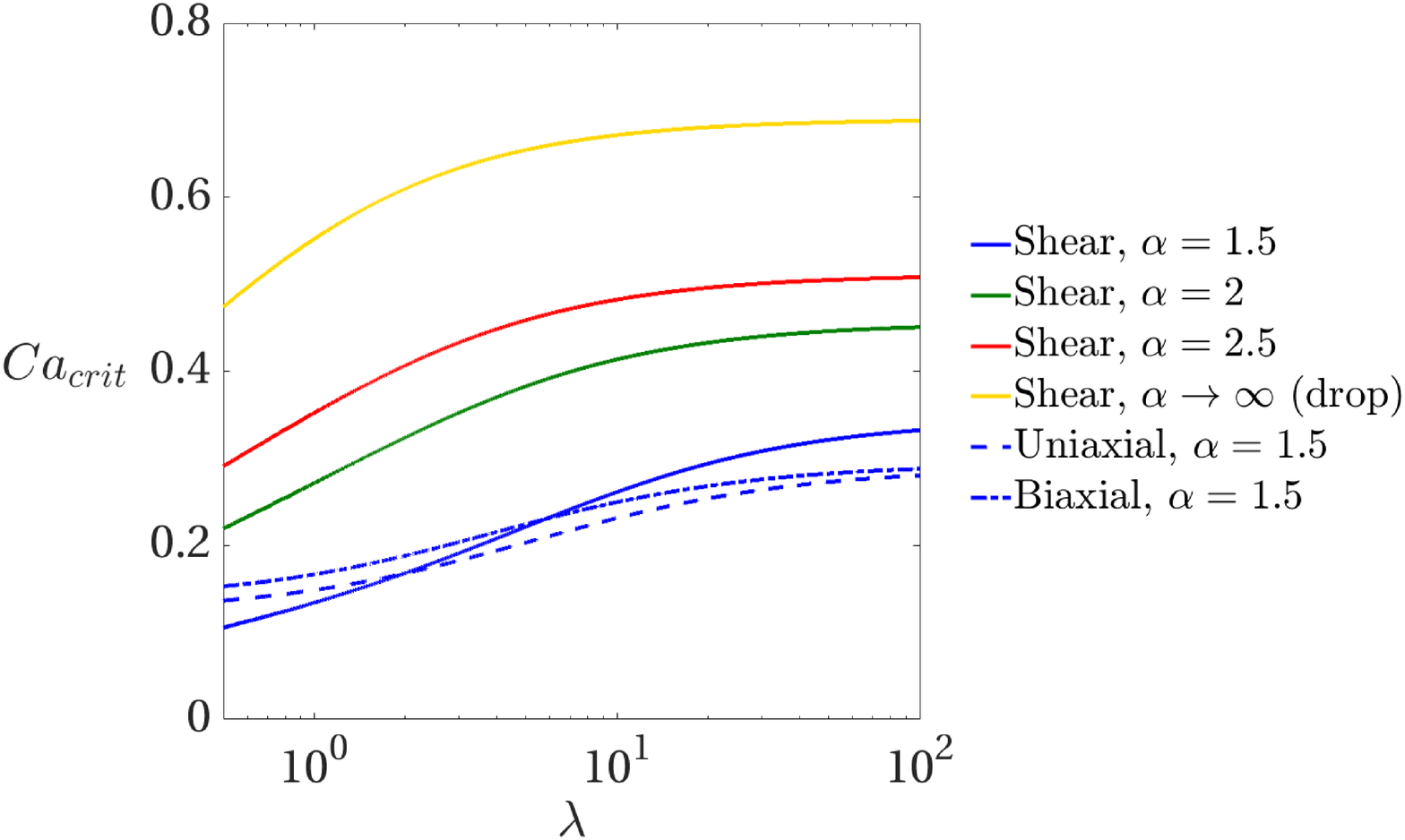}
 		\caption{The variation of the critical capillary number $Ca_{crit}$ versus the viscosity ratio $\lambda$ for various size ratios $\alpha~(=1.5, 2, 2.5)$ and in different imposed flows.}
 		\label{fig8}
 	\end{figure}

	Similar to the case of a simple shear flow, the deformation of a confining drop in extensional flows increases with decreasing $\alpha$ or $\lambda$ for a particular $Ca$, and increases with increase in $Ca$. These dependencies can also be understood by monitoring critical capillary number, $Ca_{crit}$ beyond which the confining drop breaks up ($\mathcal{D} = 1$). Therefore, $Ca_{crit}$ is plotted as a function of $\lambda$ for various $\alpha$ and various imposed flows in figure~\ref{fig8}. Irrespective of the flow type, $Ca_{crit}$ increases with increase in $\lambda$ or $\alpha$. For small viscosity ratios, $\lambda<1$, $Ca_{crit}$ for shear flows is smaller suggesting that shear flows are stronger in deforming the interface compared to the uniaxial and biaxial flows. 
	This trend reverses for large $\lambda$. As mentioned earlier, due to the difference in the deformation characteristics, uniaxial flows show lower $Ca_{crit}$ compared to biaxial flows.
	
	\subsection{Generalized shear and extensional flows}
	In this section, we expand the analysis described in the two previous sections to generalised shear and generalised extensional flows. 
	We may define a generalized shear flow \citep{graham2018} which has the velocity gradient tensor as,
	\begin{equation}\label{eq8b}
		\bm{E}^{GS}=\frac{(1+\beta)}{2}\begin{bmatrix}
			0 & 1 & 0 \\
			1 & 0 & 0\\
			0 & 0 & 0
		\end{bmatrix}, \quad  \bm{\Omega}^{GS}=\frac{(1-\beta)}{2}\begin{bmatrix}
			0 & 1& 0 \\
			-1 & 0 & 0\\
			0 & 0 & 0
		\end{bmatrix},
	\end{equation}
	where the parameter $\beta$ takes the value from $-1$ to $1$. The limiting cases of $\beta=-1$  and $\beta=1$  correspond to the pure rotational and pure straining flows, respectively. Simple shear flow is obtained when $\beta = 0$. 
	
\begin{figure}
		\centering
		\subfigure[]{
			\includegraphics[height=5cm]{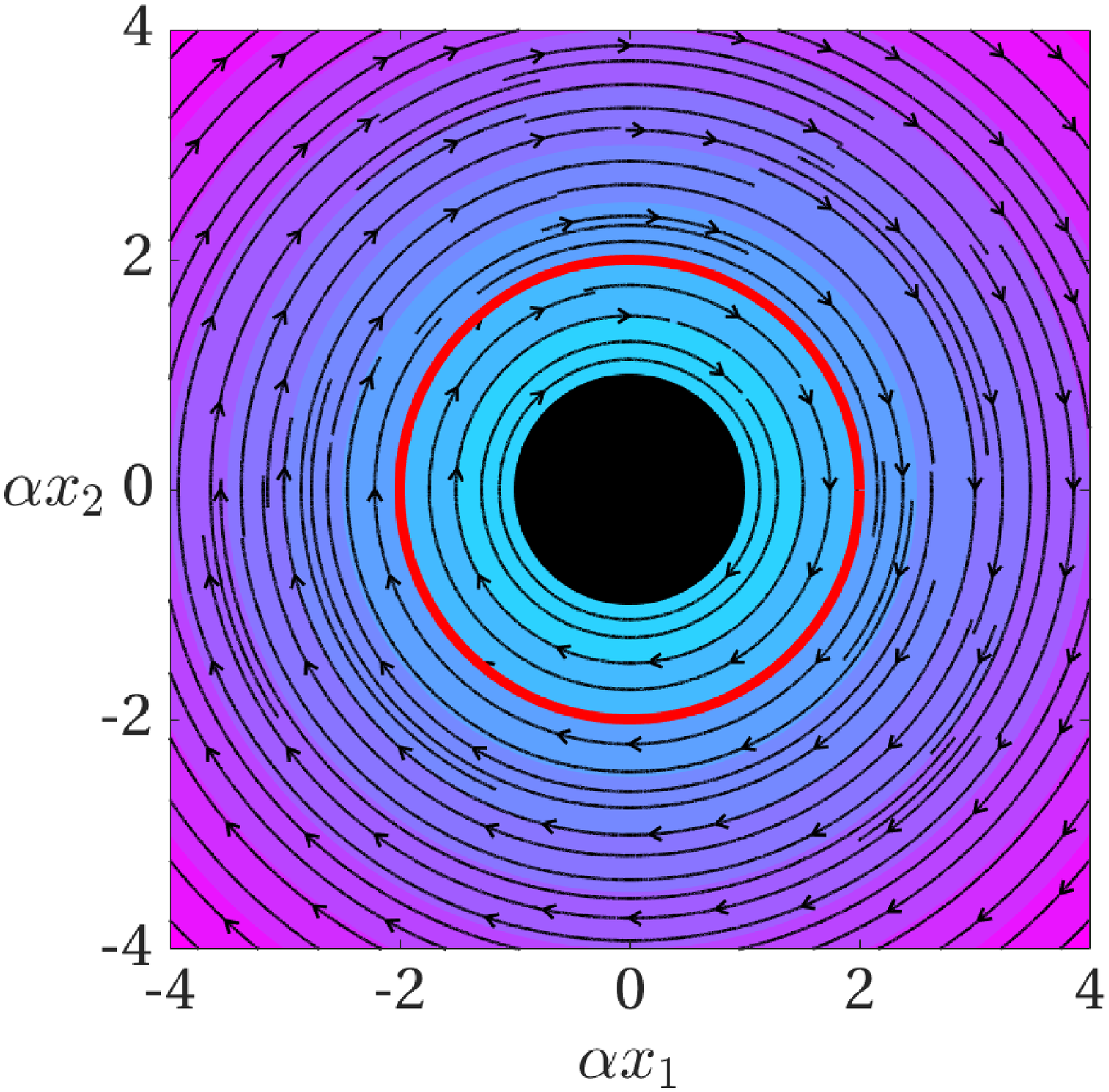}\label{fig9a}}\quad
		\subfigure[]{
			\includegraphics[height=5cm]{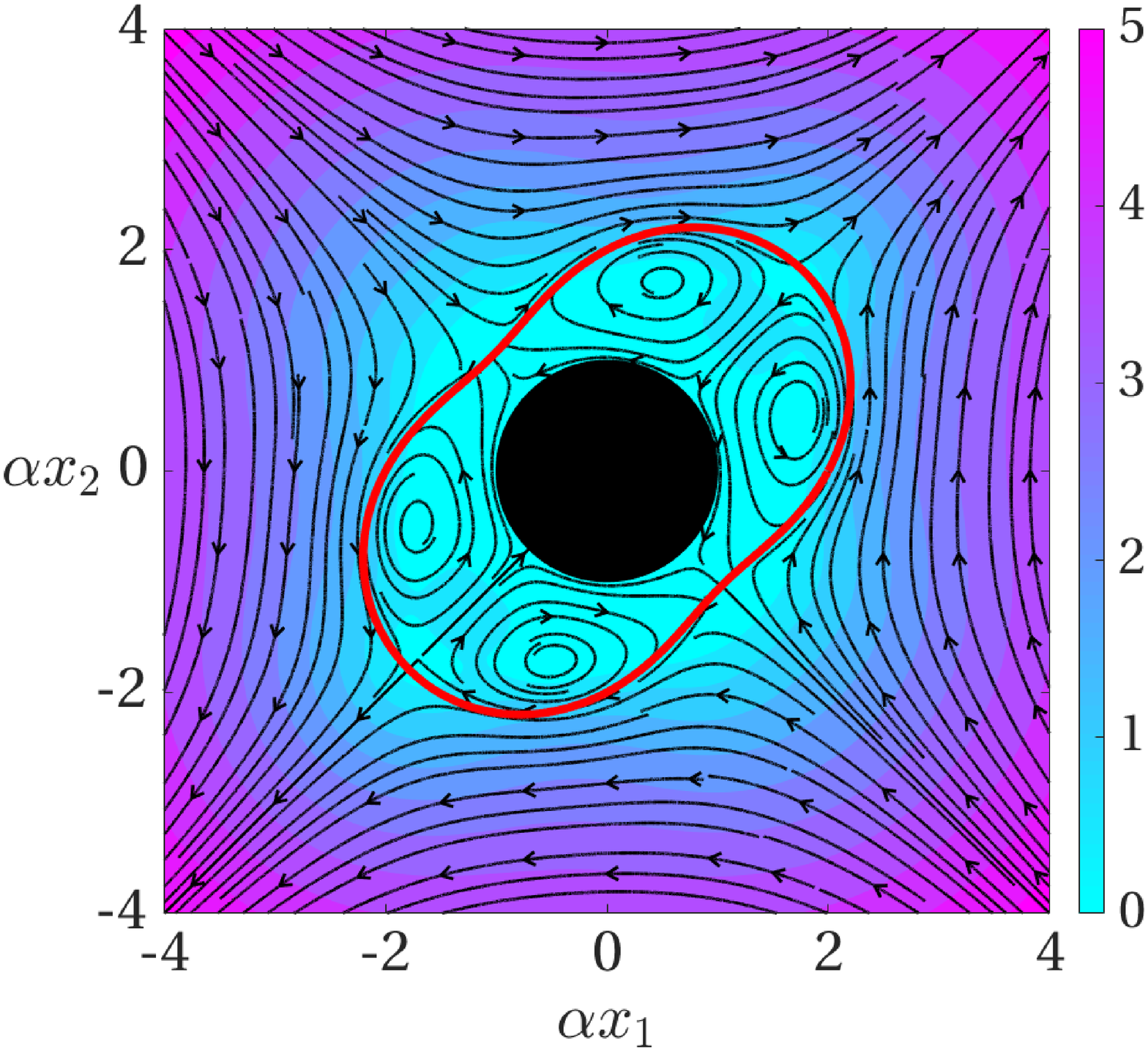}\label{fig9b}}\quad 
		\caption{Velocity field in and around a compound particle in an imposed flow (a) $\beta=-1$ (b) $\beta=1$. Here, $\alpha=2$, $Ca=0.05$ and $\lambda=1$. The color bar represents the magnitude of the velocity.}\label{fig9}
	\end{figure}

	Figure \ref{fig9} shows the steady state flow fields in and around a deformed compound particle for $\beta = -1$ and $\beta = 1$. As shown in figure \ref{fig9a}, for $\beta = -1$ which corresponds to a pure rotational flow, the velocity field has only azimuthal component of velocity, there will not be any velocity  normal to the interface, and therefore the confining drop remains spherical. As $\beta$ increases, the strength of the extensional component in the imposed flow increases (refer figure \ref{fig9b}) which results in larger deformations, with largest $D$ obtained for a purely extensional flow ($\beta = 1$).
	
\begin{figure}
		\centering
		\subfigure[]{
			\includegraphics[height=5.0cm]{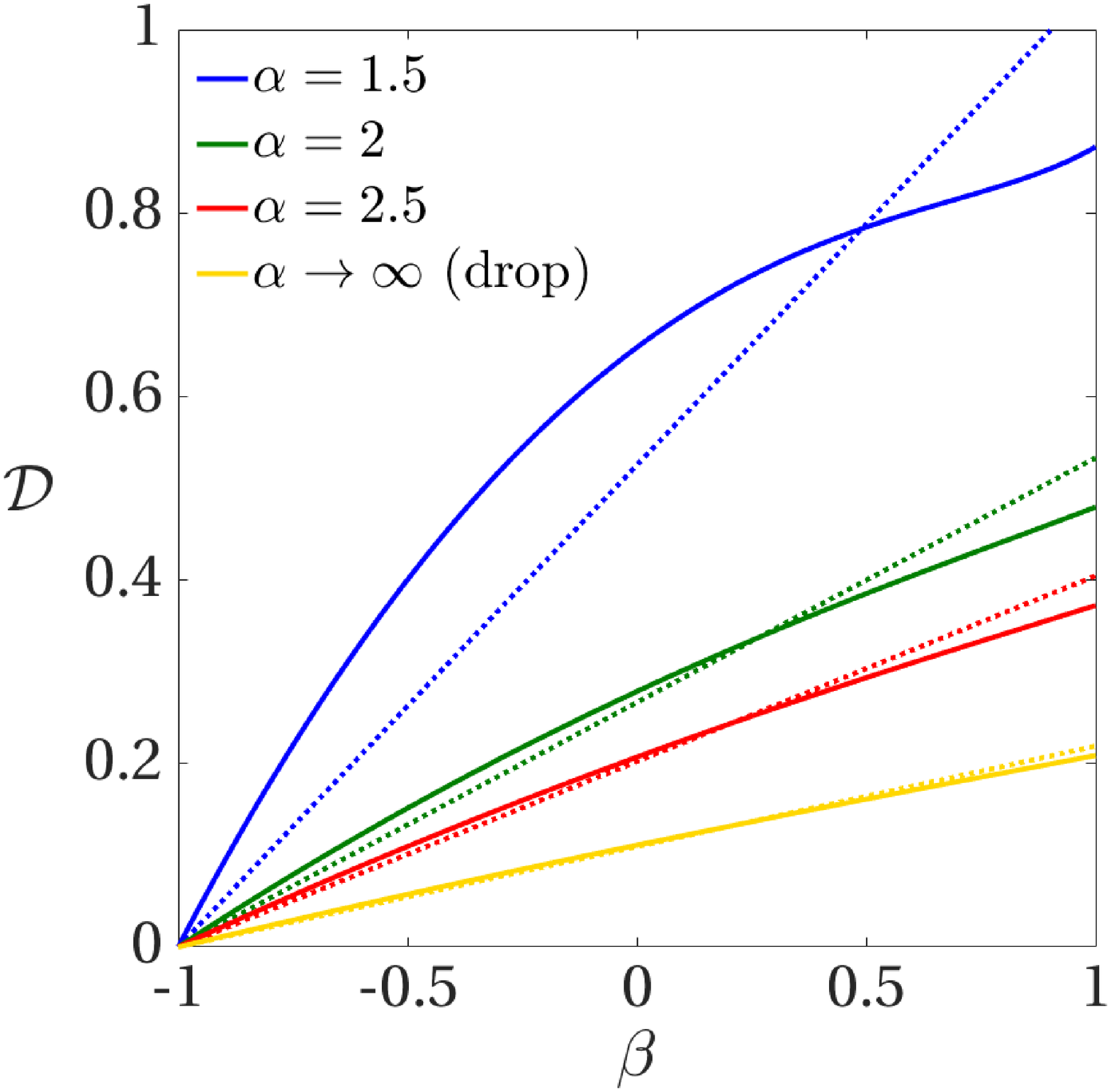}\label{fig10a}}\quad
		\subfigure[]{
			\includegraphics[height=5.0cm]{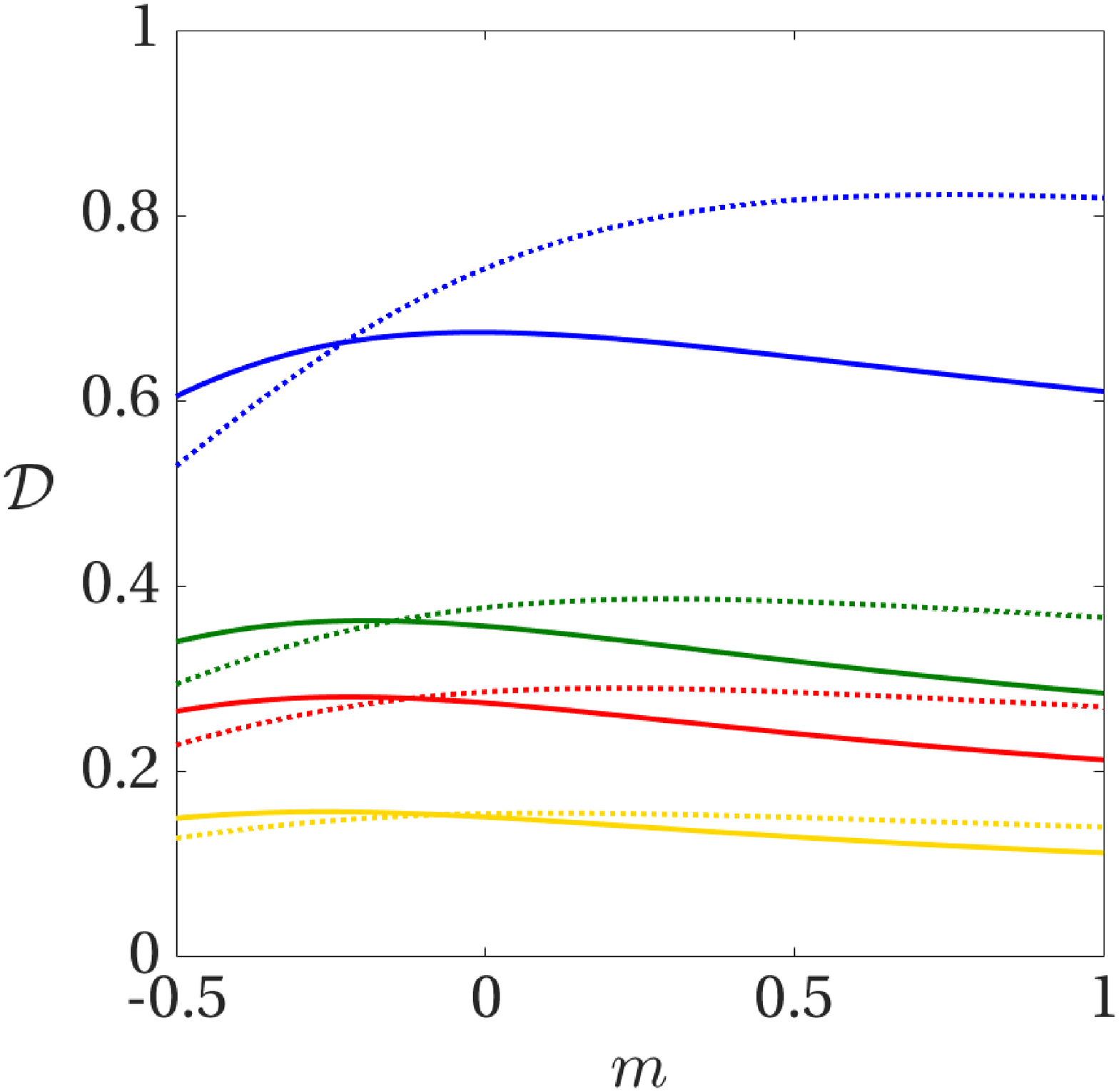}\label{fig10b}}\quad 
	\caption{The deformation parameter $\mathcal{D}$ (a) in generalised shear flows, as a function of $\beta$ and (b) in generalised extensional flows as a function of $m$. In these plots, $\lambda=1$ and $Ca=0.1$. Solid and dotted lines correspond to $\textit{O}(Ca^2)$ and $\textit{O}(Ca)$ calculations.}\label{fig10}
	\end{figure}

	Figure \ref{fig10a} summarises the deformation behaviour of the confining drop of the compound particle by plotting the dependence of deformation parameter, $\mathcal{D}$, on $\beta$. In purely rotational flows ($\beta = -1$), the confining drop doesn't deform ($\mathcal{D} = 0$), since the imposed flow corresponds to deformation free (rigid body like) rotation of the fluid. As $\beta$ increases, the strength of extensional component in the imposed flow increases which results in larger deformations, with largest $\mathcal{D}$ obtained for $\beta=1$. Figure \ref{fig10a} also shows the predicted values of $\mathcal{D}$ at $\textit{O}(Ca)$, shown as dotted lines for comparison. In the linear theory, $\mathcal{D}$ increases linearly with $\beta$ with rotational component playing no role. As mentioned earlier, the first effect of imposed vorticity is taken into account at $\textit{O}(Ca^{2})$, and thus $\mathcal{D}$ calculated at $\textit{O}(Ca^{2})$ exhibits a nonlinear behaviour with increase in $\beta$ as shown. Thus, in the rotation dominated flows ($\beta < 1$), $\textit{O}(Ca)$ calculation underpredicts the deformation of the confining drop. On the other hand, as $\beta \rightarrow 1$, $\textit{O}(Ca)$ overpredicts $\mathcal{D}$. It is also worth noting that  $\textit{O}(Ca^{2})$ corrections are significant when the hydrodynamic interaction between the encapsulated particle and the confining drop interface becomes strong. Therefore, the case of $\alpha = 1.5$ shows largest deviation in the predicted values of $\mathcal{D}$ at $\textit{O}(Ca^2)$ compared to predictions of linear theory. 
	
	A similar analysis can be performed for extensional flows as well, where, the generalized extensional flow may be defined with a velocity gradient tensor as, 
	\begin{equation}\label{eq9b}
		\bm{E}^{GE}=\frac{1 }{\sqrt{2(1+m+m^2)}}\begin{bmatrix}
			1 & 0 & 0 \\
			0 & m & 0\\
			0 & 0 & -(1+m)
		\end{bmatrix}, \quad \bm{\Omega}^{GE} = \mathbf{0},
	\end{equation}
	where, the parameter $m$ may take different values. For example, $m = -1/2$ corresponds to a uniaxial flow ($x_1$ being the extensional axis), $m =0$ corresponds to a planar extensional flow and $m =1$ corresponds to a biaxial flow ($x_3$ being the compressional axis). 
	
	Figure \ref{fig10b} shows the dependency of deformation parameter $\mathcal{D}$ on $m$. Clearly the deformation parameter is only weakly dependent on $m$, with $\mathcal{D}$ varying only slightly between uniaxial ($m = -1/2$), planar extensional ($m = 0$) and biaxial ($m = 1$) flows. However, irrespective of the type of imposed flow, $\mathcal{D}$ increases with decrease in $\alpha$. Again, predictions from $\textit{O}(Ca)$ calculations are shown as dotted lines in the same figure for comparison. It is interesting to note that, the linear theory shows a monotonic increase of $\mathcal{D}$ with $m$ while this is not the case at $\textit{O}(Ca^2)$. The implication is that linear theory falsely suggests that biaxial flows ($m = 1$) are stronger than uniaxial flows ($m = -1/2$) in deforming the confining interface \citep{chaithu2019}, while the situation reverses by taking into account of $\textit{O}(Ca^2)$ corrections. In other words, $\mathcal{D}$ at $m = 1$ is smaller than $\mathcal{D}$ at $m = -1/2$, similar to the calculations for a simple drop \citep{davis1981}, and therefore, uniaxial flows are stronger in deforming a compound particle. This effect may be understood as follows. Biaxial flow compresses the confining drop towards an oblate shape while a uniaxial flow stretches it towards a prolate shape. As deformation increases, the prolate shaped interface meets the solid inclusion before the corresponding oblate. Therefore, though biaxial flow is obtained by reversing the direction of uniaxial flow, the latter is stronger than the former to cause breakup of the confining drop. The other observation to note is that, as observed for generalised shear flows, the difference that arise due to $\textit{O}(Ca^2)$ corrections from the linear theory also increases with decrease in $\alpha$, making the $\textit{O}(Ca^2)$ calculations more relevant for compound particles.
	
	\section{Rheology of a dilute dispersion of compound particles}\label{rheology}
	So far, we analyzed the deformation dynamics of a compound particle when subjected to various imposed flows using the results derived from a domain perturbation approach. The same analysis is useful in predicting the rheological behaviour of a suspension of compound particles. In this section, we characterize the rheology of a dilute dispersion of compound particles in terms of effective shear viscosity, extensional viscosity, normal stress differences and the complex modulus. 
	
	Consider a dispersion made up of compound particles such as those shown in figure~\ref{fig1} (encapsulated particle of size $a$, confining interface of size $b = a\alpha$ and inner fluid viscosity $\hat{\mu}$) in a carrier fluid of viscosity $\mu = \lambda \hat{\mu}$. The volume fraction (based on the size of the confining drop) of the compound particles $\phi << 1$, so that the dispersion is dilute and hence any hydrodynamic interaction between the compound particles will be neglected. On subjecting this dispersion to an imposed flow described by a rate of strain tensor $\bm{E}$, we may define a volume averaged stress \citep{batchelor1970} for the dispersion as,
	\begin{equation}\label{eq1c}
		\langle \bm{\sigma}\rangle=-\langle p \rangle \bm{I}+2  \bm{E}+ \phi \bm{S},
	\end{equation}
	where $\bm{S}$ is the stresslet associated with a single compound particle. This quantity can be easily extracted from the steady state far field solution of the disturbance velocity field generated by a single compound particle (calculated in \S\ref{order1} and \S\ref{order2}),
	\begin{equation}\label{eq2c}
		\begin{aligned}
			\bm{u} = \bigg(-\frac{c_{1}}{6} \bm{E}+Ca \Big(\frac{c_{5}}{2}\big(\bm{E \cdot E} -\frac{\bm{I}}{3}(\bm{E : E})\big) +\frac{c_{8}}{4}\left(\bm{\Omega \cdot E}-\bm{E \cdot \Omega}\right) \Big) \bigg) :	\left(\frac{\bm{I x}}{r^{3}}-\frac{3\bm{x x x}}{r^{5}}\right),
		\end{aligned}
	\end{equation}
	where the last bracketed term $\left(\frac{\bm{I x}}{r^{3}}-\frac{3\bm{x x x}}{r^{5}}\right)$ is the symmetric part of the gradient of Oseen tensor. The expression for $c_{1}$ is given in appendix \ref{appA} (see (\ref{eqA5})), and the expressions for $c_{5}$ and $c_{8}$ are given in appendix \ref{appB} (see (\ref{eqB13})-(\ref{eqB14})). Therefore, as implied by the boundary integral equations for creeping flow, the stresslet can be identified as \citep{batchelor1970, leal2007, arun2012}, 
	\begin{equation}\label{eq3c}
		\begin{aligned}
			\bm{S} = 8 \upi  \bigg(-\frac{c_{1}}{6} \bm{E}+Ca \Big(\frac{c_{5}}{2}\big(\bm{E \cdot E} -\frac{\bm{I}}{3}(\bm{E : E})\big) +\frac{c_{8}}{4}\left(\bm{\Omega \cdot E}-\bm{E \cdot \Omega}\right) \Big) \bigg).
		\end{aligned}
		\end{equation}
	Substituting (\ref{eq3c}) in to (\ref{eq1c}) gives an expression for the volume averaged stress as,
	\begin{equation}\label{eq4c}
		\begin{aligned}
			\langle\bm{\sigma}\rangle=-\langle p \rangle\bm{I}+2   \bm{E}\left(1- \phi \frac{c_{1}}{2 } \right) +\phi  Ca \Bigg( 3c_{5} \left(\bm{E \cdot E} -\frac{\bm{I}}{3}(\bm{E : E})  \right)\\
			+\frac{3c_{8}}{2 }\left( \bm{\Omega \cdot E}-\bm{E \cdot \Omega}\right)\Bigg)+\textit{O}(\phi^{2}, \phi^2 Ca, \phi Ca^2),
		\end{aligned}
	\end{equation}
	which can be further analyzed to extract the rheological quantities of interest for a dilute dispersion of compound particles as discussed below.
	
	\subsection{Shear viscosity}\label{viscosity}
	Subjecting the dilute dispersion of compound particles to an imposed flow, $\bm{u}^{imp}=  x_{2} \mathbf{i}_{1}$, the effective shear viscosity of the dispersion  can be calculated from (\ref{eq4c}) as,	 
	\begin{equation}\label{eq5c}
		\begin{aligned}
			\mu_{eff}= \frac{\langle\sigma\rangle_{12}}{2E_{12}}
			= \left(1-\phi \frac{c_{1} }{2}   \right)+\textit{O}(\phi^{2}, \phi^2 Ca, \phi Ca^2 ). 
		\end{aligned}
	\end{equation}
	Thus, the first correction to the shear viscosity is $\textit{O}(\phi)$ and it depends only on the leading order stress field. In other words, $\textit{O}(Ca)$ stress field generated due to deformation of the confining drop does not contribute to shear viscosity for dilute dispersions. However, hydrodynamic interaction between compound particles can lead to further corrections, which will be accounted at $\textit{O}(\phi^2)$. Taking appropriate limits the expression derived above for shear viscosity (\ref{eq5c}) can also be deduced from the works of \cite{davis1981}, \cite{stone1990a}, \cite{mandal2016}, \cite{das2020}.

	It is interesting to look at the limiting case of (\ref{eq5c}) first.
	(i) In the limit, $\alpha \rightarrow 1$ (or $\lambda \rightarrow 0$), (\ref{eq5c}) reduces to the Einstein relation for the effective viscosity of a dilute suspension of spherical rigid particles  $\mu_{eff} = \left(1+\frac{5 }{2} \phi \right)$ \citep{einstein1906, einstein1911}.
	(ii) In the limit of $\alpha \rightarrow \infty$, (\ref{eq5c}) reduces to Taylor's relation for the effective viscosity of a dilute emulsion $\mu_{eff} =   \left(1+ \frac{5+2\lambda}{2+2\lambda}\phi \right)$  \citep{taylor1932, oldroyd1953}.
	(iii) In the limit of $\alpha \to 1$, but retaining $\textit{O}(\alpha - 1)$ terms, the effective shear viscosity of a suspension of particles that are coated with a thin fluid film of inner fluid can be calculated as,
	\begin{equation}\label{eq6c}
		\begin{aligned}
			\mu_{eff} \simeq \left(1+\phi \left( \frac{5}{2}-\frac{15}{8} (\alpha-1) \lambda+\textit{O}(\alpha-1)^{2} \right)  \right).
		\end{aligned}
	\end{equation}   
	Therefore, $\textit{O}(\alpha-1)$ correction to the effective viscosity of a dispersion is negative indicating that the viscosity of a dilute suspension of spheres can be reduced by providing a thin coating of a viscous fluid.
	(iv) In the limit of $\lambda \rightarrow \infty$,  (\ref{eq5c}) reduces to the effective viscosity of a dilute emulsion of bubbles $\mu_{eff} =   \left(1+ \phi \right)$, where the positive viscosity correction arises purely due to the interfacial tension of the confining interface.
	
	\begin{figure}
		\centering
		\subfigure[]{
			\includegraphics[height=5cm]{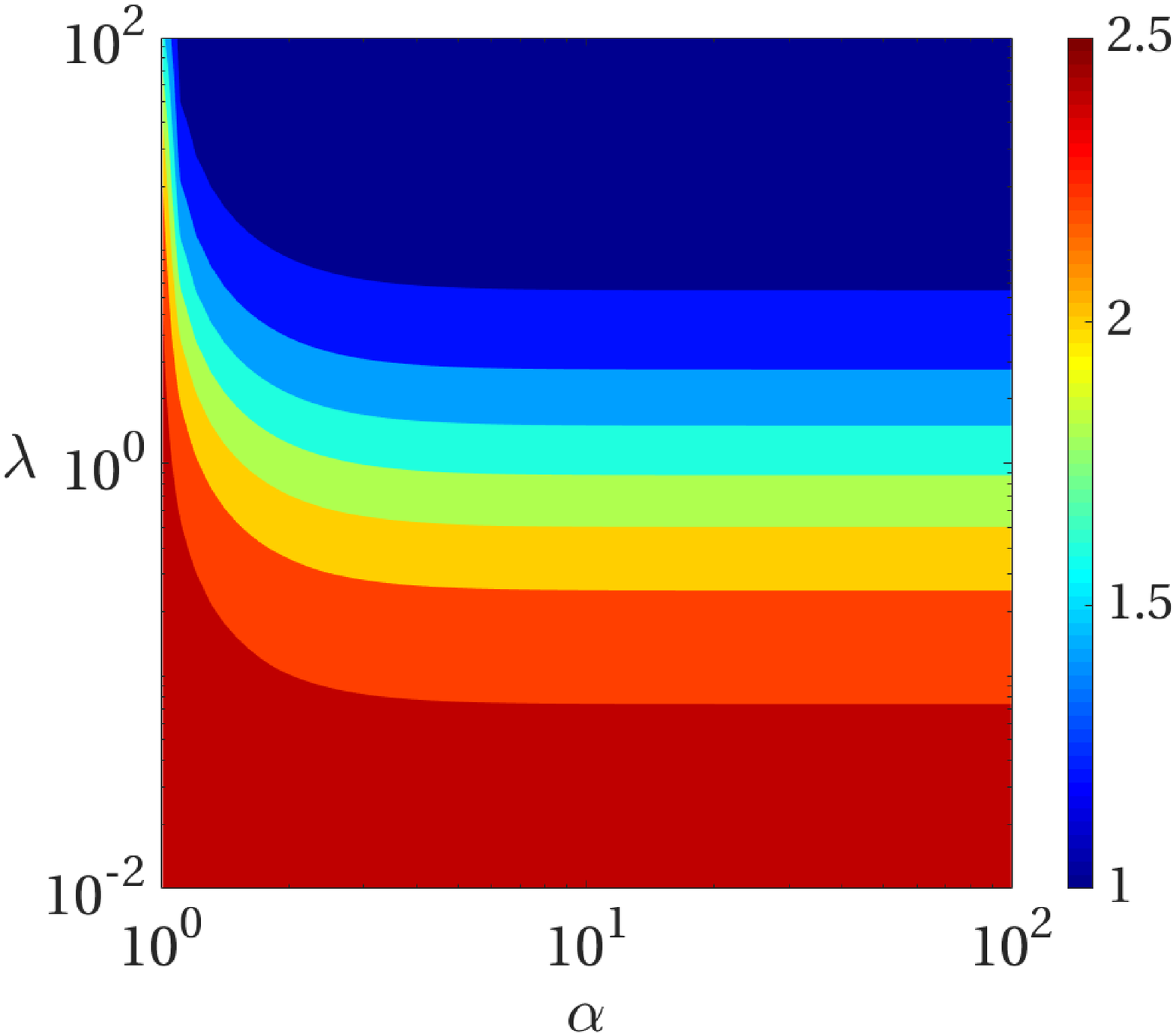}\label{fig11a}}\quad
		\subfigure[]{
			\includegraphics[height=5cm]{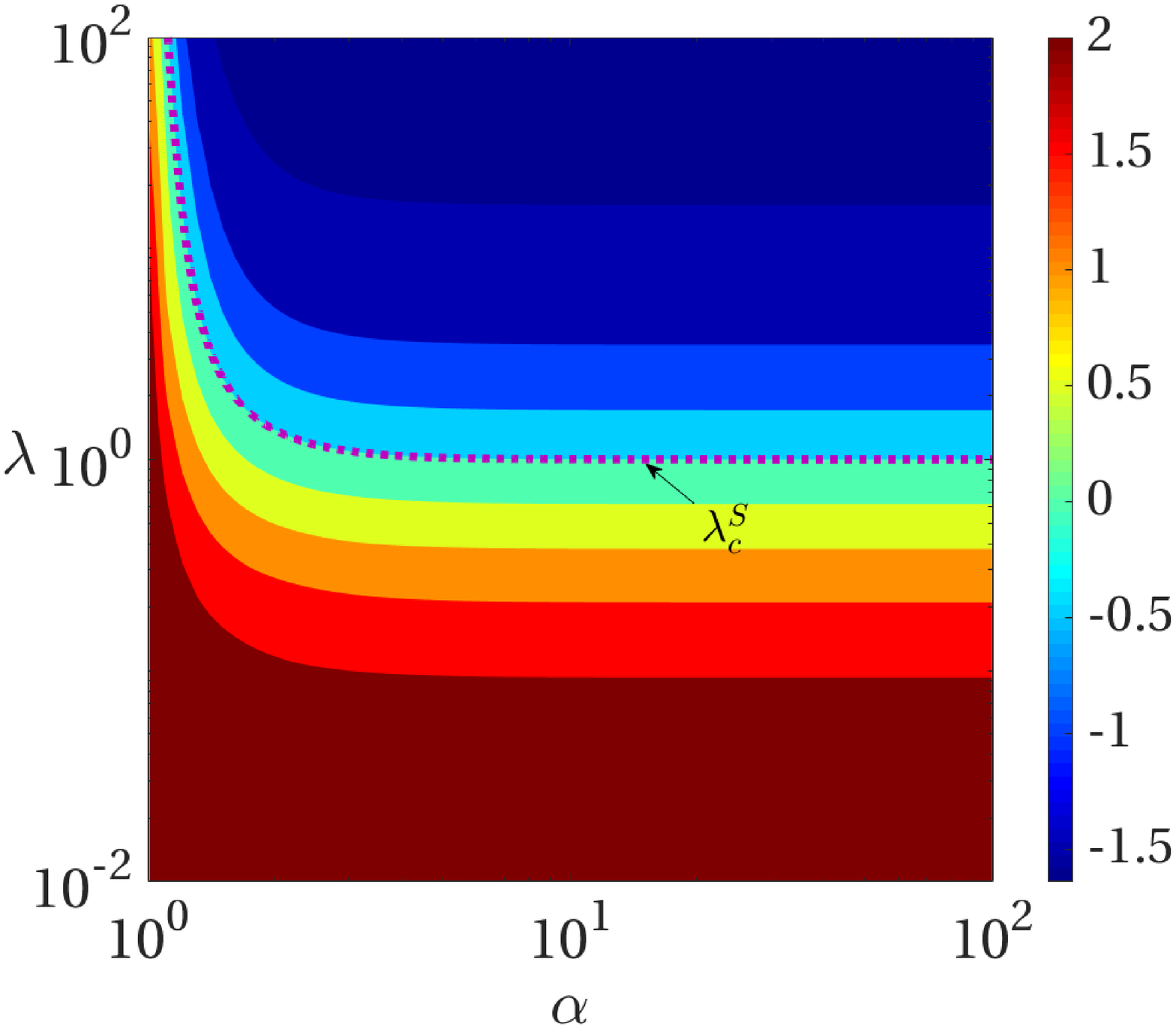}\label{fig11b}}\quad 
		\caption{Effect of the viscosity ratio $\lambda$ and size ratio $\alpha$ on the shear viscosity, plotted as contours of the enhancement factor, (a) $(\mu_{eff}-1)/\phi$ (see (\ref{eq5c})) and (b) $(\mu_{eff}^{S}-1)/\phi$ corresponding to the case of $Ca\rightarrow\infty$ (see (\ref{eq7c})).}\label{fig11}
	\end{figure}
	
	Figure~\ref{fig11a} shows the complete picture, where the enhancement factor in the effective shear viscosity $(\mu_{eff}-1)/\phi$ is plotted as a function of size ratio $\alpha$ and viscosity ratio $\lambda$. At a given viscosity ratio, $\mu_{eff}$ decreases with increase in $\alpha$ from Einstein's to Taylor's limit. Similarly, for a given size ratio, the effective viscosity decreases with increase in viscosity ratio. This behavior can be understood as follows. As the resistance to the fluid flow increases, an increase in the effective viscosity may be anticipated. The resistance to the imposed flow is maximum in the case of rigid solid spheres. This resistance decreases with increase in the thickness of the coating (confining drop size) on the rigid spheres. Therefore, the effective viscosity decreases with increase in the size of the confining drop ($\alpha$). Similarly, as the viscosity of the inner fluid decreases, the resistance to the imposed flow decreases, thus, the outer fluid can easily slip past the confining interface. Therefore, the effective viscosity decreases with increase in the viscosity ratio. It is worth noting that, as mentioned earlier, $Ca$ does not appear at $\textit{O}(\phi)$ in (\ref{eq4c}) and therefore the stress resulting from the deformation of the confining drop does not have any effect on these calculations. 
	
	Another useful rheological measure that can be extracted from (\ref{eq4c}) is the shear viscosity of the compound particles when the confining interface has zero surface tension or $Ca\rightarrow\infty$. This  corresponds to a situation where the kinematic boundary condition (\ref{eq14}) is not respected, $b_{1} = 0$ and therefore the effective shear viscosity is obtained as,
	\begin{equation}\label{eq7c}
		\mu_{eff}^{S}= \left(1-\frac{\phi c_{1}^{S}}{2}\right), 
	\end{equation}   
	where the expression for $c_{1}^{S}$ is given in the appendix \ref{appA} (see (\ref{eqA23})). In the limit of $\alpha \rightarrow \infty$, the $\mu_{eff}^{S} = \lambda \mu\big(1+\phi \frac{5(1-\lambda)}{(2+3\lambda)}\big)$, a result obtained by \cite{batchelor1972} for a dilute emulsion of drops.
	
	Figure~\ref{fig11b} illustrates the dependence of $\mu_{eff}^{S}$ on the size ratio and viscosity ratio of the compound particles in the dispersion. The dependence of $\mu_{eff}^{S}$ on $\alpha$ and $\lambda$ is similar to that of $\mu_{eff}$, however with the important difference that the effective viscosity $\mu_{eff}^{S}$ is smaller than the suspending fluid viscosity $\mu$ for a range of viscosity ratios. As seen in the figure, this reduction in viscosity occurs only beyond a viscosity ratio termed as critical viscosity ratio $\lambda_{c}^{S}$.
	Equating  $\mu_{eff}^{S}-1$ to zero, we obtain the critical viscosity ratio from (\ref{eq7c}) as,
	\begin{equation}\label{eq8c}
		\begin{aligned}
			\lambda_{c}^{S}=&-\big(3 \alpha^{10}+125 \alpha^7-336 \alpha^5+200 \alpha^3+8 -5 \big(49 \alpha^{20}+14 \alpha^{17}-1175 \alpha^{14}+2352 \alpha^{12}\\
			&-1288 \alpha^{10}-16 \alpha^7+64\big)^{1/2}\big)/(8 \big(4 \alpha^{10}-25 \alpha^7+42 \alpha^5-25 \alpha^3+4\big)),
		\end{aligned}
	\end{equation}
	beyond which a reduction in viscosity will be observed. 
	\subsection{Extensional viscosity}
	By subjecting the dilute dispersion of compound particles to a uniaxial extensional flow, we can calculate the extensional viscosity (or Trouton's viscosity) 
	$\mu_{eff}^{T}$ as the ratio of the normal stress difference to the imposed shear rate,
	\begin{equation}\label{eq9c}
		\mu_{eff}^{T}=\frac{\sigma_{33}-\sigma_{11}}{E_{33}}=\frac{\sigma_{33}-\sigma_{22}}{E_{33}}.
	\end{equation}
	Hence, the extensional viscosity of a dilute dispersion of compound particles can be obtained from (\ref{eq4c}) as
	\begin{equation}\label{eq10c}
		\begin{aligned}
			&\mu_{eff}^{T}= \Bigg(3+3\phi\bigg(-\frac{c_{1}}{2 }+Ca\frac{3c_{5}}{4 } \bigg)\Bigg)+\textit{O}(\phi^{2}, \phi^{2} Ca, \phi Ca^2 ).
		\end{aligned}
	\end{equation}
	The above expression (\ref{eq10c}) can be deduced from the work of \cite{santra2020a} on compound drops by taking the appropriate limit.
	
\begin{figure} 
		\centering
		\includegraphics[height=5cm]{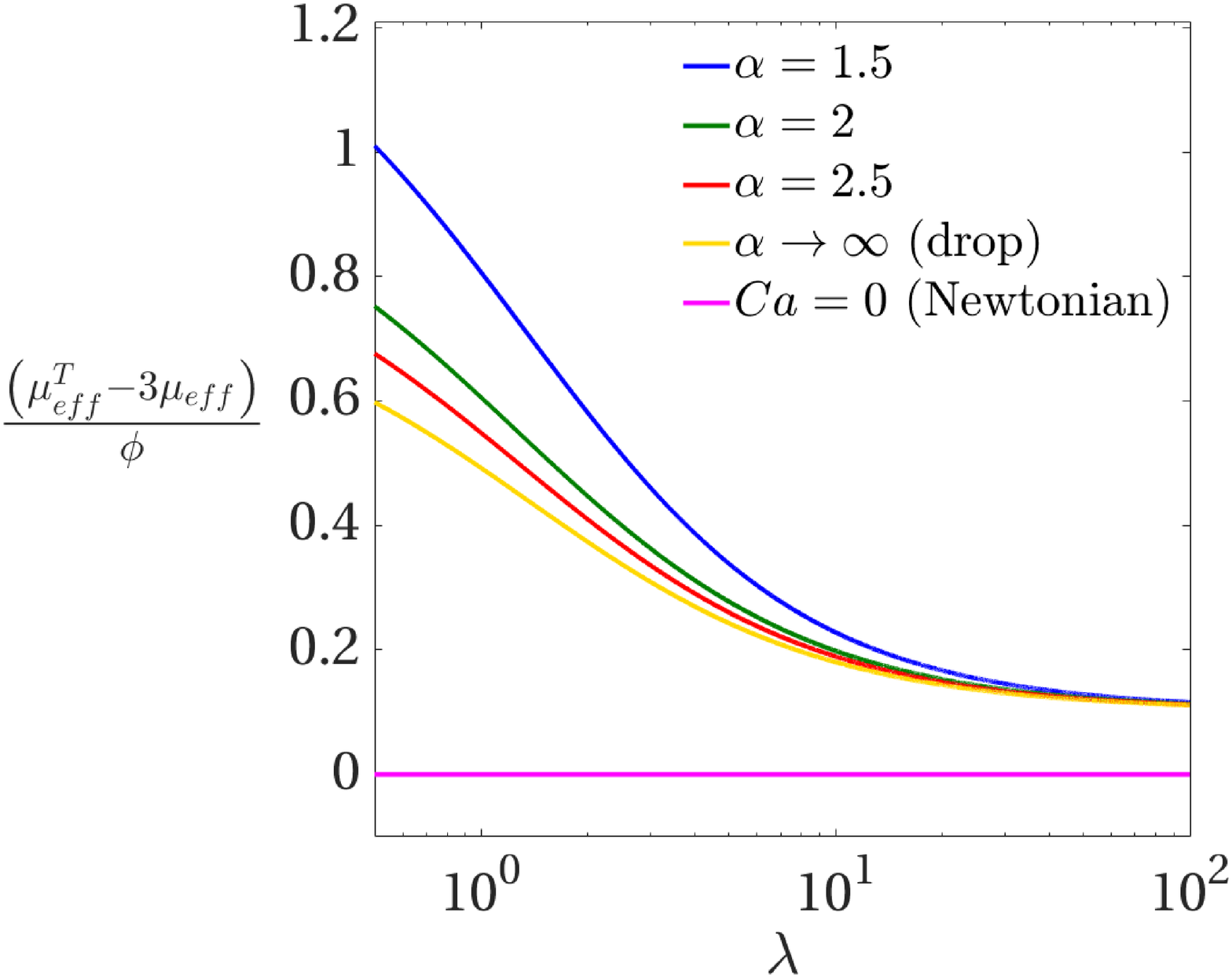}
		\caption{The variation of the Trouton viscosity versus the viscosity ratio $\lambda$ for various size ratios and $Ca=0.1$.}\label{fig12}
	\end{figure}

	The dependencies of Trouton viscosity of a dilute dispersion of compound particles on size ratio and viscosity ratio of individual compound particles are shown in figure~\ref{fig12}. Again, instead of viscosity, it is the enhancement factor $(\mu_{eff}^{T}-3\mu_{eff})/\phi$ that is plotted against $\lambda$. For a Newtonian fluid, Trouton viscosity is equal to $3\mu_{eff}$ and the enhancement factor is exactly zero. This limit is indicated by the flat line in figure~\ref{fig12}. Dispersions of compound particles show deviations from this Newtonian limit - larger deviations (and thus larger $\mu_{eff}^{T}$) are observed for smaller values of $\alpha$ and $\lambda$. As $\lambda$ increases, the enhancement factor for Trouton viscosity asymptotes to a constant, independent of $\alpha$. On the other hand, for a given $\lambda$, increase in $\alpha$ reduces Trouton viscosity. The decrease in Trouton viscosity with increase in the size ratio or with increase in viscosity ratio is same as the reasons for decrease in effective viscosity discussed in the previous section. 
	
	These observations can also be endowed by analysing equation (\ref{eq10c}). The Newtonian limit can be recovered for the special case, when $Ca = 0$ so that the confining interface is exactly spherical, and (\ref{eq10c}) reduces to give the extensional viscosity $\mu_{eff}^{T} = 3\mu_{eff}$. In general, compound particles show deviations from this Newtonian behaviour since $Ca \neq 0$ and the confining drop is deformed. Of course, as mentioned in the previous sections, the extent of deviation depends upon the specific value of $\alpha$ and $\lambda$. For large values of $\lambda$ the enhancement factor $(\mu_{eff}^{T}-3\mu_{eff})/\phi$ asymptotes to $\frac{36 Ca}{35}$ a constant, which is independent of $\alpha$ and consistent with the observations in figure~\ref{fig12}.  In the limit of $\alpha \to 1$, but retaining $\textit{O}(\alpha - 1)$ terms, the extensional viscosity of a suspension of particles that are coated with a thin fluid film of inner fluid can be calculated as,
	\begin{equation}\label{eq11c}
		\begin{aligned}
			\frac{\mu_{eff}^{T}-3\mu_{eff}}{\phi} \simeq  Ca \left( -\frac{675 (\lambda-4)}{896 (\alpha-1)}+\frac{225 \left(45 \lambda^2-130 \lambda+136\right)}{3584}+\textit{O}(\alpha-1) \right).
		\end{aligned}
	\end{equation}   
	It is clear that Trouton viscosity increases with decrease in $\alpha$. In the opposite limit, the case of simple drops without encapsulated particles, \textit{i.e.}, as $\alpha\rightarrow\infty$,  (\ref{eq10c}) provides the enhancement factor $(\mu_{eff}^{T}-3\mu_{eff})/\phi = 9 Ca \left(64 \lambda^3+732  \lambda^2+1179 \lambda+475\right)\big/ (560 (\lambda+1)^3) $. This result matches with the result of  \cite{arun2012} with slip coefficient being zero and with \cite{mandal2017} when no  surfactants are present.
	
	Hence it is clear that larger Trouton viscosity is observed for smaller values of $\alpha$ and $\lambda$. This behaviour is qualitatively similar to that of the effective shear viscosity.  However, comparison with the case of $Ca = 0$ suggests that deviations from the Newtonian limit ($\mu_{eff}^T \neq 3\mu_{eff}$) arises in the dispersions of compound particles due to the deformation of the confining interface. Consequently, a time dependent increase in the deformation of the confining drop (as discussed in \S\ref{simple}) can give rise to a time dependent increase in the extensional viscosity. Another feature to notice is that, (\ref{eq11c}) shows that the enhancement factor for $\mu_{eff}^{T}$ of a dilute dispersion of compound particles can be made negative. This is similar to the case of shear viscosity and indicates that the extensional viscosity of a dilute suspension of spheres can be reduced by providing a thin film coating of a less viscous fluid on the particles.
	
	\subsection{Normal stress differences}
	In \S\ref{example}, we have seen that an imposed flow acts to deform the confining spherical drop while the interfacial tension acts to revert it back yielding an anisotropic shape of the confining drop. Therefore, the force distribution around the compound particle will also be anisotropic, inducing normal stress differences in a dispersion of compound particles. Normal stress  differences, a characteristic of viscoelastic behaviour of a fluid, can also be determined by analyzing the volume averaged stress (\ref{eq4c}) as in the last two sections. 
	
	Subjecting the dilute dispersion of compound particles to a simple shear flow $\bm{u}^{imp}= x_{2} \mathbf{i}_{1}$, (\ref{eq4c}) renders the first and second normal stress differences $N_{1}$ and $N_{2}$ as:
	\begin{align}
		N_{1}=\sigma_{11}-\sigma_{22}=\phi Ca \left(\frac{3c_{8}}{2}\right)+\textit{O}(\phi^{2}, \phi^{2}Ca, \phi Ca^2 ) \label{eq12c},\\
		N_{2}=\sigma_{22}-\sigma_{33}=\phi Ca \left(\frac{3(c_{5}-c_{8})}{4}\right)+\textit{O}(\phi^{2}, \phi^{2}Ca, \phi Ca^2 ).\label{eq13c}
	\end{align}
	\begin{figure}
		\centering
		\subfigure[]{
			\includegraphics[height=5.0cm]{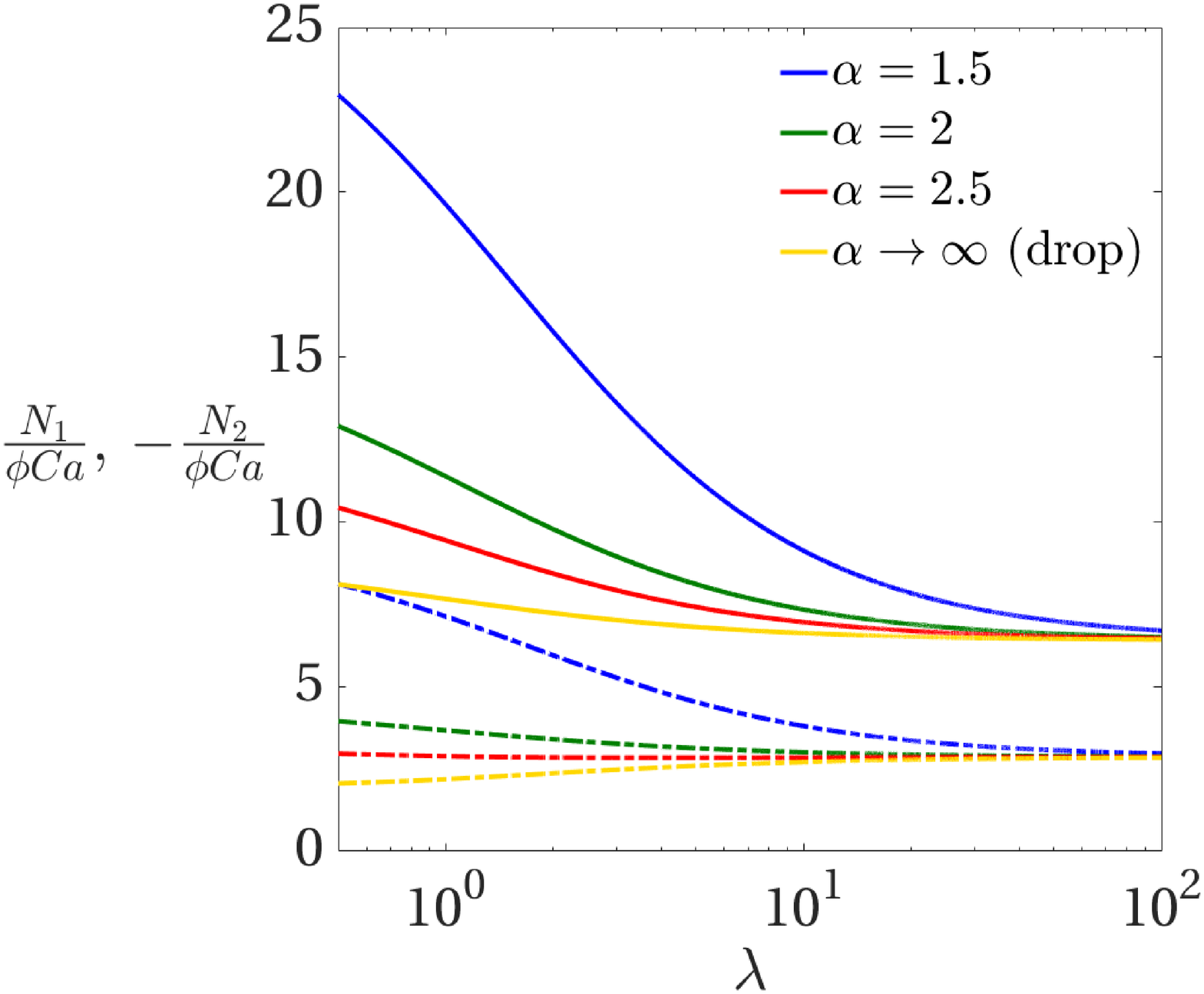}\label{fig13a}}\quad
		\subfigure[]{
			\includegraphics[height=5.0cm]{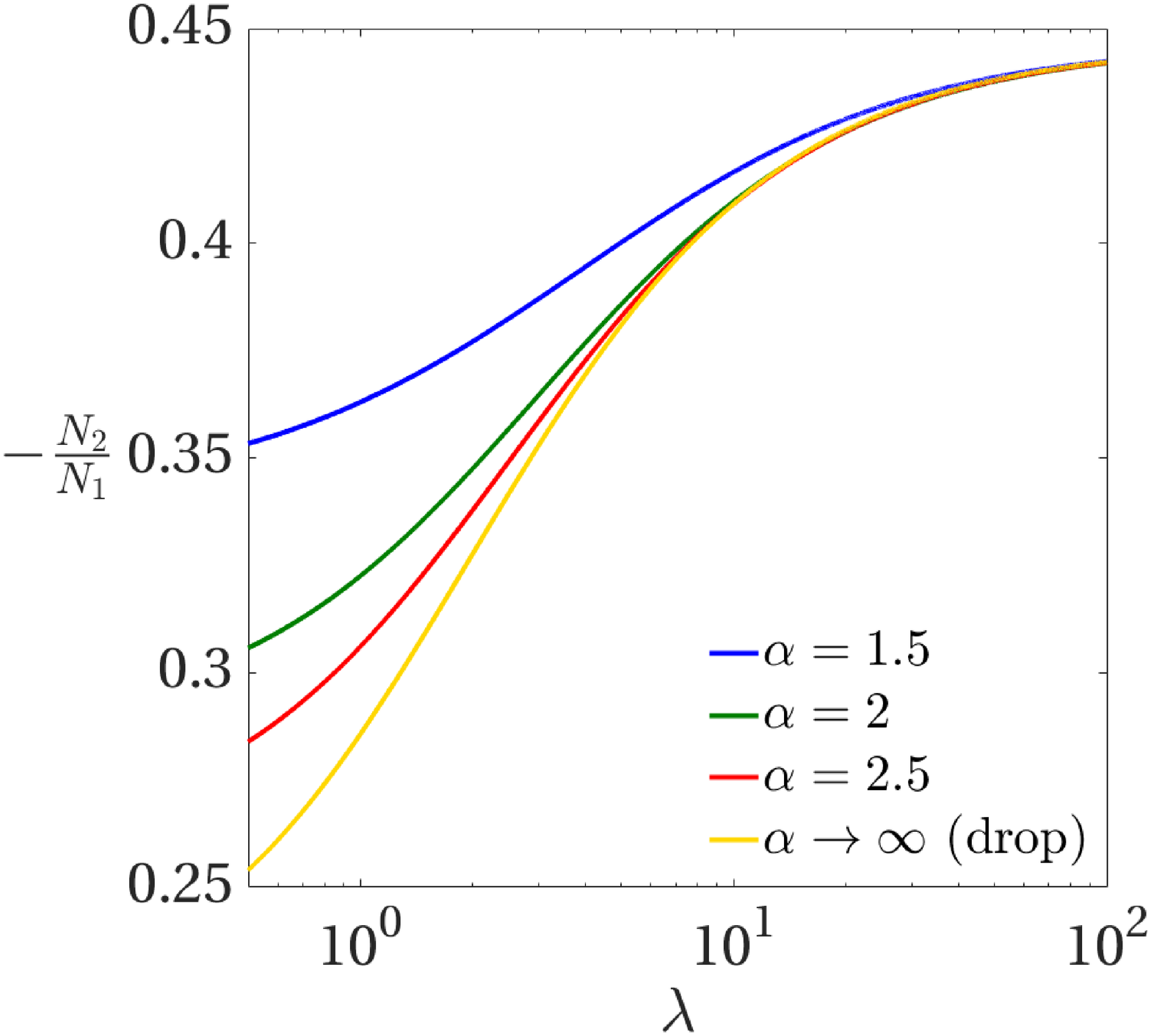}\label{fig13b}}\quad 
		\caption{Effect of the viscosity ratio $\lambda$ on the normal stress differences for different size ratios (a) $N_{1}/\phi Ca$ (solid lines) and  $N_{2}/\phi Ca$ (dash-dotted lines), (b) $-N_{2}/N_{1}$.}\label{fig13}
	\end{figure}
	
	The dependency of first and second normal stress differences on the viscosity and size ratio of compound particles is shown in figure~\ref{fig13a}. In all cases, $N_1$ is positive and $N_2$ is negative. Moreover it may be noted that both $N_{1}$ and $-N_{2}$ decrease with increase in the viscosity ratio for a given $\alpha$.
	The asymptotic values of the normal stresses as $\lambda \rightarrow \infty$ can be deduced from (\ref{eq12c}) and (\ref{eq13c}) respectively as $N_{1}/\phi Ca=32/5$ and $-N_{2}/\phi  Ca = 20/7$. For a given viscosity ratio, both $N_1$ and $-N_2$ decrease with increase in the size ratio. The two data sets are put together in figure~\ref{fig13b} which shows the ratio of normal stress differences, $-N_{2}/N_{1}$ as function of $\lambda$ for various $\alpha$. 
	This ratio increases with increase in $\lambda$ or $\alpha$. Irrespective of $\alpha$, and $\lambda$, the value of $\mod{-N_2/N_1}$ is less than $1$, irrespective of $\alpha$, and $\lambda$. This behavior can be understood by analyzing the extent of deformation of the confining drop in different planes. We have seen that the deformation of the confining drop is larger in the flow-gradient plane compared to that in the vorticity-gradient plane.  Therefore, the $N_1$ which is based on the stresses in the flow-gradient plane is always larger than the $N_2$ which is based on the stresses in the vorticity-gradient plane. The maximum value of $-N_{2}/N_{1} = 0.4464$ is approached at large $\lambda$, independent of the value of $\alpha$. 
	
	The limiting case of $\alpha\rightarrow\infty$ which corresponds to a dilute suspension of drops that do not contain encapsulated particles are also shown in figure~\ref{fig13a}. The mathematical expressions that correspond to these curves can be obtained from  (\ref{eq12c}) and (\ref{eq13c}) respectively, as 
	$N_{1}/\phi Ca = (16 \lambda+19)^2 \Big/ \big(40 (\lambda+1)^2\big)$ and $N_{2}/\phi Ca = -\Big(800 \lambda^3+1926 \lambda^2+1623 \lambda+551 \Big)\bigg/\Big(280 (\lambda+1)^3 \Big)$, and they are consistent with the works of \cite{arun2012, mandal2017}. On the other hand, in the limit $\alpha \rightarrow 1$, which corresponds to the solid particles coated with a thin film of inner fluid, we can determine the normal stress differences as
	\begin{equation}\label{eq14c}
		\begin{aligned}
			&\frac{N_{1}}{\phi  Ca}=\frac{45}{32 (\alpha-1)^2} -\frac{15 (15 \lambda-34)}{64 (\alpha-1)}+\frac{ \left(3375 \lambda^2-5460 \lambda+4412\right)}{512}+\textit{O}(\alpha-1)
		\end{aligned}
	\end{equation}
	and
	\begin{equation}\label{eq15c}
		\begin{aligned}
			&\frac{N_{2}}{\phi  Ca}=-\frac{45}{64 (\alpha-1)^2}+\frac{15 (45 \lambda-89)}{448 (\alpha-1)} -\frac{16875 \lambda^2-18720 \lambda+10484}{7168}+\textit{O}(\alpha-1).
		\end{aligned}
	\end{equation}
	The apparent large increase in $N_1$ and $-N_2$ as $\alpha\rightarrow 1$ represents the result of increased deformation of the confining interface in this limit. These expressions hold correct only when $Ca < Ca_{crit}$, \textit{i.e.}, when the thin film coated on the particle is stable (without breakup) in the imposed flow.
	
	Hence, as observed for the extensional viscosity, both $N_1$ and $N_2$ are also dependent upon extent of deformation of the confining drop. Consequently the normal stress differences for a dilute dispersion of compound particles will exhibit capillary number dependent and time dependent behaviour following the extent of deformation of the confining interface. 
	
	\subsection{Small-amplitude oscillatory shear flow (SAOS)}\label{oscillatory}
	Finally, we characterize the linear viscoelastic behavior of a dilute dispersion of compound particles by subjecting the dispersion to a small amplitude oscillatory shear flow (SAOS). In general, the rheological response of the dispersion may be linear or non-linear, and the transition from linear to non-linear response can be observed by increasing the amplitude of the shear rate at a fixed frequency. In the following, we restrict our analysis to the linear regime, where the analytical calculations are possible. 
	
	Consider the imposed oscillatory shear flow of the form
	\begin{equation}\label{eq18c}
		\bm{u}^{imp}= \exp{(i\omega t)} x_{2} \mathbf{i}_{1}, 
	\end{equation}
	where, $\omega$ is the frequency of oscillation. In the linear regime, shear stress is related to the complex modulus ($G^*$) \citep{arun2012} via,
	\begin{equation}\label{eq19c}
		\langle \sigma_{12}\rangle= \frac{G^{*}}{\omega} 2E_{12},
	\end{equation}
	where $G^{*}=G^{\prime}+ i \omega G^{\prime \prime}$, $G^{\prime}$ is the elastic or storage modulus and $G^{\prime \prime}$ is the viscous or loss modulus. Comparing (\ref{eq19c}) and the volume averaged stress, (\ref{eq4c}), we find that
	\begin{equation}\label{eq20c}
		G^{*}=\omega\left(1-\phi \frac{c_1}{2}\right).
	\end{equation}
	The constant $c_1$ in (\ref{eq20c}), associated with the velocity field, is dependent on the drop shape parameter $b_1$. 
	
	Previously, in \S\ref{simple}, we described the time evolution $b_1$, and thus the shape of the confining interface in an imposed simple shear flow. Here, we require to analyze the time evolution of the shape parameter $b_{1}$ in an imposed oscillatory shear flow, which is obtained from the leading order kinematic boundary condition, (\ref{eq17}), as
		\begin{equation}\label{eq21c}
		\frac{\partial b_{1}}{\partial t}+i \omega b_{1}=-\frac{b_{1}-b_{1}^{*}}{t_{c}}.
	\end{equation}
	In the long time limit, where the unsteady term can be neglected, so that we can obtain a frequency dependent shape parameter,
	\begin{equation}\label{eq22c}
		b_{1}=\frac{b_{1}^{*}}{1+i \omega t_{c}}.
	\end{equation}
	Following the procedure developed in \cite{arun2012}, we evaluate $c_{1}$ in terms of $b_{1}$ using (\ref{eq22c}) and substitute it in (\ref{eq20c}), to obtain the expression for the complex modulus ($G^*$) as,
	\begin{equation}\label{eq23c}
		\begin{aligned}
			G^{\prime}+i\omega &G^{\prime \prime}=\omega \bigg(1-\frac{\phi }{2}\Big(5\Big(\big( (40- 100 \alpha^3 + 100 \alpha^7 - 
			40 \alpha^{10} )\lambda - 4 (\alpha-1)^4 \big(4 + 16 \alpha + 40 \alpha^2  \\
			&+ 55 \alpha^3 + 40 \alpha^4 + 16 \alpha^5 + 4 \alpha^6 \big) \lambda^2\big)+  i \omega \big(-48 - 200 \alpha^3 + 336 \alpha^5 - 225 \alpha^7 \\
			&- 38 \alpha^{10} +\big(16 + 400 \alpha^3 - 672 \alpha^5 + 250 \alpha^7 + 6 \alpha^{10} \big)\lambda+ 8 (-1 + \alpha)^4 \big(4 + 16 \alpha\\
			& + 40 \alpha^2 + 55 \alpha^3 + 40 \alpha^4 + 16 \alpha^5 
			+ 4 \alpha^6 \big) \lambda^2\big) \Big) 
			\Big/ \Big(\big(-40 + 100 \alpha^3  - 100 \alpha^7 + 40 \alpha^{10} \big) \lambda \\
			&  +10 (-1 + \alpha)^4 \big(4 + 16 \alpha + 40 \alpha^2 + 55 \alpha^3+ 40 \alpha^4 + 16 \alpha^5 + 4 \alpha^6\big) \lambda^2  + i \omega \Big(48 \\
			& + 200 \alpha^3 - 336 \alpha^5 + 225 \alpha^7 + 38 \alpha^{10}+ \big(- 
			96+ 100 \alpha^3 - 168 \alpha^5  +75 \alpha^7+ 89 \alpha^{10}\big) \lambda \\
			& +12 (-1 + \alpha)^4 \big(4 + 16 \alpha + 40 \alpha^2 + 55 \alpha^3 + 
			40 \alpha^4 + 16 \alpha^5 + 4 \alpha^6\big) \lambda^2 \Big) \Big) \Big)\bigg).
		\end{aligned}
	\end{equation}
In the limit of $\alpha \rightarrow \infty$ which corresponds to a drop without a suspended particle, the complex modulus is given as
	\begin{equation}\label{eq23cA}
	\begin{aligned}
		G^{\prime}+i\omega G^{\prime \prime}=\omega \Bigg(1+ 5\phi \frac{4 \lambda \big(5+ 2 \lambda \big)+  i \omega \big(19 - 3\lambda - 16  \lambda^2 \big)}{ 40 \lambda\Big(1 +\lambda \Big)  + i \omega \Big( 38 +  89 \lambda +48 \lambda^2 \Big) } \Bigg),
	\end{aligned}
\end{equation}
and this matches with the expressions derived in \cite{arun2012}.
	\begin{figure}
		\centering
		\subfigure[]{
			\includegraphics[height=5cm]{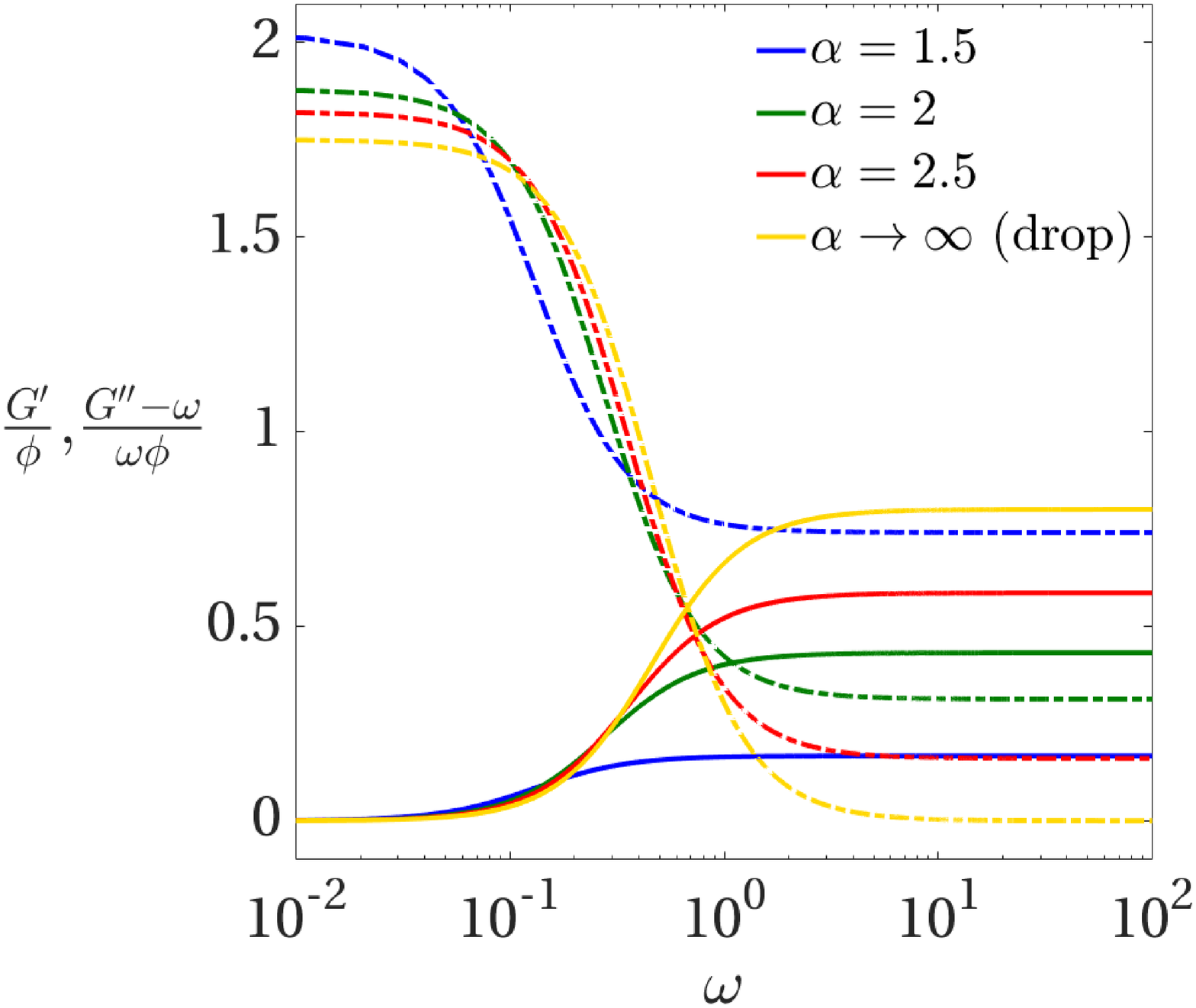}\label{fig14a}}\quad
		\subfigure[]{
			\includegraphics[height=5cm]{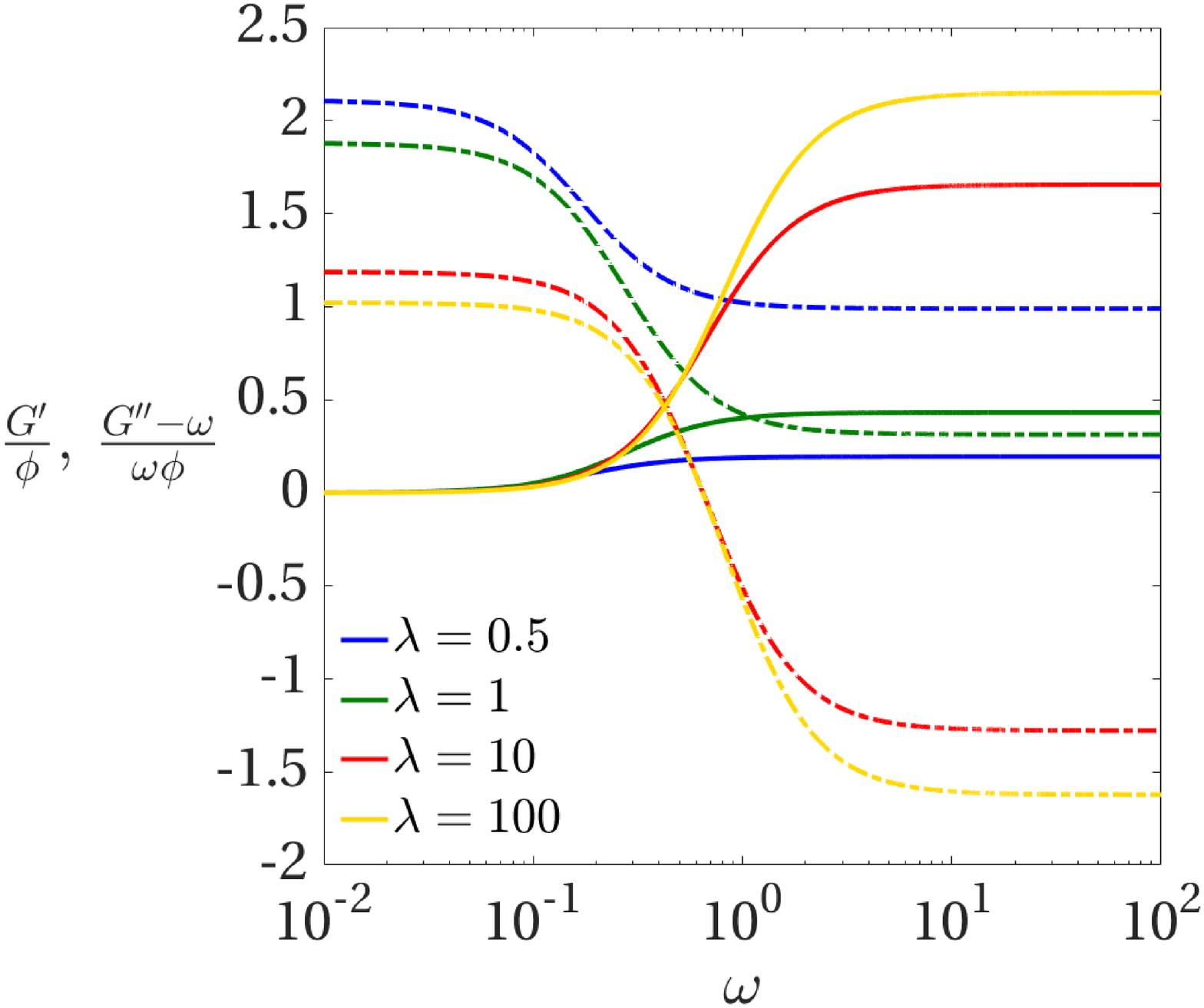}\label{fig14b}}\quad 
		\caption{The normalized components of the complex moduli $G^{\prime}/\phi$ (solid lines) and $(G^{\prime \prime}-\omega)/\phi \omega$ (dash-dotted lines) versus the frequency $\omega$ (see (\ref{eq23c})). Effect of (a) the size ratio $\alpha$ (for $\lambda=1$) and (b) the viscosity ratio $\lambda$ (for $\alpha=2$). }\label{fig14}
	\end{figure}
	
	Figure \ref{fig14} shows the variation of storage ($G^{\prime}$) and loss ($G^{\prime\prime}$) modulus with frequency ($\omega$) for different size and viscosity ratios. The value of storage modulus $G^{\prime}$, which varies  as $\omega^2$, is close to zero at low frequencies, but it increases with increase in frequency, eventually reaching a plateau at high frequencies.  On the other hand, the loss modulus $G^{\prime \prime}$ varies as $\omega$ at low and high frequencies. Figure~\ref{fig14} shows that viscous modulus, represented as $G^{\prime \prime}-\omega$  decreases with increase in frequency. It may be noticed that irrespective of size or viscosity ratio, the viscous modulus ($G^{\prime \prime}$) dominates the elastic modulus ($G^{\prime}$) at low frequencies and vice versa at high frequencies. In other words, at low frequencies, the viscous response dominates the elastic response and the dispersion behaves like a viscoelastic fluid. At high frequencies,  the elastic response dominates the viscous response and thus the dispersion behaves like a viscoelastic solid. This difference in behaviour arises, because at low frequencies, the confining interface relaxes fast enough compared to the imposed time scale, \textit{i.e.}, $t_c < \omega^{-1}$ leading to a fluid like behaviour of the dilute dispersion. But at high frequencies, the interface does not relax fast enough to follow up with changes in the imposed flow that the dispersion of compound particles will show a history dependence. 
	It can also be observed that when $\alpha$ is sufficiently small, the viscous modulus dominates the elastic modulus irrespective of frequency. Therefore, a dilute suspension of spheres coated with a thin film, always behaves like a viscoelastic fluid, provided $Ca<<1$. This analysis is valid only if (i) the time scale of the imposed flow ($1/\omega$) is much larger than the momentum diffusion time scale ($\rho a^2/\mu$) and (ii) the characteristic time scale of the droplet ($t_c$) is much smaller than the characteristic time scale of the imposed flow ($\omega^{-1}$, $G^{-1}$). Mathematically, the above conditions are, $\rho a^2/\mu <<\omega^{-1}$, $t_c<<\omega^{-1}$ and $t_c << G^{-1}$. The first inequality is maintained by the choice of Reynolds number, while the second and third inequalities are maintained by carefully selecting the values of $\omega$, and $G^{-1}$. 
	
	\section{Conclusions}\label{conclu}
	In this work, we analyzed the deformation dynamics of a single compound particle when the encapsulated particle is concentrically located inside the confining drop. Assuming that the interface deformations are small, namely $Ca<<1$ we solved the incompressible Stokes' equations analytically using a domain perturbation approach and obtained the flow field correct up to $O(Ca)$ and the deformed shape of the confining interface up to $O(Ca^2)$. Using these results, we further characterized the rheology of a dilute dispersion of compound particles in terms of effective viscosities, normal stress differences, and complex modulus. 
	
	On subjecting a compound particle to various linear flows it has been found that the presence of an encapsulated particle always enhances the deformation of the confining drop compared to a simple drop which does not contain an encapsulated particle. The enhanced deformation results from the hydrodynamic interaction between the encapsulated particle and the confining interface since the disturbance flow field developed inside the confining drop has to comply with the no-slip, no-penetration boundary conditions on the surface of the encapsulated particle and less stringent interfacial boundary conditions prescribing the continuity of velocity and stress on the confining interface. Naturally, decreasing the size of the encapsulated particle or decreasing the viscosity of the confining fluid will reduce the strength of the hydrodynamic interactions, thus resulting in less deformation of the confining interface. The extent of deformation of the confining interface was characterised by defining a deformation parameter, which also allowed us to calculate the critical capillary number $Ca_{crit}$ at which the confining interface comes in contact with the encapsulated particle resulting in a break up of the confining drop. Our analysis shows that $O(Ca^2)$ corrections to the drop shape are important and may qualitatively differ from the $O(Ca)$ predictions for compound particles. For example, $O(Ca)$ correction to the confining drop shape shows that the biaxial flow is stronger in deforming the confining interface compared to the uniaxial and shear flows at the same $Ca$. However, $O(Ca^2)$ correction shows that this is not true, instead either uniaxial or shear flows can be stronger depending upon the viscosity ratio of the confining fluid and the viscosity of the suspending medium. Similarly, in generalized extensional flows, the $O(Ca)$ correction shows a monotonic increase in deformation as we move from uniaxial to biaxial flows while the $O(Ca^2)$ correction shows a non-monotonic behaviour with deformation being least for a biaxial flow. Another feature that emerges from $O(Ca^2)$ calculation is its ability to predict the reorientation dynamics of the anisotropically deformed interface.  The elongated direction of the deformed interface rotates closer to the flow direction in an imposed simple shear flow. This reorientation can lead to a reduced deformation of the interface and therefore, in the case of generalized shear flows, it was observed that deformation of the confining drop increases as we move from vorticity dominated flows to extension dominated flows. In other words, the deformation of the confining drop of a compound particle reduces either when the strength of vorticity increases or strength of extension reduces in the imposed flow. 
	
	Using the solution obtained from the domain perturbation approach, we have analyzed the rheology of a dilute dispersion of compound particles. The different limits of the analysis, namely that of a dispersion of (i)  solid particles (ii) drops and (iii) solid particles coated with a thin fluid film have also been discussed in the context of each rheological quantity. The effective viscosity of a dilute dispersion of compound particles is found to vary between the effective viscosity of a dilute suspension of solid particles and fluid drops. While the effective viscosity is found to be independent of the drop deformation at $O(Ca)$, it decreases with increase in the size the confining drop or with decrease in the viscosity of the inner fluid. For the special case of miscible fluids, the effective viscosity can be less than the suspending fluid viscosity in the limit of large viscosity ratio. Similar to the shear viscosity, Trouton viscosity and normal stress differences decrease with increasing the size of the confining drop or decreasing the viscosity of the inner fluid (confining drop fluid). However, unlike the shear viscosity, these rheological parameters depend upon the extent of deformation of the confining interface and thus on $Ca$. It is interesting to note that for very large viscosity ratios, these rheological parameters asymptote to a constant value that depends only on $Ca$ and not on the size ratio of the encapsulated particle to the confining interface.
	
	The fact that the extensional viscosity is not equal to three times the shear viscosity indicates the non-Newtonian nature of a dilute dispersion of the compound particles, which was further characterised using a small amplitude oscillatory shear flow. In the limit of low frequencies, we found that the viscous modulus dominates the elastic modulus since the confining interface relaxes to its quasi-steady state.  However, at high frequencies, the interface deformation time scale is much smaller than the flow time scale that the elastic response dominates the viscous response. Viscous modulus is found to increase and elastic modulus is found to decrease with decrease in the size ratio of the confining drop to the particle. Similarly, we have shown that the viscous modulus increases and elastic modulus decreases  with decrease in viscosity ratio of the outer fluid to the inner fluid. 
	The present study assumes that the surface tension force is much larger than the shear forces that the capillary number is small and depends on a domain perturbation approach to proceed with analytical calculations. Numerical simulations will have to be employed to relax this assumption in order to go beyond the asymptotic limit of small capillary number and to quantitatively determine the deformation behaviour as well as rheology. Similarly, the consequences of eccentric configuration of the compound particle on the stability of the confining drop as well as rheology of the dispersion including hydrodynamic interactions between the compound particles also have to be the subject of future investigations. On the other hand, this study also inspires the investigations related to active compound particles (the encapsulated particle is active, $e.g.,$ microswimmer). Recently, \cite{chaithu2020} investigated the $O(Ca)$ deformation dynamics of an active compound particle and discussed the interesting competition that arises from activity and the imposed flows. However, higher order calculations, eccentric configurations and self propulsion of active compound particles have to be the subject of future investigations.

	\section*{Acknowledgment}
	SPT acknowledges the support by Department of Science and Technology, India via the research grant CRG/2018/000644. PKS acknowledges the support from IPDF grant by Indian Institute of Technology Madras, India.
	
	\section*{Declaration of Interests}
	The authors report no conflict of interest.
	\appendix
	\section{}\label{appA}
	This appendix provides the complete description of the velocity and pressure fields of both inner and outer fluids at leading order ($\textit{O}(1)$) when subjected to a linear flow. The pressure and velocity fields for both outer and inner fluids at leading order, same as (\ref{eq8a})-(\ref{eq11a}) but simplified after substituting for the expressions for spherical harmonics, are as follows \citep{chaithu2019}: 
	\begin{equation}\label{eqA1}
		p^{(0)}=\frac{c_{1}}{r^{5}}(\bm{x \cdot E \cdot x}),
	\end{equation}
	\begin{equation} \label{eqA2}
		\begin{aligned}
			\bm{u}^{(0)}=\left(1-\frac{6 c_{3}}{r^{5}}\right)&\bm{E\cdot x}+ \bm{\Omega \cdot x} + \left(\frac{c_{1}}{2r^{5}}+\frac{15 c_{3}}{r^{7}}\right)\bm{x \cdot (x \cdot E \cdot x)},
		\end{aligned}
	\end{equation}
	\begin{equation} \label{eqA3}
		\hat{p}^{(0)}=\left(-\frac{126 d_{3}}{5}+\frac{e_{1}}{r^{5}}\right)(\bm{x \cdot E \cdot x}),
	\end{equation}
	\begin{equation}\label{eqA4}
		\begin{aligned}
			\bm{\hat{u}}^{(0)}=&\left(d_{2}-6 d_{3}r^{2}-\frac{6 e_{3}}{r^{5}} \right) \bm{E \cdot x}+ \bm{\Omega \cdot x} + \left( \frac{12 d_{3}}{5}+\frac{\lambda e_{1}}{2r^{5}}+\frac{15 e_{3}}{r^{7}}\right)\bm{x \cdot (x \cdot E \cdot x)}.
		\end{aligned}
	\end{equation}
	In above equations (\ref{eqA1})-(\ref{eqA4}), the unknown constants $c_{i}$, $d_{i}$ and $e_{i}$ for $i=1,2,3$, are
	\begin{equation}\label{eqA5}
		\begin{aligned}
		c_{1}&=  \Big(\big(-240 -190 \alpha^{10}-1125 \alpha^7+1680 \alpha^5-1000 \alpha^3+\big(30 \alpha^{10}+1250 \alpha^7-3360 \alpha^5\\
		&+2000 \alpha^3+80\big) \lambda+\big(160 \alpha^{10}-1000 \alpha^7+1680 \alpha^5-1000 \alpha^3+160\big) \lambda^2 \big)+ b_{1} \big( \big(-128 \alpha^{10}\\
		&+800 \alpha^7-1344 \alpha^5+800 \alpha^3-128\big) \lambda^2+\big(-152 \alpha^{10}-100 \alpha^7+924 \alpha^5-800 \alpha^3\\
		&+128\big) \lambda \big) \Big) \big/ \Big( 38 \alpha^{10}+225 \alpha^7-336 \alpha^5+200 \alpha^3+48 +\big(89 \alpha^{10}+75 \alpha^7-168 \alpha^5+100 \alpha^3\\
		&-96\big) \lambda+\big(48 \alpha^{10}-300 \alpha^7+504 \alpha^5-300 \alpha^3+48\big) \lambda^2\Big), 
		\end{aligned}
	\end{equation}
	\begin{equation}\label{eqA6}
		c_{2}=0,
	\end{equation}
	\begin{equation}\label{eqA7}
		\begin{aligned}
		c_{3}&=  \Big(  \big(48+ 38 \alpha^{10}+225 \alpha^7-336 \alpha^5+200 \alpha^3+\big(-6 \alpha^{10}-425 \alpha^7+847 \alpha^5-400 \alpha^3\\
		&-16\big) \lambda+\big(-32 \alpha^{10}+200 \alpha^7-336 \alpha^5+200 \alpha^3-32\big) \lambda^2 \big) +b_{1} \big(\big(16 \alpha^{10}-100 \alpha^7+168 \alpha^5\\
		&-100 \alpha^3+16\big) \lambda^2+\big(24 \alpha^{10}+100 \alpha^7-308 \alpha^5+200 \alpha^3-16\big) \lambda \big)\Big)\Big/ \Big( 6 \big(48 \alpha^{10}-300 \alpha^7\\
		&+504 \alpha^5-300 \alpha^3+48\big) \lambda^2+6 \big(89 \alpha^{10}+75 \alpha^7-168 \alpha^5+100 \alpha^3-96\big) \lambda+6 \big(38 \alpha^{10}\\
		&+225 \alpha^7-336 \alpha^5+200 \alpha^3+48\big) \Big), 
		\end{aligned}
	\end{equation}
	\begin{equation}\label{eqA8}
		\begin{aligned}
	d_{1}&= -126 \Big( \big (  -250 \alpha ^5 \big(\alpha ^2-1\big) +250 \alpha ^5 \big(\alpha ^2-1\big) \lambda \big)-10 \alpha ^5 b_{1}  \big(2 \alpha ^5-5 \alpha ^2+8+\big(3 \alpha^5 +5 \alpha ^2   \\
	&-8\big) \lambda\big) \Big) \Big/ 5 \Big(3 \big(38 \alpha ^{10}+225 \alpha ^7-336 \alpha ^5+200 \alpha ^3+48\big)+3 \big(89 \alpha ^{10}+75 \alpha ^7 -168 \alpha ^5\\
		& +100 \alpha ^3 -96\big) \lambda +3 \big(48 \alpha ^{10} -300 \alpha ^7+504 \alpha ^5-300 \alpha ^3+48\big) \lambda ^2  \Big),
		\end{aligned}
	\end{equation}
	\begin{equation}\label{eqA9}
		\begin{aligned}
		d_{2}&= \Big(5 \alpha^3 \lambda \big( 100+ 19 \alpha^7-84 \alpha^2+\big(16 \alpha^7+84 \alpha^2-100\big) \lambda \big) -4 \alpha^3 b_{1} \lambda \big( 40+ 16 \alpha^7\\
		&-21 \alpha^2+\big(19 \alpha^7+21 \alpha^2-40\big) \lambda  \big) \Big) \Big/ \Big( 38 \alpha^{10}+225 \alpha^7-336 \alpha^5+200 \alpha^3 \\
		&+48 +\big(89 \alpha^{10}+75 \alpha^7-168 \alpha^5+100 \alpha^3-96\big) \lambda +\big(48 \alpha^{10}-300 \alpha^7+504 \alpha^5\\
		&-300 \alpha^3+48\big) \lambda^{2} \Big),
		\end{aligned}
	\end{equation}
	\begin{equation}\label{eqA10}
		\begin{aligned}
		d_{3}&=\Big( \big ( -250 \alpha ^5 \big(\alpha ^2-1\big) \lambda+250 \alpha^5 \big(\alpha ^2-1\big) \lambda ^2 \big)-10 \alpha ^5 b_{1} \lambda  \big(2 \alpha ^5-5 \alpha ^2+8+\big(3 \alpha^5\\
		&+5 \alpha ^2-8\big) \lambda\big) \Big)\Big/ \Big(3 \big(38 \alpha ^{10}+225 \alpha ^7-336 \alpha ^5+200 \alpha ^3+48\big)+3 \big(89 \alpha ^{10}+75 \alpha ^7\\
		&-168 \alpha ^5+100 \alpha ^3-96\big) \lambda +3 \big(48 \alpha ^{10}-300 \alpha ^7+504 \alpha ^5-300 \alpha ^3+48\big) \lambda ^2  \Big), 
		\end{aligned}
	\end{equation}
	\begin{equation}\label{eqA11}
		\begin{aligned}
		e_{1}=&\Big(-400-475 \alpha ^7+\big(400-400 \alpha ^7\big) \lambda +b_{1} \big(320 \alpha ^7-168 \alpha ^5+128+\big(380 \alpha ^7-252 \alpha ^5\\
		&-128\big) \lambda \big) \Big)\Big/ \Big( 38 \alpha ^{10}+225 \alpha ^7-336 \alpha ^5+200 \alpha ^3+48+\big(48 \alpha ^{10} -300 \alpha ^7+504 \alpha ^5\\
		&-300 \alpha ^3+48\big) \lambda ^2+\big(89 \alpha ^{10}+75 \alpha ^7-168 \alpha ^5+100 \alpha ^3-96\big) \lambda +48\Big),
		\end{aligned}
	\end{equation}	
	\begin{equation}\label{eqA12}
		e_{2}=0,
	\end{equation}
	\begin{equation}\label{eqA13}
		\begin{aligned}
		e_{3}=&\Big(-\big(-95 \alpha ^5-80 \big) \lambda -\big(80-80 \alpha ^5\big) \lambda ^2 +b_{1} \big(-\big(64 \alpha ^5
		-40 \alpha ^3+16\big) \lambda -\big(76 \alpha ^5\\
		&-60 \alpha ^3-16\big) \lambda ^2\big) \Big) \Big/ \Big(6 \big(38 \alpha ^{10}+225 \alpha ^7-336 \alpha ^5+200 \alpha ^3+48\big)+6 \big(89 \alpha ^{10}+75 \alpha ^7\\
		&-168 \alpha ^5+100 \alpha ^3-96\big) \lambda + 6 \big(48 \alpha ^{10}-300 \alpha ^7+504 \alpha ^5-300 \alpha ^3+48\big) \lambda ^2 \Big).
		\end{aligned}
	\end{equation}		
	The $\textit{O}(1)$ velocity field also enabled the determination of the shape of the deformed interface of the confining drop upto $\textit{O}(Ca)$. The deformed interface, which is described in terms of the shape parameter $b_1=b_{1}^{*} \Big(1-\exp\big(-\frac{t}{t_{c}}\big)\Big)$ (refer (\ref{eq15a})), where
	the time independent parameter $b_{1}^{*}$ is
\begin{equation}\label{eqA21}
	\begin{aligned}
b_{1}^{*} &= \big( -32 + 200 \alpha^3 - 231 \alpha^5 + 25 \alpha^7 + 
38 \alpha^{10} + (32 - 200 \alpha^3
+ 336 \alpha^5 - 200 \alpha^7 \\
&+ 
32 \alpha^{10}) \lambda\big)\Big/\big( 4 (-4 + 10 \alpha^3 - 10 \alpha^7 + 4 \alpha^{10}) + 
4 (4 - 25 \alpha^3 + 42 \alpha^5 - 25 \alpha^7 + 4 \alpha^{10}) \lambda\big),
\end{aligned}
\end{equation}
and the time scale, $t_{c}$, that controls the rate of deformation is
\begin{equation}\label{eqA22}
	\begin{aligned}
t_{c}&=\big(48 + 200 \alpha^3 - 336 \alpha^5 + 225 \alpha^7 + 
38 \alpha^{10} + (-96 + 100 \alpha^3 - 168 \alpha^5 + 75 \alpha^7 \\
& + 
89 \alpha^{10}) \lambda
+ (48 - 300 \alpha^3 + 504 \alpha^5 - 300 \alpha^7 + 48 \alpha^{10}) \lambda^2 \big)\Big/ \big(10 (-4 + 10 \alpha^3 \\
&- 10 \alpha^7 + 4 \alpha^{10}) \lambda +  10 (4 - 25 \alpha^3 + 42 \alpha^5 - 25 \alpha^7 + 4 \alpha^{10}) \lambda^2 \big).
\end{aligned}
\end{equation} 
When the confining interface has zero surface tension \textit{i.e.}, $Ca \rightarrow \infty$ or $b_{1}=0$, the expression for $c_{1}$ simplifies to 
\begin{equation}\label{eqA23}
	\begin{aligned}
		c_{1}^{S}&=5 \big(-38 \alpha^{10}-225 \alpha^7+336 \alpha^5-200 \alpha^3-48+ \big(6 \alpha^{10}+250 \alpha^7-672 \alpha^5 \\
		&+400 \alpha^3+16\big) \lambda+ \big(32 \alpha^{10}-200 \alpha^7+336 \alpha^5-200 \alpha^3+32\big) \lambda^2\big) \Big/ \big(38 \alpha^{10} \\
		&+225 \alpha^7-336 \alpha^5+200 \alpha^3+48+\big(89 \alpha^{10}+75 \alpha^7-168 \alpha^5+100 \alpha^3-96\big) \lambda\\
		 &+\big(48 \alpha^{10}-300 \alpha^7+504 \alpha^5-300 \alpha^3+48\big) \lambda^2\big).
	\end{aligned}
\end{equation}
The corresponding unknown constants for a drop (without an encapsulated particle) can be obtained by taking the limit $\alpha \rightarrow \infty$, and are given by 
\begin{equation}\label{eqA15}
	c_{1}=-\frac{2 \lambda+5}{\lambda+1}, \quad \quad c_{2}=0, \quad \quad	c_{3}=\frac{1}{6 \lambda+6},
\end{equation}
\begin{equation}\label{eqA16}
	d_{1}=\frac{21}{2(\lambda+1)},  \quad \quad d_{2}=-\frac{3 \lambda}{2 \lambda+2},	  \quad \quad 	d_{3}=-\frac{5 \lambda}{12 (\lambda+1)},
\end{equation}
\begin{equation}\label{eqA17}
		e_{1}=0, \quad \quad e_{2}=0, \quad \quad e_{3}=0.	
\end{equation}
Further, the steady state value of the shape parameter $b_{1}$ for a simple drop is 
\begin{equation}\label{eqA18}
	b_{1}^{*}=\frac{16 \lambda+19}{8 (\lambda+1)}.
\end{equation}
	\section{}\label{appB}
	This appendix provides the complete description of the velocity and pressure fields of both inner and outer fluids at $\textit{O}(Ca)$ when subjected to a linear flow. The solutions, same as (\ref{eq21a})-(\ref{eq24a}) but simplified after substituting for the expressions for spherical harmonics, are as follows: 
	\begin{equation}\label{eqB1}
		\begin{aligned}
			p^{(1)}&=\left(\frac{c_{5}}{r^{3}}+\frac{6c_{6}}{r^{5}}  \right)\bm{E : E}-\left( \frac{3c_{5}}{r^{5}}+\frac{60c_{6}}{r^{7}}\right)\bm{x \cdot (E \cdot E) \cdot x}+\frac{105c_{6}}{r^{9}}(\bm{x \cdot E \cdot x})^{2}\\
			&-\frac{3c_{8}}{r^{5}} \bm{x \cdot (\Omega \cdot E) \cdot x},
		\end{aligned}
	\end{equation}
	\begin{equation}\label{eqB2}
		\begin{aligned}
		\bm{u}^{(1)}&=\Big(\frac{c_{5}}{2r^{3}}+\frac{3c_{6}}{r^{5}}-\frac{3c_{16}}{r^{5}} +\frac{30 c_{23}}{r^{7}} \Big)\bm{x (E:E)}+\left(\frac{6c_{6}}{r^{5}}-\frac{6c_{16}}{r^{5}}+\frac{120c_{23}}{r^{7}} \right)\bm{(E \cdot E) \cdot x}  \\
			&-\Big( \frac{3c_{5}}{2r^{5}}+\frac{30c_{6}}{r^{7}}-\frac{15c_{16}}{r^{7}}+\frac{420c_{23}}{r^{9}}\Big)\bm{x (x \cdot (E \cdot E) \cdot x)}\\
			&-\left( \frac{15 c_{6}}{r^{7}}+\frac{420 c_{23}}{r^{9}} \right)\bm{(x \cdot E) (x\cdot E \cdot x)}+\left(\frac{105 c_{6}}{2r^{9}}+\frac{945 c_{23}}{r^{11}} \right)\bm{x (x \cdot E \cdot x)^{2}}\\
			& +\left(\frac{3c_{25}}{r^{5}} \right)(\bm{\Omega \cdot E }) \bm{\cdot x}-\left(\frac{3c_{25}}{r^{5}} \right)(\bm{E \cdot \Omega} ) \bm{\cdot x}-\left(\frac{3c_{8}}{2r^{5}}+\frac{15c_{25}}{r^{7}} \right)\bm{x (x \cdot (\Omega \cdot E) \cdot x)},
		\end{aligned}
	\end{equation}
	\begin{equation}\label{eqB3}
		\begin{aligned}
			\hat{p}^{(1)}&=\left(d_{5}r^{2}+6d_{6}r^{4}+d_{7}+\frac{e_{5}}{r^{3}}+\frac{6e_{6}}{r^{5}}  \right)\bm{E : E}-\Big(3d_{5}+60d_{6}r^{2}+\frac{3e_{5}}{r^{5}}\\
			&+\frac{60e_{6}}{r^{7}}\Big)\bm{x \cdot (E \cdot E) \cdot x}+\left(105d_{6} +\frac{105e_{6}}{r^{9}}\right)(\bm{x \cdot E \cdot x})^{2}\\
			&-\left(3d_{8}+\frac{3e_{8}}{r^{5}} \right)\bm{x \cdot (\Omega \cdot E) \cdot x},
		\end{aligned}
	\end{equation}
	\begin{equation}\label{eqB4}
		\begin{aligned}	
	    \bm{\hat{u}}^{(1)}&=\bigg(\frac{\lambda d_{5}r^2}{2}+\frac{\lambda d_{7}}{2}+d_{15}-3d_{16} r^{2}-\frac{120 d_{23} r^{4}}{7}+\frac{\lambda e_{5}}{2r^{3}}+\frac{3 \lambda e_{6}}{r^{5}}-\frac{3e_{16}}{r^{5}}\\
			&+\frac{30 e_{23}}{r^{7}} \bigg)\bm{x (E:E)}	+\Big(d_{13}-6d_{14} r^{2} -6 d_{16} r^{2} +120 d_{23} r^{4}+\frac{6 \lambda e_{6}}{r^{5}}-\frac{6e_{16}}{r^{5}}\\
			&+\frac{120e_{23}}{r^{7}} \Big )\bm{(E \cdot E) \cdot x}+\Big( -\frac{3 \lambda  d_{5}}{2}+15d_{16}+\frac{360 d_{23}r^{2}}{7}- \frac{3 \lambda e_{5}}{2r^{5}}-\frac{30 \lambda e_{6}}{r^{7}} \\
			&+\frac{15e_{16}}{r^{7}} -\frac{420e_{23}}{r^{9}}\Big)\bm{x (x \cdot (E \cdot E) \cdot x)}+\bigg(15 d_{14} -420d_{23}r^{2} -\frac{15 \lambda  e_{6}}{r^{7}}\\
			&-\frac{420 e_{23}}{r^{9}} \bigg)\bm{(x \cdot E) (x\cdot E \cdot x)}+\bigg(120d_{23}+\frac{105 \lambda  e_{6}}{2r^{9}}+\frac{945 e_{23}}{r^{11}} \bigg)\bm{x (x \cdot E \cdot x)^{2}}\\
			&+\Big(d_{18}+3d_{25}r^{2} +\frac{3e_{25}}{r^{5}} \Big)\bm{(\Omega \cdot E ) \cdot x}+\Big(d_{19}-3d_{25}r^{2}-\frac{3e_{25}}{r^{5}} \Big)\bm{(E \cdot \Omega ) \cdot x}\\
			&-\bigg(\frac{3 \lambda d_{8}}{2}+15d_{25}+\frac{3 \lambda e_{8}}{2r^{5}}+\frac{15e_{25}}{r^{7}} \bigg)\bm{x (x \cdot (\Omega \cdot E) \cdot x)}.
		\end{aligned}
	\end{equation}	
	The unknown constants are obtained on the application of boundary conditions as follows:
	\begin{enumerate}
		\item The no-slip boundary condition (see (\ref{eq18a})) on the surface of the particle gives the following set of equations:
		\begin{subequations} 
			\begin{equation}\label{B1a}
				-\frac{2\lambda d_{4} }{21  \alpha^{2} }+ \frac{ \lambda e_4  \alpha^{3}}{2}  +15 e_{12}\alpha^5=0,
			\end{equation}
			\begin{equation}\label{B1b}
				d_{11}+\frac{5 \lambda~d_{4} } {21 \alpha^{2}}- 6e_{12} \alpha^{5} =0, 
			\end{equation}
			\begin{equation}\label{B1c}
				\begin{aligned}
					\frac{\lambda~d_{5}}{2 \alpha^{2} }+\frac{\lambda~d_{7}}{2}+d_{15}-\frac{3d_{16}}{\alpha^{2}}-\frac{120 d_{23} }{7 \alpha^{4} }+\frac{\lambda~e_{5} \alpha^{3}}{2 }+ 3\lambda~e_{6} \alpha^{5} 
					+e_{15}\alpha^{3}\\
					- 3e_{16}\alpha^{5}+ 30 e_{23} \alpha^{7}=0,
				\end{aligned}
			\end{equation}
			\begin{equation}\label{B1d}
				\begin{aligned}
					d_{13}-\frac{6 d_{14}}{\alpha^{2}}  -\frac{6 d_{16}}{\alpha^{2} }+\frac{120 d_{23}}{ \alpha^{4}}+6\lambda~e_{6}\alpha^{5}
					- 6e_{16} \alpha^{5}+ 120 e_{23} \alpha^{7} =0,
				\end{aligned}
			\end{equation}
			\begin{equation}\label{B1e}
				\begin{aligned}
					-\frac{3\lambda~ d_{5} }{2 \alpha^{2}}+\frac{15d_{16}}{\alpha^{2}}+\frac{360 d_{23}}{7 \alpha^{4}}-\frac{3 \lambda~ e_{5} \alpha^{3}}{2}- 30\lambda~ e_{6}\alpha^{5}
					+  15e_{16}\alpha^{5}- 420e_{23}\alpha^{7}=0,
				\end{aligned}
			\end{equation}
			\begin{equation}\label{B1f}
				\frac{15 d_{14}}{\alpha^{2}}-\frac{420d_{23}}{\alpha^{4}} - 15\lambda~ e_{6}\alpha^{5}- 420 e_{23}\alpha^{7}=0,
			\end{equation}
			\begin{equation}\label{B1g}
				\frac{120 d_{23}}{\alpha^{4}} +\frac{105\lambda~ e_{6} \alpha^{5}}{2}+ 945 e_{23}\alpha^{7}=0,
			\end{equation}
			\begin{equation}\label{B1h}
				d_{17}+e_{17} \alpha^{3}=0,
			\end{equation}
			\begin{equation}\label{B1i}
				d_{18}-\frac{6d_{20}}{\alpha^2}+\frac{3d_{25}}{\alpha^{2}}+ e_{18}\alpha^{3}- 6e_{20}\alpha^{5}+ 3e_{25}\alpha^{5}=0,
			\end{equation}
			\begin{equation}\label{B1j}
				d_{19}-\frac{3d_{25}}{\alpha^{2}} +e_{19} \alpha^{3}- 3e_{25}\alpha^{5}=0,
			\end{equation}
			\begin{equation}\label{B1k}
				-\frac{3\lambda~ d_{8}}{2 \alpha^{2}}-\frac{15d_{25}}{\alpha^{2}} -\frac{3\lambda~ e_{8}\alpha^{3}}{2}- 15e_{25}\alpha^{5}=0,
			\end{equation}
			\begin{equation}\label{B1l}
				\frac{	15d_{20} }{\alpha^{2}}+15e_{20}\alpha^{5}=0,
			\end{equation}
			\begin{equation}\label{B1m}
				d_{22}+\frac{\lambda~ d_{10}}{2}+\frac{6d_{24} }{5\alpha^{2}}-\frac{\lambda~ e_{9}\alpha^{3}}{2}+ e_{22}\alpha^{3}-3e_{24}\alpha^{5}=0,
			\end{equation}
			\begin{equation}\label{B1n}
				d_{21}+\frac{6d_{24}}{\alpha^{2}}+6e_{24}\alpha^{5}=0,
			\end{equation}
			\begin{equation}\label{B1o}
				-\frac{12\lambda~d_{4} }{5\alpha^{2}}-\frac{3\lambda~e_{9}\alpha^{3}}{2 }-15e_{24}\alpha^{5}=0.
			\end{equation}
		\end{subequations}
		\\
		\item The boundary condition concerning the normal velocity at the confining interface (see (\ref{eq19a})) is
		\begin{align}\label{eq46}
			\left(\bm{u}\big|_{r=  (1+Ca f^{(0)})}\right)^{(1)} \cdot \bm{n}^{(0)}&+\left(\bm{u}\big|_{r=  (1+Ca f^{(0)})}\right)^{(0)} \cdot \bm{n}^{(1)} \nonumber \\ &= \left(\hat{\bm{u}}\big|_{r=  (1+Ca f^{(0)})}\right)^{(1)} \cdot \bm{n}^{(0)} 
			+\left(\hat{\bm{u}}\big|_{r=  (1+Ca f^{(0)})}\right)^{(0)} \cdot \bm{n}^{(1)},
		\end{align}
		which reduces to the following set of equations.
		\begin{subequations}
			\begin{equation}\label{B2a}
				\begin{aligned}
					\frac{c_{4}}{2  }+  9c_{12}  -d_{11}-\frac{ \lambda d_{4}  }{7}-\frac{\lambda e_{4}}{2} -9e_{12} =0, 
				\end{aligned}
			\end{equation}
			\begin{equation}\label{B2b}
				\begin{aligned}
					\frac{c_5}{2 }+3c_6 + c_{15} -3c_{16}+ 30c_{23} -\frac{\lambda d_{5}}{2}
					-\frac{\lambda d_7}{2}-d_{15}	+3d_{16} \\
					+\frac{120d_{23}}{7}-\frac{\lambda e_5}{2 }	- 3\lambda e_6 - e_{15} +3e_{16} - 30e_{23}=0,
				\end{aligned}
			\end{equation}
			\begin{equation}\label{B2c}
				\begin{aligned}
					-\frac{3c_5}{2 }- 24 c_{6}+ 9c_{16} - 300c_{23} -d_{13}+6d_{14} 
					-9d_{16}  +\frac{3\lambda d_{5}  }{2}-\frac{1200d_{23} }{7} 	+\frac{3\lambda e_5}{2 }  \\
					+ 24 \lambda e_{6} - 9e_{16}+ 300e_{23} +b_{1}\bigg(-2+12c_{3} +2   d_{2}
					-12 d_{3} - 12e_{3} \bigg) =0,
				\end{aligned}
			\end{equation}
			\begin{equation}\label{B2d}
				\begin{aligned}
					\frac{75c_{6}}{2 }+ 525c_{23} -15d_{14} +300d_{23} -\frac{75\lambda e_{6}}{2 }- 525e_{23} 
					+b_{1}\bigg(3- c_{1}	 - 48c_{3} -3  d_{2}  \\
					+\frac{114  d_{3} }{5}+ \lambda e_{1} + 48e_{3}  \bigg)=0,
				\end{aligned}
			\end{equation}
			\begin{equation}\label{B2e}
				\begin{aligned}
					-\frac{3c_{8}}{2 }- 6c_{20} - 9c_{25} +\frac{3\lambda d_{8} }{2}-d_{18}+d_{19}+6d_{20} +9d_{25}
					+\frac{3\lambda e_{8}}{2 }
					+ 6e_{20} + 9e_{25} =0,
				\end{aligned}
			\end{equation}
			\begin{equation}\label{B2f}
				\begin{aligned}
					-\frac{c_{9}}{2 }+ c_{22}  - 3c_{24} -d_{22}-\frac{\lambda d_{10}}{2}-\frac{6d_{24} }{5}+\frac{\lambda e_{9}}{2 }-e_{22} +3e_{24} =0,
				\end{aligned}
			\end{equation}
			\begin{equation}\label{B2g}
				\begin{aligned}
					-\frac{3c_{9}}{2 }- 9c_{24} -d_{21}-\frac{18d_{24} }{5}
					+\frac{3\lambda e_{9}}{2 }+ 9e_{24} =0.
				\end{aligned}
			\end{equation}
		\end{subequations}
		
		\item The boundary condition concerning the tangential component of velocity at the confining interface (see~(\ref{eq19a})) is 
		\begin{align}\label{eq47}
			\big(\bm{u} &\big|_{r= (1+Ca f^{(0)})}\big)^{(1)}-\bm{n}^{(0)} \big(\bm{u} \cdot \bm{n} \big)^{(1)}_{r=  (1+Ca f^{(0)})}
			-\bm{n}^{(1)} \big(\bm{u} \cdot \bm{n} \big)^{(0)}_{r=  (1+Ca f^{(0)})}\nonumber \\
			&=
			\big(\hat{\bm{u}} \big|_{r= (1+Ca f^{(0)})}\big)^{(1)}-\bm{n}^{(0)} \big(\hat{\bm{u}} \cdot \bm{n} \big)^{(1)}_{r=  (1+Ca f^{(0)})}
			-\bm{n}^{(1)} \big(\hat{\bm{u}} \cdot \bm{n} \big)^{(0)}_{r=  (1+Ca f^{(0)})},
		\end{align}
		which reduces to the following set of equations:
		\begin{subequations}
			\begin{equation}\label{B3a}
				\begin{aligned}
					6c_{12} +d_{11}+\frac{5\lambda d_{4} }{21}- 6e_{12} =0,
				\end{aligned}
			\end{equation}
			\begin{equation}\label{B3b}
				\begin{aligned}
					6c_{6} - 6c_{16} + 120c_{23} -d_{13}+6d_{14} +6d_{16} 
					-120d_{23} - 6\lambda e_{6} + 6e_{16} - 120e_{23}=0,
				\end{aligned}
			\end{equation}
			\begin{equation}\label{B3c}
				\begin{aligned}
					6c_{6} - 6c_{16} + 120c_{23} 
					-d_{13}+6d_{14} +6d_{16} -120d_{23} 
					- 6\lambda e_{6} 
					+ 6e_{16} - 120e_{23} \\
					-2b_{1}\left(1- 6c_{3} -d_{2}+6d_{3} + 6e_{3} \right)=0,
				\end{aligned}
			\end{equation}
			\begin{equation}\label{B3d}
				\begin{aligned}
					- 15c_{6} - 420c_{23} -15d_{14} +420d_{23} +15\lambda e_{6} + 420e_{23} +b_{1}\bigg(3+ c_{1} 
					+ 42c_{3} -3d_{2}\\
					+\frac{126d_{3} }{5}- \lambda e_{1} -42e_{3} \bigg)=0,
				\end{aligned}
			\end{equation}
			\begin{equation}\label{B3e}
				\begin{aligned}
					15c_{6} + 420c_{23} +15d_{14} -420d_{23} - 15\lambda e_{6} - 420e_{23} 
					+b_{1}\bigg(-5-c_{1}
					- 30c_{3} +5d_{2}\\-\frac{186d_{3} }{5}
					+\lambda e_{1} + 30e_{3} \bigg)=0,
				\end{aligned}
			\end{equation}
			\begin{equation}\label{B3f}
				\begin{aligned}
					c_{17} -d_{17}-e_{17} =0,
				\end{aligned}
			\end{equation}
			\begin{equation}\label{B3g}
				\begin{aligned}
					c_{18} - 6c_{20} + 3c_{25} -d_{18}+6d_{20} -3d_{25} 
					- e_{18} + 6e_{20} -3e_{25}=0,
				\end{aligned}
			\end{equation}
			\begin{equation}\label{B3h}
				\begin{aligned}
					c_{19} - 3c_{25} -d_{19}+3d_{25} - e_{19} + 3e_{25} =0,
				\end{aligned}
			\end{equation}
			\begin{equation}\label{B3i}
				\begin{aligned}
					6c_{20} - 6c_{25}  +d_{18}-d_{19}-6d_{20} +6d_{25} -6e_{20} + 6e_{25}=0,
				\end{aligned}
			\end{equation}
			\begin{equation}\label{B3j}
				\begin{aligned}
					15c_{20}  -15d_{20}- 15e_{20}  =0,
				\end{aligned}
			\end{equation}
			\begin{equation}\label{B3k}
				\begin{aligned}
					6c_{24} -d_{21}-6d_{24} - 6e_{24} =0.
				\end{aligned}
			\end{equation}
		\end{subequations}
		\item At $\textit{O}(Ca)$, the tangential stress balance (see~(\ref{eq20ab})) is obtained as,
		\begin{align}\label{eq49}
			\big(\bm{\sigma} \cdot \bm{n} \big)^{(1)}_{r=  (1+Ca f^{(0)})}- \bm{n}^{(1)}\big(\bm{\sigma}\cdot \bm{nn} \big)^{(0)}_{r= (1+Ca f^{(0)})}
			- \bm{n}^{(0)}\big(\bm{\sigma} \cdot \bm{nn} \big)^{(1)}_{r= (1+Ca f^{(0)})} =  \nonumber \\
			(1/\lambda) \bigg(\big(\hat{\bm{\sigma}} \cdot \bm{n} \big)^{(1)}_{r= (1+Ca f^{(0)})}- \bm{n}^{(1)}\big(\hat{\bm{\sigma}}\cdot \bm{nn} \big)^{(0)}_{r= (1+Ca f^{(0)})}
			- \bm{n}^{(0)}\big(\hat{\bm{\sigma}} \cdot \bm{nn} \big)^{(1)}_{r= (1+Ca f^{(0)})} \bigg),
		\end{align}
		which reduces to the following set of equations:
		\begin{subequations}
			\begin{equation}\label{B4a} 
				\begin{aligned}
					\lambda \Big( c_{4} + 48c_{12} \Big)-2d_{11}-\frac{16 \lambda d_4 }{21}- \lambda e_{4} 
					- 48e_{12} =0, 
				\end{aligned}
			\end{equation}
			\begin{equation}\label{B4b}
				\begin{aligned}
					\lambda\Bigg(- 3c_{5} - 78c_{6} + 48c_{16} - 1440c_{23}  +b_{1}\bigg( 24c_{3} 
					-4\bigg)\Bigg)
					+3\lambda d_{5} -2d_{13}\\
					+24d_{14} -6d_{16}	-\frac{5760d_{23} }{7}+ 3\lambda e_{5} + 78 \lambda e_{6} - 48e_{16} + 1440e_{23} \\
					+b_{1}\bigg(4d_{2}-24d_{3} - 24e_{3} \bigg)=0,
				\end{aligned}
			\end{equation}
			\begin{equation}\label{B4c}
				\begin{aligned}
					\lambda\Bigg(- 3c_{5} - 78c_{6} + 48c_{16} - 1440c_{23} +b_{1}\bigg(- 2c_{1} - 72c_{3} -8\bigg)\Bigg)
					+3\lambda d_{5} \\
					-2d_{13}+24d_{14}  -6d_{16}-\frac{5760d_{23} }{7}+ 3\lambda e_{5} 	+ 78\lambda e_{6} - 48e_{16} + 1440e_{23}\\ +b_{1}\bigg(8d_{2}-\frac{312d_{3} }{5}+ 2\lambda e_{1} + 72e_{3} \bigg)=0,
				\end{aligned}
			\end{equation}
			\begin{equation}\label{B4d}
				\begin{aligned}
					\lambda\Bigg( 225c_{6} + 5040c_{23} +b_{1}\bigg(8- 9c_{1} - 468c_{3} \bigg)\Bigg)
					-90d_{14} +2880d_{23}	- 225\lambda e_{6}   \\
					- 5040e_{23}	+b_{1}\left(-8d_{2}+\frac{576d_{3} }{5}+ 9\lambda e_{1} + 468e_{3} \right)=0,
				\end{aligned}
			\end{equation}
			\begin{equation}\label{B4e}
				\begin{aligned}
					\lambda\Bigg(- 225c_{6} - 5040c_{23} -b_{1}\Big(12- 7c_{1} -372c_{3} \Big)\Bigg)
					+90d_{14}-2880d_{23}	+ 225\lambda e_{6}  \\
					+ 5040e_{23} 	-b_{1}\left(-12d_{2}+\frac{768d_{3} }{5}+ 7\lambda e_{1} + 372e_{3} \right)=0,
				\end{aligned}
			\end{equation}
			\begin{equation}\label{B4f}
				\begin{aligned}
					\lambda\bigg(- 3c_{17} \bigg)+ 3e_{17} =0,
				\end{aligned}
			\end{equation}
			\begin{equation}\label{B4g}
				\begin{aligned}
					\lambda\bigg(-\frac{3c_{8}}{2 }- 3c_{18} + 24c_{20} - 24c_{25} \bigg)+\frac{3\lambda d_{8} }{2}
					-d_{18}+d_{19}+18d_{20} +3d_{25} +\frac{3\lambda e_{8}}{2 }\\+ 3e_{18} 
					- 24e_{20} + 24e_{25} =0,
				\end{aligned}
			\end{equation}
			\begin{equation}\label{B4h}
				\begin{aligned}
					\lambda\Bigg(- 6c_{20} + 3c_{18} - 24c_{25} -\frac{3c_{8}}{2 }\Bigg)+\frac{3\lambda d_{8} }{2}
					-d_{18}+d_{19}+6d_{20} +3d_{25} + 6e_{20} \\- 3e_{18} 
					+ 24e_{25} +\frac{3 \lambda e_{8}}{2 }=0,
				\end{aligned}
			\end{equation}
			\begin{equation}\label{B4i}
				\begin{aligned}
					\lambda\bigg( 3c_{8} - 18c_{20} + 48c_{25} \bigg)-3\lambda d_{8} +2d_{18}-2d_{19}
					-24d_{20} -6d_{25} \\
					- 3\lambda e_{8} + 18e_{20} - 48e_{25} = 0,
				\end{aligned}
			\end{equation}
			\begin{equation}\label{B4j}
				\begin{aligned}
					\lambda\bigg(- 75c_{20} \bigg)-30d_{20} +75e_{20} =0,
				\end{aligned}
			\end{equation}
			\begin{equation}\label{B4k}
				\begin{aligned}
					\lambda\bigg(3c_{9} + 48c_{24} \bigg)+2d_{21}+\frac{96d_{24} }{5}
					- 3\lambda e_{9} - 48e_{24} =0.
				\end{aligned}
			\end{equation}
		\end{subequations}
		\item Finally, at $\textit{O}(Ca)$, the boundary condition describing the balance of normal stress (see~(\ref{eq20aa})) at the confining interface is
		\begin{equation}\label{eq48}
			\begin{aligned}
				&\big(\bm{\sigma} \cdot \bm{n} \big)^{(1)}_{r= (1+Ca f^{(0)})} \cdot \bm{n}^{(0)}+\big(\bm{\sigma} \cdot \bm{n} \big)^{(0 )}_{r= (1+Ca f^{(0)})} \cdot \bm{n}^{(1)} -(1/\lambda )\bigg(\big(\hat{\bm{\sigma}} \cdot \bm{n} \big)^{(1)}_{r= (1+Ca f^{(0)})} \cdot \bm{n}^{(0)}\nonumber \\
				&+\big(\hat{\bm{\sigma}} \cdot \bm{n} \big)^{(0 )}_{r= (1+Ca f^{(0)})} \cdot \bm{n}^{(1)} \bigg)==  \left(\nabla \cdot \bm{n}\right)^{(2)},
			\end{aligned}
		\end{equation}
		which reduces to the following set of equations:
		\begin{subequations}
			\begin{equation}\label{B5a}
				\begin{aligned}
					\lambda\Bigg(- 3c_{4} - 72c_{12} \Bigg)-2d_{11}+\frac{\lambda d_{4} }{7}
					+ 3\lambda e_{4} +72e_{12} =4b_{2}\lambda,
				\end{aligned}
			\end{equation}
			\begin{equation}\label{B5b}
				\begin{aligned}
					\lambda\Bigg(- 3c_{5} - 30c_{6} - 4c_{15} + 24c_{16} - 360c_{23} \Bigg)
					-2\lambda d_{5}+6\lambda d_{6} -2d_{15} +18d_{16} 
					\\
					+\frac{1200d_{23}}{7} + 3\lambda e_{5} + 30\lambda e_{6} + 4e_{15} -24e_{16} 
					+360e_{23} =-\lambda\left(2b_{3}+2b_{4}\right),
				\end{aligned}
			\end{equation}
			\begin{equation}\label{B5c}
				\begin{aligned}
					\lambda\Bigg( 9c_{5} + 252c_{6} - 72c_{16} + 3600c_{23} +b_{1}\bigg(-8- 4c_{1} 
					-192c_{3} \bigg)\Bigg)-2d_{13}+6\lambda d_{5} \\	-60\lambda d_{6}
					+36d_{14} 
					-54d_{16}-\frac{12000d_{23}}{7}- 9\lambda e_{5}
					- 252\lambda e_{6}+72e_{16} - 3600e_{23}\\ +b_{1}\Big(-8d_{2}+\frac{384d_{3}}{5}+4\lambda e_{1} + 192e_{3} \Big)
					=\lambda\left(4b_{4}-8b_{5}\right),
				\end{aligned}
			\end{equation}
			\begin{equation}\label{B5d}
				\begin{aligned}
					\lambda \Bigg(-405c_{6} -6300c_{23} +b_{1}\bigg(8+13c_{1}+ 552c_{3} \bigg)\Bigg)
					-90d_{14} +105\lambda d_{6} \\+3000d_{23} + 405\lambda e_{6} 
					+ 6300e_{23} +b_{1}\Big(-8d_{2}+\frac{348d_{3} }{5}- 13\lambda e_{1} - 552e_{3} \Big)\\
					=\lambda\left(-10b_{1}^{2}+18b_{5}\right),
				\end{aligned}
			\end{equation}
			
			\begin{equation}\label{B5e}
				\begin{aligned}
					\lambda\Bigg( 9c_{8} + 48c_{20} + 72c_{25} \Bigg)+6\lambda d_{8} -2d_{18}+2d_{19}
					+36d_{20}
					+54d_{25} - 9\lambda e_{8}\\
					-48e_{20}- 72e_{25} 	=\lambda \left(4b_6\right),
				\end{aligned}
			\end{equation}
			\begin{equation}\label{B5f}
				\begin{aligned} 
					\lambda\bigg( 3c_{9} - 4c_{22} + 24c_{24} \bigg)-\lambda d_{9} -2d_{22}-\frac{36d_{24} }{5}- 3\lambda e_{9} + 4e_{22} - 24e_{24}=\\-\lambda\left(2b_{7}+2b_{8}\right),
				\end{aligned}
			\end{equation}
			\begin{equation}\label{B5g}
				\begin{aligned}
					\lambda\bigg( 9c_{9} + 72c_{24} \bigg)-3\lambda d_{9} -2d_{21}-\frac{108d_{24} }{5}
					-9\lambda e_{9}- 72e_{24} =\lambda\left(4b_{8}\right).
				\end{aligned}
			\end{equation}
		\end{subequations}
	\end{enumerate}
	A unique solution is obtained for the system of linear algebraic equations (\ref{B1a})-(\ref{B5g}) using Mathematica $11.3$, and the corresponding script file is available as a supplementary material. However, due to the complexity of the obtained expressions, which are in terms of the size ratio $\alpha$ and the viscosity ratio $\lambda$, we have not included them here. 
	
	The expressions which are used in the rheological properties are 
	\begin{align*}
	    c_{5}&=2 \Big( \big(1 + \alpha\big)^2 \big(2 + 4 \alpha + 8 \alpha^2 + 7 \alpha^3 + 8 \alpha^4 + 
				4 \alpha^5 + 2 \alpha^6\big)^2 \big(-32 - 64 \alpha - 96 \alpha^2 \\
				&+ 
				72 \alpha^3 + 240 \alpha^4 + 177 \alpha^5 + 114 \alpha^6 + 
				76 \alpha^7 + 38 \alpha^8\big)+\big(6912 + 55296 \alpha + 248832 \alpha^2 \\
				&+ 747840 \alpha^3 + 
				1628160 \alpha^4 + 2594412 \alpha^5 + 2830656 \alpha^6 + 
				1296572 \alpha^7- 2518820 \alpha^8 \\
				&- 7748880 \alpha^9 - 
				12058008 \alpha^{10} - 12764994 \alpha^{11} - 8985708 \alpha^{12} - 
				2758755 \alpha^{13} \\
				&+ 2677280 \alpha^{14} + 5392132 \alpha^{15}
				+5457336 \alpha^{16} + 4038372 \alpha^{17} + 2325360 \alpha^{18} \\
				&+ 
				1033440 \alpha^{19} + 339552 \alpha^{20} + 75456 \alpha^{21} + 9432 \alpha^{22}\big)\lambda+\big(\big(-1 + \alpha\big)^2 \big(4 + 16 \alpha \\
				&+ 40 \alpha^2 + 55 \alpha^3 + 40 \alpha^4 + 
				16 \alpha^5 + 4 \alpha^6\big) \big(-1056 - 6336 \alpha - 22176 \alpha^2 - 
				51936 \alpha^3 \\
				&- 89856 \alpha^4 - 116508 \alpha^5 - 109248 \alpha^6 - 
				59768 \alpha^7 + 18327 \alpha^8 + 81942 \alpha^9 + 95364 \alpha^{10} \\
				&+ 
				67134 \alpha^{11} + 30744 \alpha^{12} + 8784 \alpha^{13} + 1464 \alpha^{14}\big) \big)\lambda^2 
				+\big(8 \big(-1 + \alpha\big)^4 \big(4 + 16 \alpha + 40 \alpha^2 \\
					&+ 55 \alpha^3 + 40 \alpha^4 + 16 \alpha^5 + 4 \alpha^6\big)^3 \big)\lambda^3 \Big) \bigg/ 35\Big(8 \big(-1 + \alpha\big) \big(1 + \alpha\big)^3 \big(2 + 4 \alpha + 8 \alpha^2 + 7 \alpha^3\\
				& +8 \alpha^4 + 4 \alpha^5 + 2 \alpha^6\big)^3+\Big( 12 \big(-1 + \alpha\big)^2 \big(1 + \alpha\big)^2 \big(2 + 4 \alpha + 8 \alpha^2 + 
				7 \alpha^3 + 8 \alpha^4 + 4 \alpha^5
				\end{align*}
		\begin{equation}\label{eqB13}
			\begin{aligned}
				\hspace{0.8cm} 
				& + 2 \alpha^6\big)^2 \big(4 + 16 \alpha + 
				40 \alpha^2 + 55 \alpha^3 + 40 \alpha^4 
				+ 16 \alpha^5 + 4 \alpha^6\big) \Big)\lambda+\Big(6 \big(-1 + \alpha\big)^3 \big(1 + \alpha\big) \\
				&\big(2 + 4 \alpha + 8 \alpha^2 + 7 \alpha^3 + 
				8 \alpha^4 + 4 \alpha^5 + 2 \alpha^6\big) \big(4 + 16 \alpha + 40 \alpha^2 + 
				55 \alpha^3 + 40 \alpha^4 
				+ 16 \alpha^5\\
				& + 4 \alpha^6\big)^2  \Big)\lambda^{2}+\Big(\big(-1 + \alpha\big)^4 \big(4 + 16 \alpha + 40 \alpha^2 + 55 \alpha^3 + 40 \alpha^4 + 
				16 \alpha^5 + 4 \alpha^6\big)^3 \Big) \lambda^{3} \Big),
			\end{aligned}
		\end{equation}
		\begin{equation}\label{eqB14}
			\begin{aligned}
				c_{8}&=\Big( \big(-32 - 64 \alpha - 96 \alpha^2 + 72 \alpha^3 + 240 \alpha^4 + 
				177 \alpha^5 + 114 \alpha^6 + 76 \alpha^7 + 38 \alpha^8\big)^2 \\
				&+\big(16 (-1 + \alpha)^2 (4 + 16 \alpha + 40 \alpha^2 + 55 \alpha^3 + 40 \alpha^4 + 16 \alpha^5 + 4 \alpha^6) (-32 - 64 \alpha - 
				96 \alpha^2 \\
				&+ 72 \alpha^3 + 240 \alpha^4 + 177 \alpha^5 + 
				114 \alpha^6 + 76 \alpha^7 + 38 \alpha^8) \big)\lambda+\big( 64 (-1 + \alpha)^4 (4 + 16 \alpha \\
				&+ 40 \alpha^2 + 55 \alpha^3 + 
				40 \alpha^4 + 16 \alpha^5 + 4 \alpha^6)^2\big) \lambda^{3}\Big)\bigg/ 15 \Big(\big(4 (-1 + \alpha)^2 (1 + \alpha)^2 (2 + 4 \alpha + 8 \alpha^2 \\
				&+ 7 \alpha^3 + 
				8 \alpha^4 + 4 \alpha^5 + 2 \alpha^6)^2\big)
				+\big( 4 (-1 + \alpha)^3 (1 + \alpha) (2 + 4 \alpha + 8 \alpha^2 + 7 \alpha^3 + 
				8 \alpha^4 + 4 \alpha^5 \\
				&+ 2 \alpha^6) (4 + 16 \alpha + 40 \alpha^2 + 
				55 \alpha^3 + 40 \alpha^4 + 16 \alpha^5 + 4 \alpha^6)\big)\lambda+\big( (-1 + \alpha)^4 (4 + 16 \alpha \\
				&+ 40 \alpha^2 + 55 \alpha^3 + 40 \alpha^4 + 16 \alpha^5 + 4 \alpha^6)^2\big)\lambda^2 \Big).
			\end{aligned}
		\end{equation}	
	The unknown constants for a drop can be obtained by taking the limit $\alpha \rightarrow \infty$, and are as follows:
\begin{equation}\label{eqB15}	 
		c_{4}=0, \quad \quad  c_{5}=\frac{64 \lambda^3+732 \lambda^2+1179 \lambda+475}{140 (\lambda+1)^3},  \quad \quad 	c_{6}=-\frac{352 \lambda^2+1138 \lambda+855}{1080 (\lambda+1)^2}, 
\end{equation}
	\begin{equation}\label{eqB16}
 c_{8}=\frac{(16 \lambda+19)^2}{60 (\lambda+1)^2}, \quad \quad 	c_{9}=0,  \quad \quad	c_{12}=0, \quad \quad 	c_{14}=\frac{352 \lambda^2+1138 \lambda+855}{1080 (\lambda+1)^2}, 
	\end{equation}
\begin{equation}\label{eqB17}
c_{15}=0, \quad \quad		c_{16}=\frac{-1568 \lambda^3+458 \lambda^2+4915 \lambda+2565}{7560 (\lambda+1)^3}, \quad \quad c_{17}=0, 
\end{equation}
\begin{equation} \label{eqB18}
c_{18}=0, \quad \quad c_{19}=0, \quad \quad c_{20}=0, \quad  \quad c_{22}=0, \quad \quad 
\end{equation}
\begin{equation} \label{eqB19}
c_{23}=\frac{64 \lambda^2+796 \lambda+855}{15120 (\lambda+1)^2}, \quad \quad	c_{24}=0, \quad \quad c_{25}=-\frac{32 \lambda^2+86 \lambda+57}{120 (\lambda+1)^2},
\end{equation}
\begin{equation}\label{eqB20}
d_{4}=0, \quad \quad  
d_{5}=\frac{16 \lambda^2+83 \lambda+76}{10 (\lambda+1)^3}, \quad \quad
d_{6}=\frac{11 (16 \lambda+19)}{432 (\lambda+1)^2},
\end{equation}
\begin{equation}\label{eqB21}
d_{7}=\frac{\left(32 \lambda^2-554 \lambda-703\right)}{60 (\lambda+1)^2}, \quad \quad
d_{8}=-\frac{7 \left(48 \lambda^2+89 \lambda+38\right)}{40 \lambda (\lambda+1)^2}, \quad \quad
d_{9}=0, 
\end{equation}
\begin{equation} \label{eqB22}
	d_{10}=0, \quad \quad
	d_{11}=0, \quad \quad
	d_{12}=0, \quad \quad
	d_{13}=-\frac{9 \lambda \left(16 \lambda^2+3 \lambda-19\right)}{140 (\lambda+1)^3},
\end{equation}
\begin{equation}\label{eqB23}
d_{14}=-\frac{2 \lambda (16 \lambda+19)}{27 (\lambda+1)^2}, \quad \quad
d_{15}=\frac{\lambda \left(-224 \lambda^3+3942 \lambda^2+8853 \lambda+4579\right)}{840 (\lambda+1)^3}, 
\end{equation}
\begin{equation}\label{eqB24}
d_{16}=\frac{\lambda \left(400 \lambda^2+1307 \lambda+988\right)}{756 (\lambda+1)^3}, \quad \quad	d_{17}=0, \quad \quad
d_{18}=-\frac{304 \lambda^2+617 \lambda+304}{80 (\lambda+1)^2}, 
\end{equation}
\begin{equation}\label{eqB25}
	d_{19}=\frac{304 \lambda^2+617 \lambda+304}{80 (\lambda+1)^2}, \quad \quad
	d_{20}=0,  \quad \quad
	d_{21}=0,   \quad \quad
	d_{22}=0,  \quad \quad
\end{equation}
\begin{equation} \label{eqB26}
	d_{23}=-\frac{7 \lambda (16 \lambda+19)}{4320 (\lambda+1)^2}, \quad \quad
d_{24}=0, \quad \quad
d_{25}=\frac{48 \lambda^2+89 \lambda+38}{48 (\lambda+1)^2},
	\end{equation}
\begin{equation}	\label{eqB27} 
	e_{4}=0, \quad \quad  e_{5}=0,  \quad \quad 	e_{6}=0,\quad \quad  e_{8}=0, \quad \quad 	e_{9}=0,  \quad \quad e_{12}=0, 
\end{equation}
\begin{equation}\label{eqB28}
	e_{14}=0, \quad \quad
	e_{15}=0, \quad \quad	e_{16}=0, \quad \quad e_{17}=0,  \quad \quad	e_{18}=0, \quad \quad e_{19}=0,
\end{equation}
\begin{equation} \label{eqB29}
 \quad \quad e_{20}=0, \quad  \quad e_{22}=0, \quad \quad	e_{23}=0, \quad \quad	e_{24}=0, \quad \quad e_{25}=0.
\end{equation}

	\section{}\label{appC}
	The $\textit{O}(Ca)$ velocity and pressure field enabled the determination of deformed interface of the confining drop upto $\textit{O}(Ca^2)$. The steady state shape parameters  $b_{i}^{*}$ for $i=1,2,3, \cdots, 8$ describing the deformed interface shape obtained as a solution of (\ref{eq15a}) and (\ref{eq34a}) are as follows,
	\begin{subequations}\label{eqC1}
		\begin{equation}\label{B6a}
			\begin{aligned}
				b_{1}^{*}=&\Big(38 \alpha^8+76 \alpha^7+114 \alpha^6+177 \alpha^5+240 \alpha^4+72 \alpha^3-96 \alpha^2-64 \alpha-32+\big(32 \alpha^8+64 \alpha^7\\
				&+96 \alpha^6-72 \alpha^5-240 \alpha^4-72 \alpha^3+96 \alpha^2+64 \alpha+32\big) \lambda \Big) \bigg/ \Big(16 \alpha^8+32 \alpha^7+48 \alpha^6\\
				&+24 \alpha^5-24 \alpha^3-48 \alpha^2-32 \alpha-16+\big(16 \alpha^8+32 \alpha^7+48 \alpha^6-36 \alpha^5-120 \alpha^4-36 \alpha^3\\
				&+48 \alpha^2+32 \alpha+16\big) \lambda\Big),
			\end{aligned}
		\end{equation}
		\begin{equation}\label{B6b}
			b_{2}^{*}=0,
		\end{equation}
		\begin{equation}\label{B6g}
			b_{7}^{*}=0,
		\end{equation}
		\begin{equation}\label{B6h}
			b_{8}^{*}=0,
		\end{equation}
		\begin{align*}
		    b_{3}^{*}=& \Big( -64 \big(42496 \lambda^4+234320 \lambda^3+453369 \lambda^2+372790 \lambda+111245\big) \alpha^{39}-576 \big(42496 \lambda^4\\
			&+234320 \lambda^3+453369 \lambda^2
			+372790 \lambda+111245\big) \alpha^{38}-2880 \big(42496 \lambda^4+234320 \lambda^3\\
			&+453369 \lambda^2+372790 \lambda+111245\big) \alpha^{37}-240 \big(1657344 \lambda^4+9405152 \lambda^3+18663827 \lambda^2\\
			&+15697303 \lambda+4781284\big) \alpha^{36}-720 \big(1232384 \lambda^4+7595296 \lambda^3+16095009 \lambda^2\\
			&+14286389 \lambda+4554292\big) \alpha^{35}-48 \big(26899968 \lambda^4+205280032 \lambda^3+496499983 \lambda^2\\
			&+482745383 \lambda+164625464\big) \alpha^{34}-12 \big(70245888 \lambda^4
			+1085904352 \lambda^3+3355389523 \lambda^2\\
			&+3720841238 \lambda+1381097804\big) \alpha^{33}+12 \big(93618688 \lambda^4-940803488 \lambda^3-4710142331 \lambda^2\\
			&-6242960686 \lambda-2567047508\big) \alpha^{32}+12 \big(347914752 \lambda^4-96241792 \lambda^3-5315592739 \lambda^2\\
			&-9146369924 \lambda-4273714472\big) \alpha^{31}+2 \big(3497867008 \lambda^4+10367618792 \lambda^3\\
			&-23858712939 \lambda^2-69188883320 \lambda-38428447816\big) \alpha^{30}+6 \big(1544198400 \lambda^4\\		
			&+9611837144 \lambda^3+1574087091 \lambda^2-23801599868 \lambda-17269492872\big) \alpha^{29}\\
			&+12 \big(929865600 \lambda^4+8753473292 \lambda^3	+9855619355 \lambda^2-8406219104 \lambda\\
			&-10380658478\big) \alpha^{28}+12 \big(558875520 \lambda^4+11220069708 \lambda^3+22080530147 \lambda^2		\\
			&+432273432 \lambda-10846060982\big) \alpha^{27}-6 \big(3048811776 \lambda^4-15574560512 \lambda^3\\
			&-66605951073 \lambda^2-29258870932 \lambda+18371978616\big) \alpha^{26}-6 \big(10726479104 \lambda^4
		\end{align*}
		\begin{equation}\label{B6c}
			\begin{aligned}	
				\hspace{0.8cm}	
			&+9165819712 \lambda^3-73792113985 \lambda^2-63583124852 \lambda+9677259696\big) \alpha^{25}\\
			&-84 \big(1049385600 \lambda^4+3215195664 \lambda^3-3771419195 \lambda^2-6721337544 \lambda\\
			&-298106350\big) \alpha^{24}-3 \big(8342549120 \lambda^4+142861203832 \lambda^3-689646757 \lambda^2 \\
			&-217902487606 \lambda-42596875264\big) \alpha^{23}+3 \big(38671413120 \lambda^4-130717205368 \lambda^3\\
				&-139317951955 \lambda^2+198853701894 \lambda+77236736384\big) \alpha^{22}+\big(200306967680 \lambda^4\\
				&-96320752352 \lambda^3-776737688127 \lambda^2
			+388538355230 \lambda+317125553944\big) \alpha^{21}\\
			&+3 \big(34025112960 \lambda^4+118254487264 \lambda^3-299701705197 \lambda^2+25892835322 \lambda\\
			&+122748248776\big) \alpha^{20}-24 \big(4253139120 \lambda^4-27531014731 \lambda^3+28247683934 \lambda^2\\
			&+10688929290 \lambda-15658737613\big) \alpha^{19}-8 \big(25038370960 \lambda^4-68547294553 \lambda^3 \\
				&+20182463346 \lambda^2+65726917682 \lambda-42400457435\big) \alpha^{18}-24 \big(4833926640 \lambda^4\\
				&-4192223384 \lambda^3-16549718814 \lambda^2+27061462435 \lambda-11153446877\big) \alpha^{17}\\
				&+12 \big(2085637280 \lambda^4-25002975296 \lambda^3+57199582391 \lambda^2-49168871439 \lambda\\
				&+14886627064\big) \alpha^{16}+12 \big(7345699200 \lambda^4-32668209984 \lambda^3+50313530447 \lambda^2\\
				&-32704282567 \lambda+7713262904\big) \alpha^{15}+12 \big(5363239552 \lambda^4-20189130208 \lambda^3\\
				&+26264372797 \lambda^2-13596391053 \lambda+2157908912\big) \alpha^{14}+12 \big(1524405888 \lambda^4\\
				&-4991092288 \lambda^3+4164522521 \lambda^2+526377395 \lambda-1224213516\big) \alpha^{13}\\
				&-96 (\lambda-1)^2 \big(69859440 \lambda^2-252035181 \lambda+335555261\big) \alpha^{12}-96 (\lambda-1)^2 \\
				& \big(116233200 \lambda^2-391915553 \lambda+357046748\big) \alpha^{11}-192 (\lambda-1)^2 \big(48256200 \lambda^2\\
				&-180037856 \lambda+146283171\big) \alpha^{10}-128 (\lambda-1)^2 \big(54654172 \lambda^2-195640271 \lambda\\	
				&+145862299\big) \alpha^9-384 (\lambda-1)^2 \big(10872336 \lambda^2-35453585 \lambda+24761849\big) \alpha^8\\
				&-6144 (\lambda-1)^3 (182849 \lambda-473626) \alpha^7+6144 (\lambda-1)^3 (137199 \lambda-65305) \alpha^6\\
				&+3072 (\lambda-1)^3 (420312 \lambda-393305) \alpha^5+184320 (\lambda-1)^3 (4814 \lambda-4745) \alpha^4\\
				&+61440 (\lambda-1)^3 (6474 \lambda-6451) \alpha^3+122388480 (\lambda-1)^4 \alpha^2+24477696 (\lambda-1)^4 \alpha\\
				&+2719744 (\lambda-1)^4\Big) \Big/  \Big(  7560 \big(4 \alpha^{15} (\lambda+1)+12 \alpha^{14} (\lambda+1)+24 \alpha^{13} (\lambda+1)\\
				&+40 \alpha^{12} (\lambda+1)+60 \alpha^{11} (\lambda+1)+84 \alpha^{10} (\lambda+1)+112 \alpha^9 (\lambda+1)+63 \alpha^8 (\lambda+2)\\
				&-63 \alpha^7 (\lambda-2)-112 \alpha^6 (\lambda-1)-84 \alpha^5 (\lambda-1)-60 \alpha^4 (\lambda-1)-40 \alpha^3 (\lambda-1)\\
				&-24 \alpha^2 (\lambda-1)-12 \alpha (\lambda-1)-4 \lambda+4\big) \big(4 \alpha^8 (\lambda+1)+8 \alpha^7 (\lambda+1)+12 \alpha^6 (\lambda+1)\\
				&+\alpha^5 (6-9 \lambda)-30 \alpha^4 \lambda-3 \alpha^3 (3 \lambda+2)+12 \alpha^2 (\lambda-1) +8 \alpha (\lambda-1)+4 (\lambda-1)\big)^3\Big),
			\end{aligned} 
		\end{equation}
		\begin{align*}
			b_{4}^{*}=&\Big( 64 \big(5632 \lambda^4+172976 \lambda^3+497595 \lambda^2+495874 \lambda+165623\big) \alpha^{39}+576 \big(5632 \lambda^4\\
			&+172976 \lambda^3+497595 \lambda^2+495874 \lambda+165623\big) \alpha^{38}+2880 \big(5632 \lambda^4+172976 \lambda^3\\
			&+497595 \lambda^2+495874 \lambda+165623\big) \alpha^{37}+3840 \big(13728 \lambda^4+438359 \lambda^3+1289855 \lambda^2\\
			&+1310941 \lambda+445717\big) \alpha^{36}+5760 \big(20416 \lambda^4+727418 \lambda^3+2265585 \lambda^2+2411032 \lambda\\
			&+852449\big) \alpha^{35}+96 \big(1782528 \lambda^4+83603144 \lambda^3+289430855 \lambda^2+331649806 \lambda\\
			&+124039567\big) \alpha^{34}+24 \big(4654848 \lambda^4+498202304 \lambda^3	+2061470930 \lambda^2+2617627771 \lambda
	\end{align*}
		\begin{equation}\label{B6d}
			\begin{aligned}	
				\hspace{0.8cm} 
					&+1049642422\big) \alpha^{33}-24 \big(6203648 \lambda^4-537419200 \lambda^3-3076478428 \lambda^2-4502383661 \lambda\\
			&-1968848734\big) \alpha^{32}-24 \big(23054592 \lambda^4-208943120 \lambda^3-3642490127 \lambda^2-6708406504 \lambda\\
					&-3294389716\big) \alpha^{31} -4 \big(231786368 \lambda^4+5253575056 \lambda^3-15723428046 \lambda^2\\
			&-50222966023 \lambda-29390861480\big) \alpha^{30}-12 \big(102326400 \lambda^4+6005697376 \lambda^3\\
					&+3000736362 \lambda^2-15768544147 \lambda-12756648516\big) \alpha^{29}-12 \big(123235200 \lambda^4\\
			&+11232194216 \lambda^3+18897345407 \lambda^2-6606228617 \lambda-14055906056\big) \alpha^{28}\\
				&-12 \big(74067840 \lambda^4+13220365896 \lambda^3+38692505507 \lambda^2+12825312663 \lambda\\
			&-11908017656\big) \alpha^{27}+6 \big(404059392 \lambda^4-11850873440 \lambda^3-104060118027 \lambda^2\\
			&-79631012494 \lambda+9589050444\big) \alpha^{26}+6 \big(1421581568 \lambda^4+24874046176 \lambda^3\\
			&-94108977451 \lambda^2-132137006798 \lambda-14892109620\big) \alpha^{25}+84 \big(139075200 \lambda^4\\
				&+4679717352 \lambda^3-2861407541 \lambda^2-11492134269 \lambda-3240689992\big) \alpha^{24}\\
				&+3 \big(1105639040 \lambda^4+154943915656 \lambda^3+78101613437 \lambda^2-299534341642 \lambda\\
				&-147708963616\big) \alpha^{23}-3 \big(5125127040 \lambda^4-86830205416 \lambda^3-220888876117 \lambda^2\\
				&+203240984922 \lambda+185541623696\big) \alpha^{22}-\big(26546706560 \lambda^4+116624551312 \lambda^3\\
				&-914317166391 \lambda^2+238068605906 \lambda+596370449488\big) \alpha^{21}-3 \big(4509352320 \lambda^4\\
				&+146137570864 \lambda^3-318169826517 \lambda^2-23001186938 \lambda+192868280896\big) \alpha^{20}\\
				&+12 \big(1127338080 \lambda^4-43675058606 \lambda^3+62416654663 \lambda^2+24437539647 \lambda\\
				&-44306473784\big) \alpha^{19}+4 \big(6636676640 \lambda^4-87155248058 \lambda^3+72308373729 \lambda^2\\
				&+126800078581 \lambda-118589880892\big) \alpha^{18}+12 \big(1281281760 \lambda^4-3961926256 \lambda^3\\
				&-22074945957 \lambda^2+58745421917 \lambda-33989831464\big) \alpha^{17}-12 \big(276409760 \lambda^4\\
				&-15599853704 \lambda^3+53112136847 \lambda^2-64878887247 \lambda+27090194344\big) \alpha^{16}\\
				&-12 \big(973526400 \lambda^4-21330157224 \lambda^3+57007769687 \lambda^2-55629793927 \lambda\\
				&+18978655064\big) \alpha^{15}-12 \big(710790784 \lambda^4-16289253952 \lambda^3+40973881537 \lambda^2\\
				&-36337428729 \lambda+10942010360\big) \alpha^{14}-12 \big(202029696 \lambda^4-8039633056 \lambda^3\\
				&+19841323937 \lambda^2-16417896865 \lambda+4414176288\big) \alpha^{13}+96 (\lambda-1)^2 \big(9258480 \lambda^2\\
				&+217308843 \lambda-6504508\big) \alpha^{12}+96 (\lambda-1)^2 \big(15404400 \lambda^2-206253617 \lambda\\
				&+283193567\big) \alpha^{11}+96 (\lambda-1)^2 \big(12790800 \lambda^2-359007952 \lambda+373731177\big) \alpha^{10}\\
				&+128 (\lambda-1)^2 \big(7243324 \lambda^2-251120789 \lambda+247988215\big) \alpha^9+384 (\lambda-1)^2 \big(1440912 \lambda^2\\
				&-57486503 \lambda+56197841\big) \alpha^8+6144 (\lambda-1)^3 (24233 \lambda-1860499) \alpha^7\\
				&-3072 (\lambda-1)^3 (36366 \lambda+1490029) \alpha^6-12288 (\lambda-1)^3 (13926 \lambda+103519) \alpha^5\\
				&-184320 (\lambda-1)^3 (638 \lambda+997) \alpha^4-61440 (\lambda-1)^3 (858 \lambda-313) \alpha^3\\
				&-16220160 (\lambda-1)^4 \alpha^2-3244032 (\lambda-1)^4 \alpha-360448 (\lambda-1)^4 \Big)\Big/ \Big(  7560 \big(4 \alpha^{15} (\lambda+1)\\
				&+12 \alpha^{14} (\lambda+1)+24 \alpha^{13} (\lambda+1)+40 \alpha^{12} (\lambda+1)+60 \alpha^{11} (\lambda+1)+84 \alpha^{10} (\lambda+1)\\
				&+112 \alpha^9 (\lambda+1)+63 \alpha^8 (\lambda+2)-63 \alpha^7 (\lambda-2)-112 \alpha^6 (\lambda-1)-84 \alpha^5 (\lambda-1)\\	&-60 \alpha^4 (\lambda-1)-40 \alpha^3 (\lambda-1)-24 \alpha^2 (\lambda-1)-12 \alpha (\lambda-1)-4 \lambda+4\big) \big(4 \alpha^8 (\lambda+1)\\	
				&+8 \alpha^7 (\lambda+1)+12 \alpha^6 (\lambda+1)+\alpha^5 (6-9 \lambda)-30 \alpha^4 \lambda-3 \alpha^3 (3 \lambda+2)+12 \alpha^2 (\lambda-1)\\
				&+8 \alpha (\lambda-1)+4 (\lambda-1)\big)^3\Big),
			\end{aligned}
		\end{equation}
		\begin{equation}\label{B6e}
			\begin{aligned}	
			b_{5}^{*}=&\Big(16 \alpha^{32} \big(10496 \lambda^3+34976 \lambda^2+38749 \lambda+14269\big)+96 \alpha^{31} \big(10496 \lambda^3+34976 \lambda^2\\
			&+38749 \lambda+14269\big)+336 \alpha^{30} \big(10496 \lambda^3  +34976 \lambda^2+38749 \lambda+14269\big)\\
			&+12 \alpha^{29} \big(608768 \lambda^3+2170608 \lambda^2+2557567 \lambda+996552\big)+24 \alpha^{28} \big(356864 \lambda^3\\
				&+1615184 \lambda^2+2247841 \lambda+991996\big)+84 \alpha^{27} \big(20992 \lambda^3+447472 \lambda^2+901583 \lambda\\
			&+477218\big)+6 \alpha^{26} \big(-2041472 \lambda^3+3253848 \lambda^2+14838647 \lambda+9667362\big)\\
			&+\alpha^{25} \big(-23574016 \lambda^3-5621216 \lambda^2+87927052 \lambda+72889040\big)-12 \alpha^{24} \big(2721088 \lambda^3\\
			&+3488388 \lambda^2-5786731 \lambda-7135500\big)-28 \alpha^{23} \big(2256640 \lambda^3+4338832 \lambda^2-1358443 \lambda\\
			&-3818374\big)-48 \alpha^{22} \big(2095264 \lambda^3+5193926 \lambda^2+507618 \lambda-2934783\big)\\
				&-12 \alpha^{21} \big(2812928 \lambda^3+29423776 \lambda^2+15765577 \lambda-14225356\big)+12 \alpha^{20} \big(19055488\lambda^3\\
				& -28102800 \lambda^2-40707277 \lambda+13868214\big)+3 \alpha^{19} \big(169124672 \lambda^3-34919120 \lambda^2\\
				&-260418423 \lambda+29358246\big)+2 \alpha^{18} \big(164317504 \lambda^3+243615600 \lambda^2-412825725 \lambda \\
				&-51367954\big)-3 \alpha^{17} \big(132325696 \lambda^3-415223088 \lambda^2+165936613 \lambda+123211954\big)\\
				&-56 \alpha^{16} \big(15144416 \lambda^3-22416176 \lambda^2-2715671 \lambda+9987431\big)-192 \alpha^{15} \big(2067589 \lambda^3\\
				&-304152 \lambda^2-4622303 \lambda+2858866\big)+16 \alpha^{14} \big(20539688 \lambda^3-73281544 \lambda^2+76422679 \lambda\\
				&-23680823\big) +48 \alpha^{13} \big(10570292 \lambda^3-25338188 \lambda^2+18138625 \lambda-3370729\big)\\
				&+12 \alpha^{12} \big(19055488 \lambda^3-37006552 \lambda^2+16767015 \lambda+1184049\big)-24 \alpha^{11} \big(1406464 \lambda^3\\
				&-7162656 \lambda^2+9984645 \lambda-4228453\big)-12 \alpha^{10} \big(8381056 \lambda^3-25744384 \lambda^2+26305175 \lambda\\
				&-8941847\big)-896 \alpha^9 (\lambda-1)^2 (70520 \lambda-87377)-384 \alpha^8 (\lambda-1)^2 (85034 \lambda-139235)\\
				&-512 \alpha^7 (\lambda-1)^2 (46043 \lambda-70165)-768 \alpha^6 (\lambda-1)^2 (15949 \lambda-21409)\\
				&+10752 \alpha^5 (\lambda-1)^2 (164 \lambda-99)+8564736 \alpha^4 (\lambda-1)^3+7305216 \alpha^3 (\lambda-1)^3\\
				&+3526656 \alpha^2 (\lambda-1)^3+1007616 \alpha (\lambda-1)^3+167936 (\lambda-1)^3 \Big)\Big/ \\
				&\Big(  432 (\alpha-1)^3 \big(4 \alpha^7 (\lambda+1)+12 \alpha^6 (\lambda+1)+24 \alpha^5 (\lambda+1)+15 \alpha^4 (\lambda+2)-15 \alpha^3 (\lambda-2)\\
				&-24 \alpha^2 (\lambda-1)-12 \alpha (\lambda-1)-4 \lambda+4\big)^2 \big(4 \alpha^{15} (\lambda+1)+12 \alpha^{14} (\lambda+1)+24 \alpha^{13} (\lambda+1)\\
				&+40 \alpha^{12} (\lambda+1)+60 \alpha^{11} (\lambda+1)+84 \alpha^{10} (\lambda+1)+112 \alpha^9 (\lambda+1)+63 \alpha^8 (\lambda+2)\\
				&-63 \alpha^7 (\lambda-2)-112 \alpha^6 (\lambda-1)-84 \alpha^5 (\lambda-1)-60 \alpha^4 (\lambda-1)-40 \alpha^3 (\lambda-1)\\
				&-24 \alpha^2 (\lambda-1)-12 \alpha (\lambda-1)-4 \lambda+4\big)\Big),
			\end{aligned}
		\end{equation}
		\begin{align*}
		    	b_{6}^{*}=&\Big( 2 \alpha^{18} (3 \lambda+2) (16 \lambda+19)^2+4 \alpha^{17} (3 \lambda+2) (16 \lambda+19)^2+6 \alpha^{16} (3 \lambda+2) (16 \lambda+19)^2\\
				&-3 \alpha^{15} \big(4352 \lambda^3+2304 \lambda^2-7689 \lambda-5092\big)-60 \alpha^{14} \big(512 \lambda^3+464 \lambda^2-539 \lambda-437\big)
				\\&-6 \alpha^{13} \big(2688 \lambda^3+2696 \lambda^2-2781 \lambda-2603\big)+3 \alpha^{12} \big(19488 \lambda^3-9004 \lambda^2-14031 \lambda\\
				&+3547\big)+8 \alpha^{11} \big(14232 \lambda^3-6781 \lambda^2-10059 \lambda+2608\big)-8 \alpha^{10} \big(4044 \lambda^3-7402 \lambda^2\\
				&-303 \lambda+3661\big)-120 \alpha^9 \big(1488 \lambda^3-1439 \lambda^2-711 \lambda+662\big)-12 \alpha^8 \big(2696 \lambda^3\\
				&-903 \lambda^2-1907 \lambda+114\big)+96 \alpha^7 (\lambda-1)^2 (1186 \lambda+799)+96 \alpha^6 (\lambda-1)^2 (609 \lambda+431)\\
				&-48 \alpha^5 (\lambda-1)^2 (336 \lambda-1)-1280 \alpha^4 (\lambda-1)^2 (24 \lambda+1)-128 \alpha^3 (\lambda-1)^2 (102 \lambda+23)\\
					&+4608 \alpha^2 (\lambda-1)^3+3072 \alpha (\lambda-1)^3+1536 (\lambda-1)^3 \Big)
				\Big/	\Big(20 (\alpha-1)^4 \lambda \big(4 \alpha^7 (\lambda+1)
		\end{align*}
		\begin{equation}\label{B6f}
			\begin{aligned}
			\hspace{0.8cm}
			&+12 \alpha^6 (\lambda+1)+24 \alpha^5 (\lambda+1)+15 \alpha^4 (\lambda+2)-15 \alpha^3 (\lambda-2)-24 \alpha^2 (\lambda-1)\\
				&-12 \alpha (\lambda-1)-4 \lambda+4\big)^2 \Big).
			\end{aligned}
		\end{equation}
	\end{subequations}
The corresponding steady state shape parameters for a drop without the encapsulated particle can be obtained by taking the limit $\alpha \rightarrow \infty$, and are as follows:
	\begin{equation}\label{eqC2}
			b_{1}^{*}=\frac{16 \lambda+19}{8 (\lambda+1)},
	\end{equation}
\begin{equation}\label{eqC3}
	b_{2}^{*}=0, \quad \quad b_{7}^{*}=0, \quad \quad b_{8}^{*}=0,
\end{equation}
\begin{equation}\label{eqC4}
	b_{3}^{*}=-\frac{42496 \lambda^3+191824 \lambda^2+261545 \lambda+111245}{30240 (\lambda+1)^3},
\end{equation}
\begin{equation}\label{eqC5}
	b_{4}^{*}=\frac{5632 \lambda^3+167344 \lambda^2+330251 \lambda+165623}{30240 (\lambda+1)^3},
\end{equation}
\begin{equation}\label{eqC6}
b_{5}^{*}=\frac{10496 \lambda^2+24480 \lambda+14269}{1728 (\lambda+1)^2},
\end{equation}
\begin{equation}\label{eqC7}
	b_{6}^{*}=\frac{(3 \lambda+2) (16 \lambda+19)^2}{160 \lambda (\lambda+1)^2}.
\end{equation}

	\bibliographystyle{jfm}
	\bibliography{JFM_template.bib}

\providecommand{\noopsort}[1]{}\providecommand{\singleletter}[1]{#1}%
\begin{thebibliography}{74}
\expandafter\ifx\csname natexlab\endcsname\relax\def\natexlab#1{#1}\fi
\def\au#1{#1} \def\ed#1{#1} \def\yr#1{#1}\def\at#1{#1}\def\jt#1{\textit{#1}}
  \def\bt#1{#1}\def\bvol#1{\textbf{#1}} \def\vol#1{#1} \def\pg#1{#1}
  \def\publ#1{#1}\def\arxiv#1{#1}\def\org#1{#1}\def\st#1{\textit{#1}}

\bibitem[Bai~Chin \& Dae~Han(1979)]{bai1979}
{\sc \au{Bai~Chin, H.} \& \au{Dae~Han, C.}} \yr{1979}  \at{Studies on droplet
  deformation and breakup. {I}. {D}roplet deformation in extensional flow}.
  \jt{J. Rheol.}  \bvol{23}~(5),  \pg{557--590}.

\bibitem[Barnes {\em et~al.\/}(1989)Barnes, Hutton \& Walters]{barnes1989}
{\sc \au{Barnes, H.~A.}, \au{Hutton, J.~F.} \& \au{Walters, K.}} \yr{1989} {\em
  An {I}ntroduction to {R}heology\/}.  \publ{Elsevier}.

\bibitem[Batchelor(1970)]{batchelor1970}
{\sc \au{Batchelor, G.~K.}} \yr{1970}  \at{The stress system in a suspension of
  force-free particles}.  \jt{J. Fluid Mech.}  \bvol{41}~(3),  \pg{545--570}.

\bibitem[Batchelor \& Green(1972)]{batchelor1972}
{\sc \au{Batchelor, G.~K.} \& \au{Green, J.~T.}} \yr{1972}  \at{The
  determination of the bulk stress in a suspension of spherical particles to
  order c2}.  \jt{J. Fluid Mech.}  \bvol{56},  \pg{401--427}.

\bibitem[Bazhlekov {\em et~al.\/}(1995)Bazhlekov, Shopov \&
  Zapryanov]{bazhlekov1995}
{\sc \au{Bazhlekov, I.~B.}, \au{Shopov, P.~J.} \& \au{Zapryanov, Z.~D.}}
  \yr{1995}  \at{Unsteady motion of a type-{A} compound multiphase drop at
  moderate {R}eynolds numbers}.  \jt{J. Colloid Interface Sci.}
  \bvol{169}~(1),  \pg{1--12}.

\bibitem[Bird {\em et~al.\/}(1987)Bird, Armstrong \& Hassager]{bird1987}
{\sc \au{Bird, R.~B.}, \au{Armstrong, R.~C.} \& \au{Hassager, O.}} \yr{1987}
  {\em Dynamics of {P}olymeric {L}quids. {V}ol. 1: {F}luid {M}echanics\/}.
  \publ{John Wiley and Sons Inc}.

\bibitem[Chaithanya \& Thampi(2019)]{chaithu2019}
{\sc \au{Chaithanya, K. V.~S.} \& \au{Thampi, S.~P.}} \yr{2019}  \at{Dynamics
  and stability of a concentric compound particle – a theoretical study}.
  \jt{Soft Matter}  \bvol{15},  \pg{7605--7615}.

\bibitem[Chaithanya \& Thampi(2020)]{chaithu2020}
{\sc \au{Chaithanya, K. V.~S.} \& \au{Thampi, S.~P.}} \yr{2020}
  \at{Deformation dynamics of an active compound particle in an imposed shear
  flow - a theoretical study}.  \jt{J. Phys. D: Appl. Phys.}  \bvol{53}~(31),
  \pg{314001}.

\bibitem[Chen {\em et~al.\/}(2013)Chen, Liu \& Shi]{chen2013}
{\sc \au{Chen, Y.}, \au{Liu, X.} \& \au{Shi, M.}} \yr{2013}  \at{Hydrodynamics
  of double emulsion droplet in shear flow}.  \jt{Appl. Phys. Lett.}
  \bvol{102}~(5),  \pg{051609}.

\bibitem[Chen {\em et~al.\/}(2015{\natexlab{{\em a\/}}})Chen, Liu, Zhang \&
  Zhao]{chen2015b}
{\sc \au{Chen, Y.}, \au{Liu, X.}, \au{Zhang, C.} \& \au{Zhao, Y.}}
  \yr{2015{\natexlab{{\em a\/}}}}  \at{Enhancing and suppressing effects of an
  inner droplet on deformation of a double emulsion droplet under shear}.
  \jt{Lab Chip}  \bvol{15}~(5),  \pg{1255--1261}.

\bibitem[Chen {\em et~al.\/}(2015{\natexlab{{\em b\/}}})Chen, Liu \&
  Zhao]{chen2015a}
{\sc \au{Chen, Y.}, \au{Liu, X.} \& \au{Zhao, Y.}} \yr{2015{\natexlab{{\em
  b\/}}}}  \at{Deformation dynamics of double emulsion droplet under shear}.
  \jt{Appl. Phys. Lett.}  \bvol{106}~(14),  \pg{141601}.

\bibitem[Choe {\em et~al.\/}(2018)Choe, Park, Park \& Lee]{choe2018}
{\sc \au{Choe, G.}, \au{Park, J.}, \au{Park, H.} \& \au{Lee, J.~Y.}} \yr{2018}
  \at{Hydrogel biomaterials for stem cell microencapsulation}.  \jt{Polym.}
  \bvol{10}~(9),  \pg{997}.

\bibitem[Das {\em et~al.\/}(2020)Das, Mandal \& Chakraborty]{das2020}
{\sc \au{Das, S.}, \au{Mandal, S.} \& \au{Chakraborty, S.}} \yr{2020}
  \at{Interfacial viscosity-dictated morpho-dynamics of a compound drop in
  linear flows}.  \jt{Phys. Fluids}  \bvol{32}~(6),  \pg{062006}.

\bibitem[Davis \& Brenner(1981)]{davis1981}
{\sc \au{Davis, A. M.~J.} \& \au{Brenner, H.}} \yr{1981}  \at{Emulsions
  containing a third solid internal phase}.  \jt{J. Engng. Mech. Div. ASCE}
  \bvol{107}~(3),  \pg{609--621}.

\bibitem[Einstein(1906)]{einstein1906}
{\sc \au{Einstein, A.}} \yr{1906}  \at{Eine neue bestimmung der molek{\"u}
  ldimensionen}.  \jt{Ann. Phys.}  \bvol{324}~(2),  \pg{289--306}.

\bibitem[Einstein(1911)]{einstein1911}
{\sc \au{Einstein, A.}} \yr{1911}  \at{Berichtigung zu meiner arbeit : Eine
  neue bestimmung der molek{\"u} ldimensionen}.  \jt{Ann. Phys.}
  \bvol{339}~(3),  \pg{591--592}.

\bibitem[Gasperini {\em et~al.\/}(2014)Gasperini, Mano \& Reis]{gasperini2014}
{\sc \au{Gasperini, L.}, \au{Mano, J.~F.} \& \au{Reis, R.~L.}} \yr{2014}
  \at{Natural polymers for the microencapsulation of cells}.  \jt{J. R. Soc.
  Interface}  \bvol{11}~(100),  \pg{20140817}.

\bibitem[Golemanov {\em et~al.\/}(2008)Golemanov, Tcholakova, Denkov,
  Ananthapadmanabhan \& Lips]{golemanov2008}
{\sc \au{Golemanov, K.}, \au{Tcholakova, S.}, \au{Denkov, N.~D.},
  \au{Ananthapadmanabhan, K.~P.} \& \au{Lips, A.}} \yr{2008}  \at{Breakup of
  bubbles and drops in steadily sheared foams and concentrated emulsions}.
  \jt{Phys. Rev. E}  \bvol{78}~(5),  \pg{051405}.

\bibitem[Grace(1982)]{grace1982}
{\sc \au{Grace, H.~P.}} \yr{1982}  \at{Dispersion phenomena in high viscosity
  immiscible fluid systems and application of static mixers as dispersion
  devices in such systems}.  \jt{Chem. Eng. Commun.}  \bvol{14}~(3-6),
  \pg{225--277}.

\bibitem[Graham(2018)]{graham2018}
{\sc \au{Graham, M.~D.}} \yr{2018} {\em Microhydrodynamics, {B}rownian motion,
  and {C}omplex fluids\/}.  \publ{Cambridge University Press}.

\bibitem[Hamedi \& Babadagli(2010)]{hamedi2010}
{\sc \au{Hamedi, S.~Y.} \& \au{Babadagli, T.}} \yr{2010}  \at{Effects of
  nano-sized metals on viscosity reduction of heavy oil/bitumen during thermal
  applications}.  \bt{In {\em Canadian Unconventional Resources and
  International Petroleum Conference\/}},  \pg{p. 137540}.  \publ{Society of
  Petroleum Engineers}.

\bibitem[Harper(1982)]{harper1982}
{\sc \au{Harper, J.~F.}} \yr{1982}  \at{Surface activity and bubble motion}.
  \jt{Appl. Sci. Res.}  \bvol{38}~(1),  \pg{343--352}.

\bibitem[Hua {\em et~al.\/}(2014)Hua, Shin \& Kim]{hua2014}
{\sc \au{Hua, H.}, \au{Shin, J.} \& \au{Kim, J.}} \yr{2014}  \at{Dynamics of a
  compound droplet in shear flow}.  \jt{Int. J. Heat Fluid Fl.}  \bvol{50},
  \pg{63--71}.

\bibitem[Jansen {\em et~al.\/}(2001)Jansen, Agterof \& Mellema]{jansen2001}
{\sc \au{Jansen, K. M.~B.}, \au{Agterof, W. G.~M.} \& \au{Mellema, J.}}
  \yr{2001}  \at{Droplet breakup in concentrated emulsions}.  \jt{J. Rheol.}
  \bvol{45}~(1),  \pg{227--236}.

\bibitem[Jia {\em et~al.\/}(2020)Jia, Wang \& Fan]{jia2020}
{\sc \au{Jia, L.}, \au{Wang, R.} \& \au{Fan, Y.}} \yr{2020}  \at{Encapsulation
  and release of drug nanoparticles in functional polymeric vesicles}.
  \jt{Soft Matter}  \bvol{16}~(12),  \pg{3088--3095}.

\bibitem[Johnson(1981)]{johnson1981}
{\sc \au{Johnson, R.~E.}} \yr{1981}  \at{Stokes flow past a sphere coated with
  a thin fluid film}.  \jt{J. Fluid Mech.}  \bvol{110},  \pg{217--238}.

\bibitem[Johnson \& Sadhal(1983)]{johnson1983}
{\sc \au{Johnson, R.~E.} \& \au{Sadhal, S.~S.}} \yr{1983}  \at{Stokes flow past
  bubbles and drops partially coated with thin films. {P}art 2. {T}hin films
  with internal circulation-a perturbation solution}.  \jt{J. Fluid Mech.}
  \bvol{132},  \pg{295--318}.

\bibitem[Johnson \& Sadhal(1985)]{johnson1985}
{\sc \au{Johnson, R.~E.} \& \au{Sadhal, S.~S.}} \yr{1985}  \at{Fluid mechanics
  of compound multiphase drops and bubbles}.  \jt{Annu. Rev. Fluid Mech.}
  \bvol{17}~(1),  \pg{289--320}.

\bibitem[Kawano \& Hashimoto(1997)]{kawano1997}
{\sc \au{Kawano, S.} \& \au{Hashimoto, H.}} \yr{1997}  \at{A numerical study on
  motion of a sphere coated with a thin liquid film at intermediate {R}eynolds
  numbers}.  \jt{Trans. ASME, J. Fluids Eng.}  \bvol{119},  \pg{397--403}.

\bibitem[Kim \& Dabiri(2017)]{kim2017}
{\sc \au{Kim, S.} \& \au{Dabiri, S.}} \yr{2017}  \at{Transient dynamics of
  eccentric double emulsion droplets in a simple shear flow}.  \jt{Phys. Rev.
  Fluids}  \bvol{2}~(10),  \pg{104305}.

\bibitem[Leal(2007)]{leal2007}
{\sc \au{Leal, L.~G.}} \yr{2007} {\em Advanced {T}ransport {P}henomena: {F}luid
  {M}echanics and {C}onvective {T}ransport {P}rocesses\/}.  \publ{Cambridge
  University Press}.

\bibitem[Li \& Pozrikidis(1997)]{li1997}
{\sc \au{Li, X.} \& \au{Pozrikidis, C.}} \yr{1997}  \at{The effect of
  surfactants on drop deformation and on the rheology of dilute emulsions in
  {S}tokes flow}.  \jt{J. Fluid Mech.}  \bvol{341},  \pg{165--194}.

\bibitem[Loewenberg(1998)]{loewenberg1998}
{\sc \au{Loewenberg, M.}} \yr{1998}  \at{Numerical simulation of concentrated
  emulsion flows}.  \jt{J. Fluid Eng.}  \bvol{120},  \pg{824--832}.

\bibitem[Loewenberg \& Hinch(1996)]{loewenberg1996}
{\sc \au{Loewenberg, M.} \& \au{Hinch, E.~J.}} \yr{1996}  \at{Numerical
  simulation of a concentrated emulsion in shear flow}.  \jt{J. Fluid Mech.}
  \bvol{321},  \pg{395--419}.

\bibitem[Mandal {\em et~al.\/}(2017)Mandal, Das \& Chakraborty]{mandal2017}
{\sc \au{Mandal, S.}, \au{Das, S.} \& \au{Chakraborty, S.}} \yr{2017}
  \at{Effect of marangoni stress on the bulk rheology of a dilute emulsion of
  surfactant-laden deformable droplets in linear flows}.  \jt{Phys. Rev.
  Fluids}  \bvol{2}~(11),  \pg{113604}.

\bibitem[Mandal {\em et~al.\/}(2016)Mandal, Ghosh \& Chakraborty]{mandal2016}
{\sc \au{Mandal, S.}, \au{Ghosh, U.} \& \au{Chakraborty, S.}} \yr{2016}
  \at{Effect of surfactant on motion and deformation of compound droplets in
  arbitrary unbounded {S}tokes flows}.  \jt{J. Fluid Mech.}  \bvol{803},
  \pg{200--249}.

\bibitem[Mietus {\em et~al.\/}(2002)Mietus, Matar, Lawrence \&
  Briscoe]{mietus2002}
{\sc \au{Mietus, W. G.~P.}, \au{Matar, O.~K.}, \au{Lawrence, C.~J.} \&
  \au{Briscoe, B.~J.}} \yr{2002}  \at{Droplet deformation in confined shear and
  extensional flow}.  \jt{Chem. Eng. Sci.}  \bvol{57}~(7),  \pg{1217--1230}.

\bibitem[Mori(1978)]{mori1978}
{\sc \au{Mori, Y.~H.}} \yr{1978}  \at{Configurations of gas-liquid two-phase
  bubbles in immiscible liquid media}.  \jt{Int. J. Multiph. Flow}
  \bvol{4}~(4),  \pg{383--396}.

\bibitem[Mulligan \& Rothstein(2011)]{mulligan2011}
{\sc \au{Mulligan, M.~K.} \& \au{Rothstein, J.~P.}} \yr{2011}  \at{Deformation
  and breakup of micro-and nanoparticle stabilized droplets in microfluidic
  extensional flows}.  \jt{Langmuir}  \bvol{27}~(16),  \pg{9760--9768}.

\bibitem[Oldroyd(1953)]{oldroyd1953}
{\sc \au{Oldroyd, J.~G.}} \yr{1953}  \at{The elastic and viscous properties of
  emulsions and suspensions}.  \jt{Proc. R. Soc. Lond. A}  \bvol{218}~(1132),
  \pg{122--132}.

\bibitem[Pal(1996{\natexlab{{\em a\/}}})]{pal1996a}
{\sc \au{Pal, R.}} \yr{1996{\natexlab{{\em a\/}}}}  \at{Effect of droplet size
  on the rheology of emulsions}.  \jt{AIChE J.}  \bvol{42}~(11),
  \pg{3181--3190}.

\bibitem[Pal(1996{\natexlab{{\em b\/}}})]{pal1996b}
{\sc \au{Pal, R.}} \yr{1996{\natexlab{{\em b\/}}}}  \at{Multiple o/w/o emulsion
  rheology}.  \jt{Langmuir}  \bvol{12}~(9),  \pg{2220--2225}.

\bibitem[Pal(2007)]{pal2007}
{\sc \au{Pal, R.}} \yr{2007}  \at{Rheology of double emulsions}.  \jt{J.
  Colloid Interface Sci.}  \bvol{307}~(2),  \pg{509--515}.

\bibitem[Pal(2011)]{pal2011}
{\sc \au{Pal, R.}} \yr{2011}  \at{Rheology of simple and multiple emulsions}.
  \jt{Curr. Opin. Colloid Interface Sci.}  \bvol{16}~(1),  \pg{41--60}.

\bibitem[Patlazhan {\em et~al.\/}(2015)Patlazhan, Vagner \&
  Kravchenko]{Patlazhan2015}
{\sc \au{Patlazhan, S.}, \au{Vagner, S.} \& \au{Kravchenko, I.}} \yr{2015}
  \at{Steady-state deformation behavior of confined composite droplets under
  shear flow}.  \jt{Phys. Rev. E}  \bvol{91}~(6),  \pg{063002}.

\bibitem[Qu \& Wang(2012)]{qu2012}
{\sc \au{Qu, X.} \& \au{Wang, Y.}} \yr{2012}  \at{Dynamics of concentric and
  eccentric compound droplets suspended in extensional flows}.  \jt{Phy.
  Fluids}  \bvol{24}~(12),  \pg{123302}.

\bibitem[Ramachandran \& Leal(2012)]{arun2012}
{\sc \au{Ramachandran, A.} \& \au{Leal, L.~G.}} \yr{2012}  \at{The effect of
  interfacial slip on the rheology of a dilute emulsion of drops for small
  capillary numbers}.  \jt{J. Rheol.}  \bvol{56}~(6),  \pg{1555--1587}.

\bibitem[Reigh {\em et~al.\/}(2017)Reigh, Zhu, Gallaire \& Lauga]{reigh2017}
{\sc \au{Reigh, S.~Y.}, \au{Zhu, L.}, \au{Gallaire, F.} \& \au{Lauga, E.}}
  \yr{2017}  \at{Swimming with a cage: low-{R}eynolds-number locomotion inside
  a droplet}.  \jt{Soft Matter}  \bvol{13}~(17),  \pg{3161--3173}.

\bibitem[Rushton \& Davies(1983)]{rushton1983}
{\sc \au{Rushton, E.} \& \au{Davies, G.~A.}} \yr{1983}  \at{Settling of
  encapsulated droplets at low {R}eynolds numbers}.  \jt{Int. J. Multiph. Flow}
   \bvol{9}~(3),  \pg{337--342}.

\bibitem[Russel {\em et~al.\/}(1989)Russel, Saville \& Schowalter]{russel1989}
{\sc \au{Russel, W.~B.}, \au{Saville, D.~A.} \& \au{Schowalter, W.~R.}}
  \yr{1989} {\em Colloidal {D}ispersions\/}.  \publ{Cambridge university
  press}.

\bibitem[Sadhal {\em et~al.\/}(1997)Sadhal, Ayyaswamy \& Chung]{sadhal1997}
{\sc \au{Sadhal, S.~S.}, \au{Ayyaswamy, P.~S.} \& \au{Chung, J.~N.}} \yr{1997}
  {\em Transport {P}henomena with {D}rops and {B}ubbles\/}.  \publ{Springer
  Science \& Business Media}.

\bibitem[Sadhal \& Johnson(1983)]{sadhal1983}
{\sc \au{Sadhal, S.~S.} \& \au{Johnson, R.~E.}} \yr{1983}  \at{Stokes flow past
  bubbles and drops partially coated with thin films. {P}art 1. {S}tagnant cap
  of surfactant film-exact solution}.  \jt{J. Fluid Mech.}  \bvol{126},
  \pg{237--250}.

\bibitem[Sadhal \& Oguz(1985)]{sadhal1985}
{\sc \au{Sadhal, S.~S.} \& \au{Oguz, H.~N.}} \yr{1985}  \at{Stokes flow past
  compound multiphase drops: the case of completely engulfed drops/bubbles}.
  \jt{J. Fluid Mech.}  \bvol{160},  \pg{511--529}.

\bibitem[Sagis \& {\"O}ttinger(2013)]{sagis2013}
{\sc \au{Sagis, L. M.~C.} \& \au{{\"O}ttinger, H.~C.}} \yr{2013}  \at{Dynamics
  of multiphase systems with complex microstructure. {I}. {D}evelopment of the
  governing equations through nonequilibrium thermodynamics}.  \jt{Phys. Rev.
  E}  \bvol{88}~(2),  \pg{022149}.

\bibitem[Santra {\em et~al.\/}(2020{\natexlab{{\em a\/}}})Santra, Das \&
  Chakraborty]{santra2020}
{\sc \au{Santra, S.}, \au{Das, S.} \& \au{Chakraborty, S.}}
  \yr{2020{\natexlab{{\em a\/}}}}  \at{Electrically modulated dynamics of a
  compound droplet in a confined microfluidic environment}.  \jt{J. Fluid
  Mech.}  \bvol{882},  \pg{A23}.

\bibitem[Santra {\em et~al.\/}(2020{\natexlab{{\em b\/}}})Santra, Panigrahi,
  Das \& Chakraborty]{santra2020a}
{\sc \au{Santra, S.}, \au{Panigrahi, D.~P.}, \au{Das, S.} \& \au{Chakraborty,
  S.}} \yr{2020{\natexlab{{\em b\/}}}}  \at{Shape evolution of compound droplet
  in combined presence of electric field and extensional flow}.  \jt{Phys. Rev.
  Fluids}  \bvol{5}~(6),  \pg{063602}.

\bibitem[Smith {\em et~al.\/}(2004)Smith, Ottino \& Cruz]{smith2004}
{\sc \au{Smith, K.~A.}, \au{Ottino, J.~M.} \& \au{Cruz, M.~O.}} \yr{2004}
  \at{Encapsulated drop breakup in shear flow}.  \jt{Phys. Rev. Lett.}
  \bvol{93}~(20),  \pg{204501}.

\bibitem[Somerville {\em et~al.\/}(2020)Somerville, Law, Rey, Vogel, Archer \&
  Buzza]{somerville2020}
{\sc \au{Somerville, W. R.~C.}, \au{Law, A.~D.}, \au{Rey, M.}, \au{Vogel, N.},
  \au{Archer, A.~J.} \& \au{Buzza, D. M.~A.}} \yr{2020}  \at{Pattern formation
  in two-dimensional hard-core/soft-shell systems with variable soft shell
  profiles}.  \jt{Soft Matter}  \bvol{16}~(14),  \pg{3564--3573}.

\bibitem[Song {\em et~al.\/}(2010)Song, Xu \& Yang]{song2010}
{\sc \au{Song, Y.}, \au{Xu, J.} \& \au{Yang, Y.}} \yr{2010}  \at{Stokes flow
  past a compound drop in a circular tube}.  \jt{Phys. Fluids}  \bvol{22}~(7),
  \pg{072003}.

\bibitem[Soni {\em et~al.\/}(2018)Soni, Thaokar \& Juvekar]{soni2018}
{\sc \au{Soni, P.}, \au{Thaokar, R.~M.} \& \au{Juvekar, V.~A.}} \yr{2018}
  \at{Electrohydrodynamics of a concentric compound drop in an {AC} electric
  field}.  \jt{Phys. Fluids}  \bvol{30}~(3),  \pg{032102}.

\bibitem[Srinivasan \& Shah(2014)]{srinivasan2014}
{\sc \au{Srinivasan, A.} \& \au{Shah, S.~N.}} \yr{2014}  \at{Surfactant-based
  fluids containing copper-oxide nanoparticles for heavy oil viscosity
  reduction}.  \bt{In {\em SPE Annual Technical Conference and Exhibition\/}},
  \pg{p. 170800}.  \publ{Society of Petroleum Engineers}.

\bibitem[Stone(1994)]{stone1994}
{\sc \au{Stone, H.~A.}} \yr{1994}  \at{Dynamics of drop deformation and breakup
  in viscous fluids}.  \jt{Annu. Rev. Fluid Mech.}  \bvol{26}~(1),
  \pg{65--102}.

\bibitem[Stone \& Leal(1990)]{stone1990a}
{\sc \au{Stone, H.~A.} \& \au{Leal, L.~G.}} \yr{1990}  \at{Breakup of
  concentric double emulsion droplets in linear flows}.  \jt{J. Fluid Mech.}
  \bvol{211},  \pg{123--156}.

\bibitem[Tadros(2011)]{tadros2011}
{\sc \au{Tadros, T.~F.}} \yr{2011} {\em Rheology of {D}ispersions: {P}rinciples
  and {A}pplications\/}.  \publ{John Wiley \& Sons}.

\bibitem[Taylor(1932)]{taylor1932}
{\sc \au{Taylor, G.~I.}} \yr{1932}  \at{The viscosity of a fluid containing
  small drops of another fluid}.  \jt{Proc. R. Soc. Lond. A}  \bvol{138}~(834),
   \pg{41--48}.

\bibitem[Taylor(1934)]{taylor1934}
{\sc \au{Taylor, G.~I.}} \yr{1934}  \at{The formation of emulsions in definable
  fields of flow}.  \jt{Proc. R. Soc. Lond. A}  \bvol{146}~(858),
  \pg{501--523}.

\bibitem[Vlahovska {\em et~al.\/}(2009)Vlahovska, B{\l}awzdziewicz \&
  Loewenberg]{vlahovska2009}
{\sc \au{Vlahovska, P.~M.}, \au{B{\l}awzdziewicz, J.} \& \au{Loewenberg, M.}}
  \yr{2009}  \at{Small-deformation theory for a surfactant-covered drop in
  linear flows}.  \jt{J. Fluid Mech.}  \bvol{624},  \pg{293--337}.

\bibitem[Wen {\em et~al.\/}(2015)Wen, Yu, Zhu, Jiang \& Qin]{wen2015}
{\sc \au{Wen, Hui}, \au{Yu, Yue}, \au{Zhu, Guoli}, \au{Jiang, Lei} \& \au{Qin,
  Jianhua}} \yr{2015}  \at{A droplet microchip with substance exchange
  capability for the developmental study of {C}. elegans}.  \jt{Lab Chip}
  \bvol{15}~(8),  \pg{1905--1911}.

\bibitem[Wisdom {\em et~al.\/}(2013)Wisdom, Watson, Qu, Liu, Watson \&
  Chen]{wisdom2013}
{\sc \au{Wisdom, K.~M.}, \au{Watson, J.~A.}, \au{Qu, X.}, \au{Liu, F.},
  \au{Watson, G.~S.} \& \au{Chen, C.~H.}} \yr{2013}  \at{Self-cleaning of
  superhydrophobic surfaces by self-propelled jumping condensate}.  \jt{Proc.
  Natl. Acad. Sci.}  \bvol{110},  \pg{7992--7997}.

\bibitem[Xu {\em et~al.\/}(2013)Xu, Liu, Zhao \& Li]{xu2013}
{\sc \au{Xu, M.}, \au{Liu, H.}, \au{Zhao, H.} \& \au{Li, W.}} \yr{2013}
  \at{How to decrease the viscosity of suspension with the second fluid and
  nanoparticles?}  \jt{Sci. Rep.}  \bvol{3},  \pg{3137}.

\bibitem[Zhang {\em et~al.\/}(2015)Zhang, Zhao, Li, Xu \& Liu]{zhang2015}
{\sc \au{Zhang, J.}, \au{Zhao, H.}, \au{Li, W.}, \au{Xu, M.} \& \au{Liu, H.}}
  \yr{2015}  \at{Multiple effects of the second fluid on suspension viscosity}.
   \jt{Sci. Rep.}  \bvol{5},  \pg{16058}.

\bibitem[Zhang {\em et~al.\/}(2017)Zhang, Wang, Zhou, Zhang, Gong, Gou, Xu \&
  Ma]{zhang2017}
{\sc \au{Zhang, Q.}, \au{Wang, T.}, \au{Zhou, Q.}, \au{Zhang, P.}, \au{Gong,
  Y.}, \au{Gou, H.}, \au{Xu, J.} \& \au{Ma, B.}} \yr{2017}  \at{Development of
  a facile droplet-based single-cell isolation platform for cultivation and
  genomic analysis in microorganisms}.  \jt{Sci. Rep.}  \bvol{7},  \pg{41192}.

\bibitem[Zhao \& Macosko(2002)]{zhao2002}
{\sc \au{Zhao, R.} \& \au{Macosko, C.~W.}} \yr{2002}  \at{Slip at
  polymer--polymer interfaces: {R}heological measurements on coextruded
  multilayers}.  \jt{J. Rheol.}  \bvol{46}~(1),  \pg{145--167}.

\bibitem[Zhou {\em et~al.\/}(2006)Zhou, Yue \& Feng]{zhou2006}
{\sc \au{Zhou, C.}, \au{Yue, P.} \& \au{Feng, J.~J.}} \yr{2006}  \at{Formation
  of simple and compound drops in microfluidic devices}.  \jt{Phys. Fluids}
  \bvol{18}~(9),  \pg{092105}.

\end{thebibliography}
	
\end{document}